\documentclass[twocolumn]{aastex7}
\usepackage{amsmath,amstext}
\usepackage[T1]{fontenc}
\usepackage{subfigure}

\usepackage{sidecap}
\usepackage{enumitem}
\usepackage{appendix}
\usepackage[export]{adjustbox}
\usepackage{caption}
\usepackage{subcaption}
\usepackage{float}
\usepackage{accents}
\usepackage{xfrac}
\newcommand{\myvec}[1]{\accentset{\rightharpoonup}{#1}}

\begin{document}

\begin{flushright}
        \includegraphics[width=6cm]{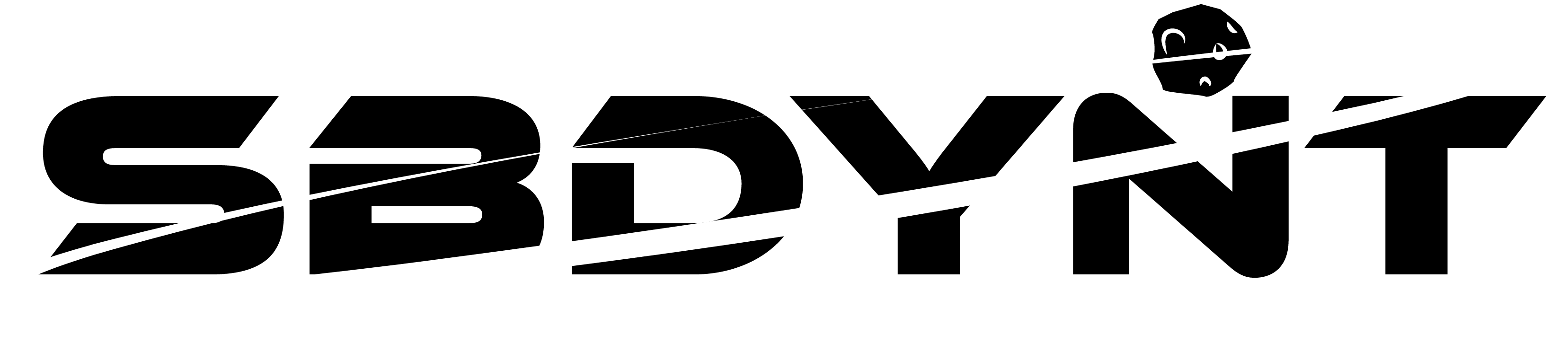} \end{flushright}
    
    \vspace*{-0.5cm} 

\title{Small Body Dynamics with SBDynT: Proper Elements and Chaos Analysis} 

\author[0000-0003-4051-2003]{Dallin Spencer}
\affiliation{Brigham Young University, Department of Physics and Astronomy, N283 ESC, Provo, UT 84602, USA}
\email{djspenc@byu.edu}

\author[0000-0001-8736-236X]{Kat Volk}
\affiliation{Planetary Science Institute, 1700 E Fort Lowell Rd STE 106, Tucson, AZ 85719, USA}
\email{kvolk@psi.edu}

\author[0000-0003-1080-9770]{Darin Ragozzine}
\affiliation{Brigham Young University, Department of Physics and Astronomy, N283 ESC, Provo, UT 84602, USA}
\email{darin_ragozzine@byu.edu}

\author[0000-0002-1226-3305]{Renu Malhotra}
\affiliation{University of Arizona, Lunar and Planetary Laboratory, 1629 E University Blvd, Tucson, AZ 85721, USA}
\email{malhotra@arizona.edu}

\shorttitle{}

\begin{abstract}
The Small Body Dynamics Tool (SBDynT) is software written for the community of Solar System small body researchers to perform dynamical classification, characterization, and investigation. SBDynT provides advanced simulation analysis capabilities that make it straightforward to determine mean motion resonance occupation, proper orbital elements, and a variety of {stability} indicators. These calculations can be performed for small bodies that are known, newly discovered, or simulated; observational uncertainties can be incorporated through the use of dynamical clones. In this paper, we describe the methods for producing proper orbital elements and {stability} indicators, which serve as essential tools for characterizing dynamical stability and long-term evolution. Through extensive validation, we demonstrate that this code offers a robust open-source framework for investigating the dynamics of Solar System small bodies with high accuracy. We also aim for computational efficiency allowing SBDynT to provide dynamical information for the several-fold increases in small bodies expected in the LSST era.  

\end{abstract}


\section{Introduction}\label{s:intro}

\setcounter{footnote}{0} 

In the study of solar system small bodies (SSSBs), several methods can be used to explore their orbital dynamics. 
Modern methods include high-accuracy and fast ``n-body integration'' which use well-understood gravitational interactions to propagate orbits over billions of years. 
Several integrators are used for the study of solar system dynamics and evolution, such as \textsc{mercury} \citep{mercury}, \textsc{SWIFT} \citep{swift}, and \textsc{{\sc rebound}} \citep{rebound}. 
However, a simulation of an object's orbit  typically requires additional interpretation from a detailed dynamical analysis to extract meaningful scientific insights about its behavior and evolution. 
Although open source simulators have been available for some time, generally accessible codes capable of performing a wide range of dynamical analyses for SSSBs have largely been lacking.
With the upcoming Vera C. Rubin Observatory's Legacy Survey of Space and Time (LSST) expected to discover over 37,000 new trans-Neptunian objects (TNOs) and approximately 5 million asteroids \citep{eggl2019,Kurlander:2025}, the SSSB community faces an unprecedented influx of new data. 
Precise homogeneous astrometry over the 10-year survey will lead to significantly improved orbit quality in addition to quantity. 
This prompts the development of standardized, robust software capable of performing efficient and reliable dynamical analyses, eliminating the need for individual researchers to write custom code for each study \citep{Hsieh:2019}. 
We aim to provide the Small Body Dynamics Tool (SBDynT) \citep{Spencer:2024,Volk:2024}, an efficient, user-friendly, open-source software tool for the analysis and study of SSSB dynamical histories using {\sc rebound} \footnote{\url{https://github.com/small-body-dynamics/SBDynT}}. 

{An early version of SBDynT was discussed and presented in \cite{Volk:2025}. 
After previous Alpha and Beta builds were recently completed,  this paper introduces the major release of SBDynT version 1.0,  as available on the GitHub repository, containing all of the following features:}
\begin{itemize}
    \item Easily initialized, high quality Solar System simulations producing past and future trajectories for known SSSBs, including clones sampling orbit uncertainties
    \item Resonance occupation and identification with machine learning \citep{Volk:2025, Volk:2025dda}
    \item Proper orbital elements
    \item {Stability}/Chaos indicators.
\end{itemize}

In this paper, we will discuss the methods and results for computing proper orbital elements (Section~\ref{sec:proper_elements}) and {stability} indicators (Section~\ref{sec:chaos}) over varying timescales for small body orbits using SBDynT. 
The process of identifying resonance occupation with SBDynT is discussed separately \citep{Volk:2025}. 
We will compare our generated proper elements to published catalogs, and discuss any outlying regions of proper element space with differences between the catalogs and SBDynT (Section~\ref{sec:results}). 
We will also demonstrate the versatility of the {stability} indicators provided by SBDynT by building and comparing generated stability maps in Classical TNO space (Section~\ref{sec:chaos_results}).

\section{Proper Orbital Elements}\label{sec:proper_elements}

SSSB trajectories are almost always well-described as slightly perturbed heliocentric orbits.
\footnote{SBDynT focuses on long-term (thousands to billions of years) gravitational trajectories of SSSBs approximated as massless points orbiting the Sun. SBDynT is not directly designed for studying natural or artificial satellites, moons, or SSSB binaries. Very close encounters or collisions between objects (e.g., when SSSBs masses may perturb orbits \cite{Bernstein:2025} and \cite{Fuentes-Munoz:2025}) could lead to long-term inaccuracies. It is not natively designed to study the motion of dust or other objects which have non-negligible non-gravitational forces important for the motion (such as precise orbits of outgassing comets). When precise (ephemeris quality) or short-term orbits (e.g., for interstellar objects) are required, other resources are more appropriate such as ASSIST \citep{Holman:2023}. While SBDynT does not currently handle these more specific science cases with high accuracy and fidelity, it could easily be used as a first analysis to identify objects in
 need of more customized modeling.}

Osculating heliocentric orbital elements describe the instantaneous Keplerian orbit that a theoretical body would follow if (suddenly) only the influence of the Sun's gravity dictated the orbit of that object.
While orbits may be described using various orbital elements and angles, the osculating elements we consider here are the semi-major axis ($a$), eccentricity ($e$), inclination ($I$), argument of periapse ($\omega$), longitude of the ascending node ($\Omega$), and the mean anomaly ($M$). 
Together, they represent a snapshot of the orbit's shape and orientation at a specific epoch; together with the mass of the Sun, they represent a useful coordinate transformation from the SSSBs heliocentric position and velocity since all but $M$ are conserved in the unperturbed Keplerian case.
Over time, an object's osculating orbit evolves primarily due to additional forces such as gravitational interactions with other bodies in the Solar system with the strongest effect usually on the orientation angles $\omega$ and $\Omega$, slower effects on $e$ and $I$, and often minimal effects on $a$. Even the relatively rapid changes in orientation angles take thousands of orbital periods and are thus called ``secular'' in the sense of long period cycles of time. 

Tracking an object's orbit into the deep past or future allows for a variety of unique scientific insights.
For example, statistically significant clusters of objects can indicate a past non-gravitational interaction. 
A common example is ejecta from an asteroid collision; at the time of the collision large numbers of present-day SSSBs had very similar orbital elements due to have nearly identical positions and very similar velocities.
Backwards integration of these asteroids shows a tight clustering in $a$, $e$, $I$, $\omega$, and $\Omega$; conversely, identifying SSSBs where such elements cluster at a specific point in the past can be used to identify and characterize such clusters known as collisional families (e.g., \cite{Hirayama:1918}, \cite{Nesvorny:2006},  \cite{Marcus:2011}, \cite{Nesvorny:2024}). Such families can be potentially primordial \cite{Ferrone:2023}, older \cite{Carruba:2016}, young \cite{Knezevic:2006}, or ``very young'' \cite{Nesvorny:2026} and can include thousands of objects or even just two in a ``pair.'' 

Advances in computational abilities and tools (including SBDynT) enable direct searches in backwards integrations for clusters, families, pairs, or other unusual orbital dynamics. 
While this may become more common in the future, at present the state-of-the-art is to use a dynamically meaningful proxy: clusters in proper orbital elements. 
As discussed in more detail below, proper orbital elements approximately remove the forcing effects of the planets and are theoretically constant; in the real solar system, they are often nearly conserved away from mean-motion resonances (e.g., commensurabilities between the orbital period of an SSSB and the orbital period of a planet). 

The first asteroid families were discovered as clusters in, effectively, analytical proper orbital elements \citep{Hirayama:1918}.
It can be shown the the small differences in orbital elements at the time of family formation are amplified by planetary perturbations which cause the osculating elements to significantly diverge on secular timescales.
Thus, removing the forced terms results in families that remain clustered in proper elements for much longer timescales. 

Determining proper elements and using them to identify small body families is a continuously active area of research in the SSSB community \citep[e.g.,][]{Nesvorny:2024}. 
While it may be possible to identify families more accurately through direct inspection of the n-body integrations \citep[e.g.,][]{Marcus:2011}, identifying clusters in proper elements is likely more straightforward computationally and is the current state-of-the-art.
The goal for the accuracy and precision of proper element identification is driven primarily by the desire to identify and characterize asteroid families. 
Furthermore, proper elements are, themselves, somewhat of an approximation to the actual historical orbits of SSSBs. 
Thus, different algorithmic choices are possible in the calculation of proper elements and a SBDynT's implementation of proper element calculations can be considered successful if the precision is comparable to state-of-the-art catalogs and if these catalogs (which use somewhat different methods) are generally consistent within these uncertainties. We explicitly explore these criteria in Section \ref{sec:results}. 

In any case, the accurate identification of clusters and families provides significant insight into formation and evolution of SSSBs, particularly through collisions. \cite{Zappala:1990}
Families also provide insights into surface processing, non-gravitational dynamical evolution (e.g., from Yarkovsky and YORP effects), and the number density of SSSBs as a function of time. \cite{Bottke:2006, Vokrouhlicky:2006, Bolin:2017}
Proper elements can also be used to understand secular resonances which sculpt dynamical pathways throughout the solar system. \cite{O'Brien:2007}
Overall, the ability to mathematically suppress the already known perturbations by planetary forcing helps uncover more interesting phenomena. 

We present an update to the SBDynT code aimed to make proper element calculations open source, easy to use, and computationally efficient. 
We begin by reviewing the theory of proper elements in Section \ref{sec:properelementtheory}. 
We then explain how we use {\sc REBOUND} to compute SSSB orbits sufficient for accurate proper element calculation (Section \ref{sec:sim_ic}). The calculation of ``synthetic'' proper elements using Fourier methods is discussed in detail in Section \ref{sec:calc_proper}. 
Proper orbital elements are often assigned an ``uncertainty'' based on how well they are conserved quantities in time which we also follow (Section \ref{sec:proper_uncertainties}). 
Finally, we note a mostly theoretical distinction between proper elements and ``mean elements'' that is relevant for many SSSBs in Section \ref{sec:mean_elements}.

\subsection{Proper Element Theory}
\label{sec:properelementtheory}

The time evolution of orbital elements can be approximated as a system of coupled normal modes in which proper elements are something like the initial conditions. A useful starting point is to transform to different set of variables. While the orbital elements ($a$, $e$, $I$, $\omega$, $\Omega$, $M$) are useful and intuitive, singularities are sometimes introduced when modeling perturbations on orbital evolution. For example, the inclination of an object evolving to $0^{\circ}$ can cause $\Omega$ to instantaneously change by $180^{\circ}$ to keep the inclination positive.
For this reason, an additional set of orbital elements, sometimes referred to as the ``equinoctial elements'', are often considered in perturbation theory.
We use the modified equinoctial elements \citep{Arsenault:1970}, defined by: 
\begin{equation}\label{hvec}
\begin{split}
    h(t) &= e(t)*\sin\bigl(\varpi(t)\bigr) \\
    k(t) &= e(t)*\cos\bigl(\varpi(t)\bigr)
\end{split}
\end{equation}
and
\begin{equation}\label{pvec}
\begin{split}
    p(t) &= \sin\bigl(I(t)\bigr)*\sin\bigl(\Omega(t)\bigr) \\
    q(t) &= \sin\bigl(I(t)\bigr)*\cos\bigl(\Omega(t)\bigr)
\end{split}
\end{equation}
where $\varpi$ is the longitude of periapse given by $\varpi = \omega + \Omega$. 

The $h$ and $k$  in Equation \ref{hvec} are two-component vectors generated from $e$ and $\varpi$ that describe the orientation of the orbital ellipse, while $p$ and $q$ in Equation \ref{pvec}, generated from $I$ and $\Omega$, similarly describe the orientation of the orbit plane. 
These two-component vectors jointly represent the orientation of the orbital ellipse and the orbital plane.
In addition, singularities that occur with the original orbital elements at $e = 0$, or $i = 0^\circ,90^\circ$ are avoided with the equinoctial elements, preserving stability in computations of the orbital evolution over time.
As two-component vectors, these equinoctial elements can jointly represent the full time evolution of the orbital shape by joining them into complex vectors:
\begin{equation}\label{complex_ei}
\begin{split}
    \myvec{e} &= k + \mathbf{i}h \\
    \myvec{I} &= q + \mathbf{i}p
\end{split}
\end{equation}
which can be represented in simplified exponential form as:
\begin{equation}
    \label{eulerf}
\begin{split}
    \myvec{e} &= e(t)\times \text{e}^{\mathbf{i}\, \varpi(t) }\\ 
    \myvec{I} &= \sin\bigl(I(t)\bigr)\times \text{e}^{\mathbf{i}\, \Omega(t) }
\end{split}
\end{equation}
where $\mathrm{e}$ and $\mathbf{i}$ are Euler's constant and the imaginary number, respectively. We will refer to $\myvec{e}$ and $\myvec{I}$ throughout the text as the eccentricity and inclination vectors; these vectors capture variations in the orbit, such as nodal or apsidal precession and changes in eccentricity and inclination.

In orbital dynamics theory, deviations from Keplerian motion are often described using the ``disturbing function,'' similar to the  perturbation of a Hamiltonian. 
The disturbing function is found by first producing the familiar 2-body central force equation of gravity for the body whose motion is being studied, and then adding additional forces to the system.
The disturbing potential can then be expanded out (e.g., using Legendre polynomials) to high-order in small parameters relevant to a specific system \citep[e.g.,][]{mandd1999}. 
For solar system bodies not in mean motion resonances, the main non-trivial perturbations can be determined by expanding the disturbing function to second-order in $e$ and $I$ and first-order in the planet/Sun mass ratio and ignoring the fast changing terms involving the mean anomaly, which tend to average out.
The result is an analytical expression for long-term ``secular'' changes to the orbital elements and is known as (second-order) Laplace-Lagrange secular theory.

In this theory, the time evolution of the orbital shape and plane of a small body can  be represented by first-order differential equations that are coupled between the planets and the small body. 
The resulting solution for the evolution of $\myvec{e}$ and $\myvec{I}$ is a linear combination of eigenmodes that depend on the masses and orbital properties of the $N$ included bodies:
\begin{equation}\label{eq:lltheory}
\begin{split}
\myvec{e_j} & = \sum_{i=1}^{N} e_{ji} \times \text{e}^{\mathbf{i}(g_i t + \beta_i)}\\
\myvec{I_j} & = \sum_{i=1}^{N} I_{ji} \times \text{e}^{\mathbf{i}(s_i t + \gamma_i)}
\end{split}
\end{equation}
where $N$ is the number of planets, the $g_i$ and $s_i$ are secular planetary frequencies and where the $\beta_i$, and $\gamma_i$ are phases (see \cite{mandd1999} for full details).
What is significant about this formulation is that the secular evolution of the orbital plane and orientation are uncoupled from each other  (At higher orders, these are no longer true, but the coupling effect is smaller.) Furthermore, there is no secular evolution of the semi-major axis at all.

Using the nature of linear combinations, the individual terms for each body in the system can be isolated, and it is possible to produce a singular term corresponding to the properties of just the small body's orbit.
The isolated term in the $\myvec{e}$ and $\myvec{I}$ evolution  that depends on the small body's orbital properties is called the ``free'' or proper term, and all additional terms which depend on the planets are the ``forced'' terms.
While the forced terms contribute to the evolution of the osculating elements over time, and their amplitude depends on the object’s initial free parameters (e.g., distance to a perturber), their temporal variation is governed solely by the evolution of the planetary perturber.
In contrast, the free or proper term represents the body’s fundamental orbital parameters as set after its most recent non-secular perturbation. The proper terms are fixed in time; the only temporal variation of the orbit in this  approximation  is the constant proper precession of the angles $\varpi$ and $\Omega$. 
As such, we can write the proper orbital angles as $\varpi_p(t) = gt + \varpi_{p,0}$ and $\Omega_p(t) = st + \Omega_{p,0}$ where $g$ and $s$ are the proper precession rates, which are typically separate from the planetary secular rates $g_i$ and $s_i$; the amplitudes of $\myvec{e_p}$ and $\myvec{I_p}$ remain fixed over time.

In the Solar System, the secular timescales that affect a small body's osculating $\myvec{e}$ and $\myvec{I}$ are primarily determined by the $g_i$ and $s_i$ planetary secular frequencies (Equation~\ref{eq:lltheory}) and typically span from tens of thousands to millions of years.
The shortest of these significant secular periods is associated with Saturn’s dominant nodal precession frequency, $s_6$, which has a period of $\sim$50,000 years.
Some shorter-period variations may arise from linear combinations of secular frequencies, so dominant secular evolution generally is defined on timescales of $t_{sec} > 10,000$ years.
In the case of small-body orbital evolution, these secular frequencies have the the largest associated amplitudes in or near what we call secular resonances.

Secular resonances differ from mean-motion resonances in 2 significant ways.
First, while mean-motion resonances can have significant effects on the order of just a couple of orbital periods, the influence of secular resonances is felt on longer timescales.
Second, mean-motion resonances are primarily defined by the relative orbital periods (and therefore semi-major axes) of the small body relative to a planet; even at the highest order, the shape of mean-motion resonances in orbital space are confined to a relatively small width around a specific semi-major axis.
Secular resonances are functions of the precession rates of the small-body, $g$ and $s$, which are themselves functions of the amplitude of the planetary forced terms.
At first-order, these forced terms are primarily functions of semi-major axis, but at second-order, $g$ and $s$ become functions of eccentricity and inclination as well. 
As such, the shape of secular resonances in orbital space is much more complex.
The combined effects of mean-motion and secular resonances are strongly seen in the dynamical architecture of the asteroid belt and the TNOs, and are important features to consider when computing small body proper motion. 

Proper orbital elements can be calculated either analytically \citep[e.g.][]{Milani:1998} using secular perturbation theory \citep[e.g.][]{Brouwer:1950,mandd1999}, or they can be calculated with the use of a numerical n-body orbit propagation \citep{Knezevic:2000}. 
In this second approach, ``synthetic'' proper elements are calculated from numerical simulations of an object's orbit by numerically identifying the planetary forced terms from the simulation, and then removing those forced terms from the SSSB's orbital evolution.
While both methods are effective, synthetic proper elements have been found to be more reliable than their (even higher-order) analytical counterparts. 
\cite{Knezevic:2019} found that both methods provide similar answers, with analytical proper elements always falling well within the computational uncertainties of the synthetic elements. 
However, they found the analytical method to produce less accurate \textit{g} and \textit{s} frequencies than the synthetic method. 
Perhaps most importantly, they also found that the synthetic method produced the smallest errors in the proper elements for asteroid family members, which is a primary use of proper elements in dynamical studies; this cemented the superiority of synthetic proper elements over analytical ones. 

For these reasons, SBDynT uses the synthetic method to calculate proper orbital elements, which we will describe in detail below. 
We adopt the approach outlined by \citet{Knezevic:2000}, calculating synthetic proper elements guided by the application of secular planetary motion theory.

\subsection{Orbit Simulation}\label{sec:sim_ic}

Computing  synthetic proper elements requires simulating the orbit of a given small-body over a timescale appropriate to the secular evolution of that body, which we do using the {\sc rebound} n-body integrator \citep{rebound}.
SBDynT provides tools to initialize a {\sc rebound} simulation using observed small body orbits from the JPL Small-Body Database\footnote{\url{https://ssd.jpl.nasa.gov/tools/sbdb\_lookup.html\#/}} via API query; this includes the option to query the orbit-fit uncertainty covariance matrix to generate clones spanning the observational uncertainties. Orbits can also be provided directly by the user.\footnote{The ability to easily calculate proper elements for arbitrary objects is a major advantage of SBDynT over the current use of catalogs or significant user dynamical expertise.} All small bodies (and any clones) are treated as a massless particles in the simulation, and should be given the ``asteroid'' or ``tno'' object\_type by the user, which will determine the default run properties to be used in the analysis.
Users may use custom run properties for their analysis, but we will focus on the default settings in this section, which have been selected to be effective for the analysis of these objects. 

As part of the initialization process, SBDynT queries the precise positions and velocities of the major planets to be included in the simulation at the epoch of the object's best-fit orbit from JPL Horizons\footnote{\url{https://ssd.jpl.nasa.gov/horizons/}} \citep{Horizons}; we query for the heliocentric positions of each planet's system barycenter (e.g., the barycenter and mass of Jupiter plus its regular moons rather than just Jupiter itself). 
The masses of each included planet system is taken from JPL's DE441 ephemerides (the current basis for Horizons queries; \citealt{DE440}). 
For major planets that are \textit{not} going to be included in the integrations (e.g., the terrestrial planets are often excluded from simulations of TNOs to save computational time), SBDynT queries Horizons for those planets' positions and velocities; the Sun's position is modified to be the barycenter of the Sun and those missing planets, and their mass is added to the Sun.  The user can choose which planets to include, ranging from just the gas giants (the default for TNOs within SBDynT), the seven largest planets (excluding Mercury), or all of the planets. 
We note that excluding Mercury does not significantly affect the computed proper elements of either the asteroids or the TNOs, and doing so allows the simulation to be much more efficient; as such, the default setting for asteroid simulations within SBDynT is to exclude Mercury \footnote{{We compare the proper orbital elements reported for 200 numbered asteroid simulations which were integrated with and without Mercury, with all other simulation variables remaining equal. We find that the reported proper $e,I$ for the asteroids between the two simulations produce differences which equate to $\approx20\%$ of the numerical uncertainty itself. In other words, the choice to not include Mercury in the simulation has a smaller effect on the reported proper elements than the choice of simulation resolution, while speeding up the integration by a factor of $\approx 2.5$. Systematic uncertainties which are of even smaller order than the loss of Mercury (such as the uncertainty in the 
planetary positions, or happenstance aliasing effects caused by alignment with the orbital periods of the planets) can also be considered negligible in comparison to the numerical uncertainties of this method.}}.

SBDynT integrates the simulation for 10 Myr in the case of asteroids and 150 Myr for TNOs; these choices are several times longer than the first-order secular evolution timescales for these regions. 
{The integration timestep is selected to be $T/40$ years, where T is the orbital period of the shortest period particle, or planet, contained in the simulation; e.g. a TNO simulation containing the 4 giant planets would have a timestep of $T_{Jup}/40\approx0.2965$ years.}

By default, the integration is split into two halves, producing both a backwards and a forwards integration.
The motivation for using integrations in both directions is primarily to retrieve accurate proper $\omega,\Omega$ angles, which are defined at the $t=0$ epoch.
As we discuss below in Section~\ref{sec:calc_proper}, the filtering method performed by SBDynT inherently produces edge effects in the time-series of the filtered proper elements; this can cause the obtained proper $\omega$ and $\Omega$ angles at the start or end of the time-series to be artificially shifted from their correct locations.
By placing the present day at the center of the integrated time period, we avoid the issue of edge effects, and retrieve more accurate proper $\omega,\Omega$ values. 

The option is left open to the user to perform only a forwards or backwards integration if they wish, and the proper element computation accommodates for whichever type of integrations is provided. 
The state of all the bodies in the simulation are saved to a {\sc rebound} SimulationArchive binary file \citep{Rein:2017} every 500 years for asteroids and every 5,000 years for TNOs by default. 
This output cadence is chosen to ensure sufficient resolution for most purposes, though even higher resolution runs may be preferred when computing proper elements in certain cases, which we will discuss in the next section.
SBDynT's default choice is to use the Mercurius integrator within {\sc rebound}, which combines the efficiency of the Wisdom-Holman method \cite{Wisdom:1991} implemented in WHFast \citep{whfast} for most of the orbital evolution with the accuracy of IAS15 \citep{ias15} during close encounters. 

In practice, proper elements are less meaningful for small bodies experiencing close encounters with planets{, so we include a flag indicating close encounters and potential planet-crossing orbits (discussed in greater detail in Section-\ref{sec:mean_elements})}.
{SBDynT is meant to be a general purpose tool, and users might choose to simulate real or synthetic small body populations that contain a mix of dynamical behaviors.
As such, using a hybrid integrator by default even for proper element calculations ensures any potential close encounters are simulated accurately, and produce final orbits that best represent the real trajectories of the small body.
Users who do not need to accurately compute close encounter interactions for their test particles have the option to modify their integrator and use WHFast directly \footnote{{When using the WHFast integrator, users also have the option to turn the WHFast fast mode off, which speeds up the integration speed significantly, as outlined in }}. 
This would be a faster computational option if the user is integrating a set of objects and is uninterested in the final result of the objects which experience close encounters.}

All of the run properties mentioned above, such as the output cadence, length of integration, and choice of integrator are able to be modified by the user if they wish.
Examples of how to run custom integrations for proper element calculation are described in various example notebooks contained in the SBDynT GitHub repository.

\subsection{Calculating Synthetic Proper Orbital Elements}\label{sec:calc_proper}

Once an integration is complete, we calculate time series of $\myvec{e}$ and $\myvec{I}$ (Equation~\ref{complex_ei}) for the small body and all the planets contained in the simulation.
For the proper element calculation, we start from the simulated osculating heliocentric orbital elements rather than barycentric orbital elements (which is otherwise SBDynT's default) for consistency with prior catalogs.
We then compute the complex Fourier transforms on those arrays, $\mathcal{F}(\myvec{e})$ and $\mathcal{F}(\myvec{I})$, using the python \texttt{numpy} \citep{numpy} library's Fast Fourier Transform (FFT) functionality, a symmetrized algorithm for computing a Discrete Fourier Transform. The FFT decomposes each time series into a discrete set of frequency bins, which are determined by the sampling cadence and total integration time of the series, and returns the corresponding complex amplitudes for each bin, which contains both the strength and phase of each frequency.
This Fourier analysis enables us to identify the forcing terms associated with the gravitational influences of all the major planets included in the simulation as notable peaks in the signal. 

To properly compute the forcing terms we want to remove, it is necessary to first numerically identify the planetary secular frequencies in our simulation.
While the secular frequencies of the planets have been defined by analytical secular theory \citep[e.g.][]{Brouwer:1950}, these frequencies can vary slightly from the analytically defined values in a numerical simulation \citep{Knezevic:2000}.
This is primarily due to 2 reasons: (1) the output from the numerical simulation may not be resolved enough to capture the precise secular frequency, shifting it to the center of the nearest bin, and/or (2) excluding some planets from the simulation can slightly shift the precession frequencies of the remaining planets even when the overall evolution of the small body is not meaningfully affected by the missing planets.
In other words, very small changes in the simulated solar system compared to the actual solar system can cause very subtle changes to the full n-body solution, so we need to identify the secular frequencies from the simulations themselves; this ensures that the identified secular frequencies correspond to the actual forced terms found in the simulated small body's orbital evolution.

In the simplest case, secular frequencies can be identified as peaks in the power spectrum of the Fourier transforms $\mathcal{F}(\myvec{e_i})$ and $\mathcal{F}(\myvec{I_i})$.
This is the primary way SBDynT identifies the planetary secular frequencies from the planet's individual power spectra, though steps are taken to reduce the influence of massive planets like Jupiter on the frequency identification for smaller planets like Mars.
More details on the identification of the planetary secular frequencies are included in Appendix~\ref{freq_id_app}.

For a small body, the peaks associated with planets' secular frequencies can be used to identify the relative strengths of the individual frequencies in the body's evolution. Conceptually, the proper $g$ and $s$ frequencies for the small body should always correspond to the frequency with the highest amplitude signal in the spectrum.
However, there are some limitations with the Discrete Fourier Transform method that cause this to not always be the case.

One such effect is ``spectral leakage,'' which can cause the power of a signal to be spread among several neighboring frequency bins (i.e., the signal produces a sinc function in Fourier space, rather than the expected Dirac delta function), lowering the amplitude of the true central frequency.
This effect applies to both the small body's proper frequency and the planetary frequencies present in the signal, and the magnitude of spectral leakage is a function of the distance of the true frequency from the center of the nearest discrete frequency bin. The width of the leakage is therefore inconsistent, varying for every frequency in the signal.
Additionally, if the SSSB is near a secular resonance, ($g \approx g_i$ or $s \approx s_i$), the amplitude of the planetary frequency can often appear dominant over the true nearby proper frequency, making them difficult to separate.

Multiple steps are taken to mitigate these problems.
To identify the small body's proper frequency, we first do a `soft' filter of the planetary secular frequencies from the power spectrum; this is done with a simple division of the planetary secular frequency and the adjoining bins by an empirically determined scalar. 
This soft filter of the planetary frequencies is less comprehensive than the filter used to compute the proper elements, and is merely done to separate the SSSB proper $g,s$ frequencies from the planetary frequencies that may appear in the signal.
Once this is done, the signal is further binned by summing up the power for local regions near the $g,s$ frequencies within the power spectra to account for spectral leakage. The peak frequency from the discretized power spectrum is then accepted as the proper frequency.
A more comprehensive description of the full method used by SBDynT to identify the proper frequencies is found in Appendix~\ref{freq_id_app}.

Once the dominant planet and small body secular frequencies have been identified, the forced amplitude terms can be filtered from the small body's signal, leaving only the free or proper amplitudes. 
The challenge of computing synthetic proper orbital elements is to filter out the forced terms without changing the proper amplitudes. This is challenging for several reasons:
\begin{enumerate}
    \item The frequency resolution for a Discrete Fourier Transform is determined by the length of the time series and the time separation $\Delta t$ between the saved values. 
    Poor time series resolution (large $\Delta t$) can make precise signal identification difficult, but a high resolution (small $\Delta t$) is more memory intensive and computationally inefficient; similarly, the total length of the integration is finite and limited by computational considerations. 
    It is important that the filtering done by SBDynT be optimized to balance accuracy and efficiency.
    \item For small bodies near or in secular resonances, it is particularly important to ensure that the filtering does not remove power associated with the proper frequency just because that proper frequency is commensurate with a planetary frequency. In these cases, the planetary frequency can be close enough to the proper frequency that it should be considered part of the proper motion of the small body. 
    This is particularly important when spectral leakage causes the proper frequency to be spread out in frequency space.
\item There are non-linear (e.g., higher-order) terms associated with both the proper and forced frequencies that contribute to the orbital evolution of the small body. The process of filtering must be sure to preserve the non-linear proper terms, while filtering out the non-linear forced terms.
    See Appendix \ref{freq_list_app} for further discussion on where these terms come from and how we identify them.  
    \item Some non-linear terms that contribute significantly to the forced orbital evolution have such long periods that very few cycles occur within the length of the simulation; the most relevant example is the $g_8 + s_8$ frequency for TNOs, which has an $\approx65$ Myr period. 
Long-period terms like these can inflate the secular average terms at long-period in the Fourier transform, as well as increase the effect of spectral leakage for the dominant terms. 
\end{enumerate}

\begin{figure*}[t]
    \centering
    \hspace*{0cm}\vspace*{0cm}\includegraphics[width=1\textwidth, keepaspectratio]{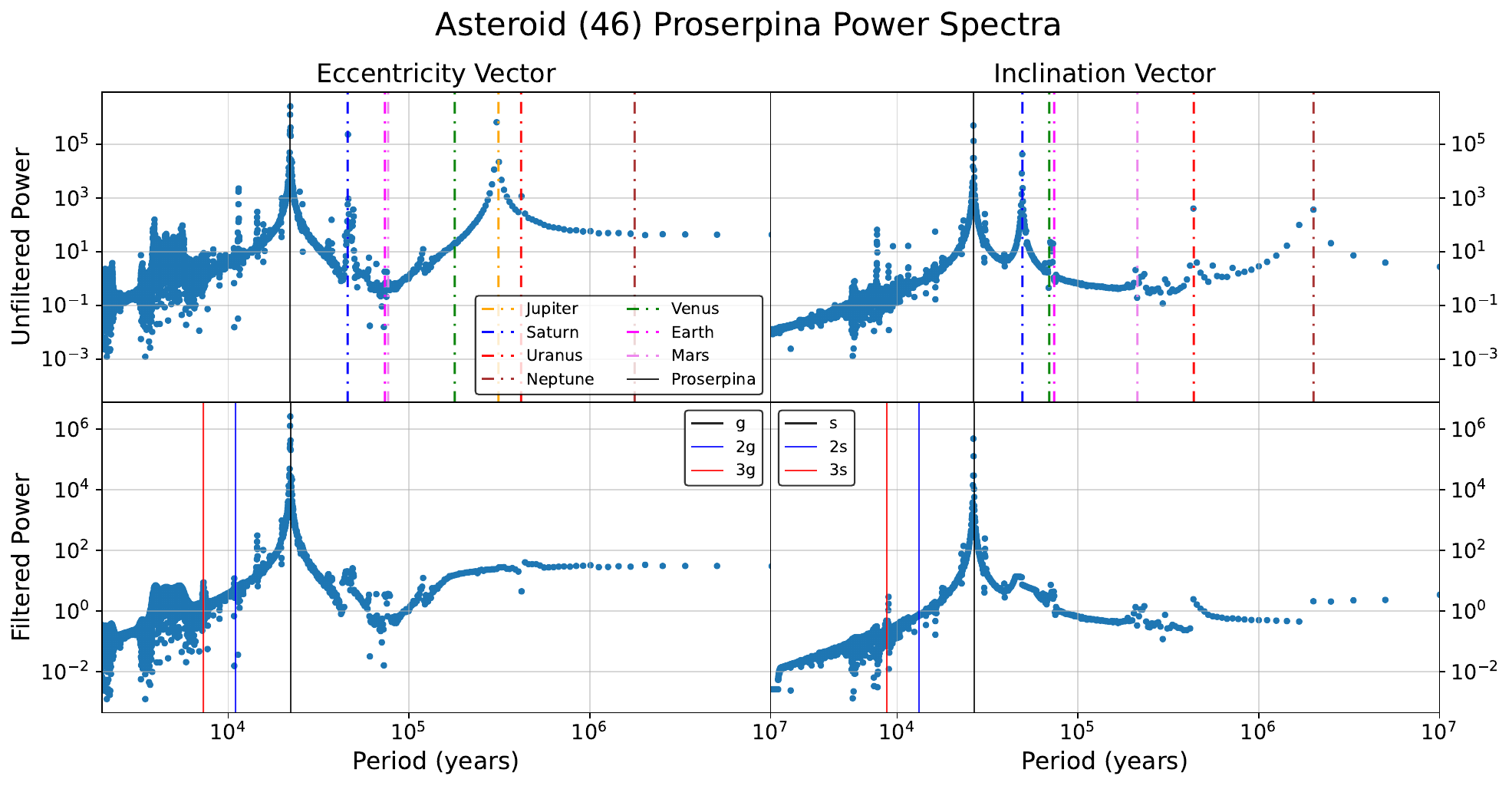}
    \caption{The transformed $\protect\myvec{e}$ and $\protect\myvec{I}$ vectors in frequency space for Asteroid (46) Proserpina before (top panels) and after (bottom panels) filtering. 
    In the top panels, vertical lines indicate forcing frequencies from the 7 planets included in the simulation as well as the peak proper frequency of the asteroid itself. Peaks corresponding to these frequencies are removed by the filter, as desired for the computation of proper elements.
    In the bottom panels, the proper terms corresponding to the free motion of Proserpina are shown; these terms are preserved in our filtering. 
    There are many additional linear and non-linear forced terms which are also handled are not shown explicitly in this figure. } 
    \label{fig:FreqSpace}
\end{figure*}

To account for these difficulties, SBDynT follows a specific algorithm to protect frequencies that correspond to the SSSB's proper terms, while also averaging or smoothing the contributed power across the rest of the signal, effectively preserving the proper amplitudes of the secular motion. 
This is in contrast to some methods for filtering out forced terms by zeroing out the power at the planetary secular frequencies, which risks removing power in the signal which corresponds to the proper motion, again due to spectral leakage.
SBDynT first identifies the linear and non-linear secular terms for the small body to protect, and for the planets, which will be removed; a description of the terms SBDynT specifically identifies can be found in Appendix \ref{freq_list_app}. 
We follow the method of \cite{Milani:1994} and include in our filter all of the linear through 4th-order secular terms that could be relevant to the small bodies in our filter.
SBDynT produces a mask of the signal for the proper frequency and nearby bins to avoid any loss of the proper motion contained in these regions. 
The mask protects the dominant term by a width of decimal-exponent\footnote{The decimal-exponent (dex) is a useful measure of distance in frequency space, which is represented primarily on a log scale. 
As an example, using a dex = 0.2 to produce a mask centered at a frequency $f_i = 10^{-3}$ $yr^{-1}$ would have a frequency range of $\left (f_i  - 10^{log_{10}(f_i)-0.2},f_i  + 10^{log_{10}(f_i)+0.2} \right )$, or $(10^{-3.2}, 10^{-2.8})$ $yr^{-1}$.} (dex) = 0.20, and the higher order terms by a width of dex = 0.01; these widths were chosen empirically to reduce the computed proper element uncertainties, and improve the computed proper elements as compared to the published catalogs.   

There still exist cases where a secular planetary frequency lies within this empirically determined mask, indicating that the small body lies near a planetary secular resonance.
To account for these nearby frequencies, a check is done to identify any nearby planetary secular resonance frequencies that fall into the protected mask. If any forced frequencies are identified as falling into the mask, a small number of bins centered at the planetary frequency are removed from the mask, allowing the filter be applied to those frequencies.
When frequencies are close enough to the proper frequency that the object is in a secular resonance, those frequencies are not filtered out at all, and are considered part of the proper motion of the small body.

Once the identification of protected and filtered frequencies is complete, one last step exists before filtering is applied: accounting for any unwanted short-period terms in the simulation. 
In most cases, the amplitudes of the short-period terms are small, and there is no need to account for them during the filtering process.
However, there are cases, such as objects near mean-motion resonances, where short-period terms are quite significant and ought to be filtered out. 
This is especially prevalent for objects with one or more short-period terms near the Nyquist frequency; this can artificially enhance the size of some non-secular frequencies which do not represent the proper motion of the object.

However, the danger of filtering out short-period terms is that they can overlap with higher-order proper terms that we wish to keep.
To avoid this, SBDynT first determines the shortest-period significant proper secular frequency present in the object's evolution. 
For our 4th-order model, this is primarily the $4g$ or $4s$ frequency which is dependent on the proper motion of the small body itself.
We note that these values are rarely shorter than the $2g_6-2s_6$ frequency, which is the fastest relevant 4th order secular frequency among the Solar System planets, and occurs with a period of $\sim11,800$ years.
SBDynT places a ceiling on the power of every term that has a period $< \sfrac{1}{10}$ of the shortest-period secular term; the ceiling is set equal to the power of the spectrum at $f_i = (4g\;\;or\;\;4s)\times 10$, or the lower limit of the short-period frequencies.

We note that in previous studies, the short-period terms are handled differently than they are in our method.
When computing proper elements for asteroids, \cite{Knezevic:2000} reports that their OrbFit code uses a digital filter on an initial set of ``input'' orbital elements  sampled at a rate of 2 years and returns a set of filtered ``output'' orbital elements  sampled at a rate of 200 years.
Their proper element computation is then performed on this set of output orbital elements.
Due to how finely sampled their sample of orbital elements is, their digital filter is very effective at capturing short-period terms present in the simulation. 
Implementing this in the present form of SBDynT and {\sc REBOUND} would have significant other negative consequences in terms of efficiency and/or memory, so we do not follow their method of digital filtering the orbital elements arrays; we will show later that our filtering method still produces proper elements comparable to those reported by OrbFit.

In contrast to SBDynT and \cite{Knezevic:2000}'s handling of the short-period terms, \cite{Nesvorny:2024} states that they do not filter out short-period terms at all, because the effect is found to be negligible in their testing. 
We will discuss our findings on the impact of the short-period filter on the proper elements later in Section \ref{sec:results}.

Once the short-period filter has been applied, SBDynT applies a gaussian convolution filter using Scipy's \texttt{gaussian\_filter1d} \citep{scipy} to the log of the power spectrum.
The kernel of the gaussian filter, which describes how widely the filter should sample nearby bins for averaging, is selected based on the size of the proper frequency which is saved.
An object with a longer-period proper frequency is thus filtered with a larger kernel, and a shorter period proper frequency is given a small kernel.

A benefit to  this method is that it can account for  higher-order terms that typically are not specifically filtered out by previous methods, as the gaussian kernel is applied across the entire power spectrum, with the exception of any protected frequencies.
The protected regions near the proper frequency and its associated higher-order terms are then placed back into the signal using the previously-built protected mask. 
The filtered power spectrum is then representative of the amplitude of the free or proper motion that is contained in the signal. 
An example of this filtering method is shown in Figure~\ref{fig:FreqSpace}, where the peak frequencies of the perturbing planets can be identified in the Fourier space of the unfiltered $\myvec{e}$ and $\myvec{I}$ vectors for Asteroid (46) Proserpina, an object selected at random.
in the final filtered vectors (bottom panels), the protected frequencies are indicated, with the rest of the spectrum appearing to be largely smoothed out.

An inverse Fourier transform is then used to convert the filtered Fourier-space vector $\hat{f}$ into the filtered eccentricity or inclination vector, $\myvec{e}_{filtered}$ or $\myvec{I}_{filtered}$.
These vectors can be decomposed into the small body's filtered time-evolution in $e$, $I$, $\omega$ and $\Omega$. 
At this point, while this filtering of the planetary secular frequencies from the $\myvec{e}$ and $\myvec{I}$ vectors effectively removed the dominant linear frequencies from the vector, non-linear secular frequencies which appear only in the eccentricity or inclination evolution may still remain.
These non-linear secular frequencies which result from combinations of the proper $g,s$ frequencies and the $g_i,s_i$ planetary frequencies are phase modulated when the $\myvec{e}, \myvec{I}$ vectors are formed, causing the signal from these non-linear frequencies to be contained within the proper signal of the vectors themselves and thereby avoid being filtered out.

While the residual amplitude of these non-linear terms is generally quite small, and has no impact on the final proper reported proper elements if the full period of the non-linear term is captured by the simulation, these terms can produce larger proper element uncertainties for near-resonant objects, as we will discuss in Section-\ref{sec:proper_uncertainties}.
To avoid this, SBDynT runs an identical filter as previously described on the filtered eccentricity and inclination time-arrays, with the added requirement to protect the 0-bin value, or the mean value, of the time arrays.
Details describing this additional filter and its impact on secularly resonant objects are described in Appendix-\ref{sec:add_filter}.

\begin{figure}[ht]
    \centering
    \vspace*{-0.45cm}\hspace*{-1.00cm}\includegraphics[width=1.15\linewidth]{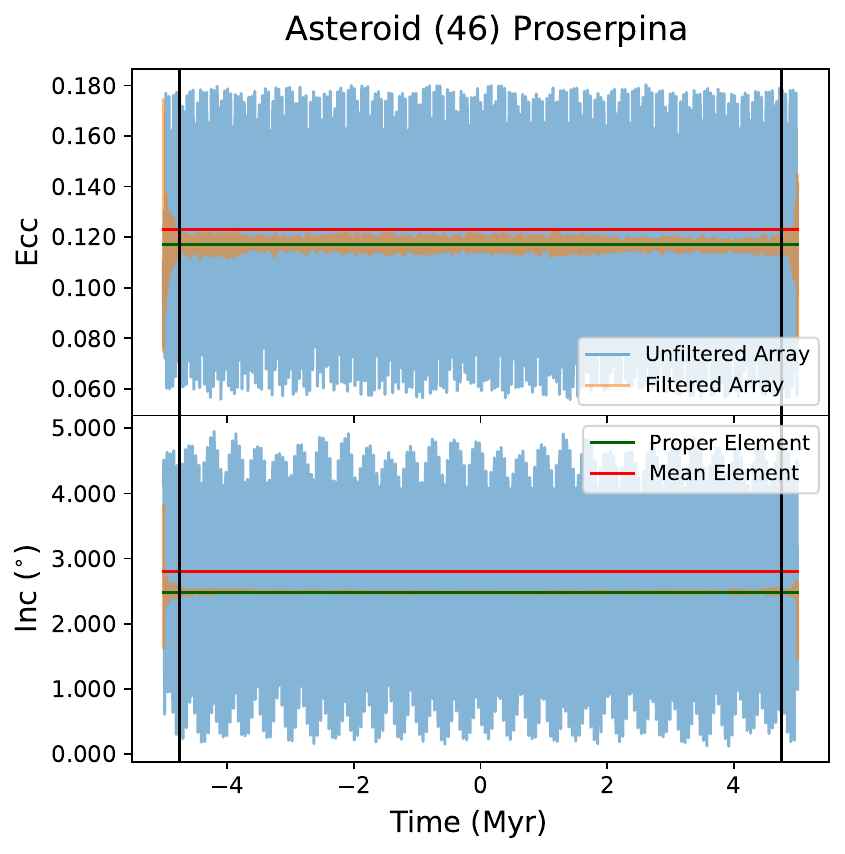}
    \caption{Eccentricity and Inclination time arrays for Asteroid (46) Proserpina, before (blue) and after (orange) SBDynT filtering. Black lines indicate the 2.5\% cutoff for producing the proper element from the mean of the filtered value. The red and green horizontal lines indicate the mean of the osculating elements and the proper elements, respectively.}
    \label{fig:eccandinc}
\end{figure}

Typical to signal filtering, there is often visible contamination at the edges of the resulting reconstructed time series (see Figure~\ref{fig:eccandinc}).
Thus, while SBDynT could report the mean value of this final array as the proper element, SBDynT applies two final steps to compute the final resulting proper element.

First, the first 2.5\% of the filtered time arrays are removed, as seen in Figure-\ref{fig:eccandinc}, to disregard the edge effects present.
However, edge effects vary in magnitude, and some edge effects may still exist in the array, which corrupt the mean value.

Then, to minimize any remaining edge-effect values within the filtered time-arrays, a Hann window function is applied to the filtered $e,I$ time arrays.
This window function effectively decreases the weight of variation near the edges of the time-array, and increases the weight of the motion near the center, which is expected to be more numerically stable after filtering.
The sum of the windowed $e,I$ arrays is then divided by the sum of the Kaiser window itself, thereby producing an effective de-weighted mean of the time-array; the resulting value is the reported proper element by SBDynT.

SBDynT also reports proper $\omega$ and $\Omega$ angles, which are defined as the values of these angles at $t=0$ in the filtered $\myvec{e},\myvec{I}$ arrays, with $\omega$ computed as the difference of the proper $\Omega-\varpi$.
As such, the reported proper $\omega$ and $\Omega$ angles are dependent on the defined $t=0$, and for a default run will correspond to the epoch of the best-fit orbit used to initialize the simulation.
We note that because of the edge effects introduced by filtering, accurate proper$\omega, \Omega$ angles can only be defined for integrations where the $t=0$ time is neither the beginning or end of the simulation, and is why SBDynT has the default setting for simulations to integrate both backwards and forwards in time.

Typically, best-fit orbits for real objects contained in the MPC are defined with respect to some epoch between the present day and the J2000 epoch; the evolution of the slow-moving orbital angles for orbits on this timescale is negligible, and can typically be ignored in the interest of understanding the present day orbital shape and orientation.
Users who run simulations of synthetic particles should be aware of what their $t=0$ time represents, however.
If the user wishes to represent some past or future state of the solar system, such as when searching for or modeling the evolution of collisional families or primordial small body populations, the resulting proper angles will instead represent the angles at that defined epoch with respect to the locally forced plane.

Computing the proper semi-major axis of an SSSB's orbit is much simpler; \cite{Knezevic:2000} found that secular planetary influence on semi-major axis evolution are all high-order terms, and for the most part only short-period effects can contribute on a large-scale to the semi-major axis evolution \citep{Milani:1987}. 
Our testing indicates something similar, with short-period variation dominating the greater orbital evolution, centered on the mean.
For this reason, SBDynT applies the same gaussian\_filter1d to the semi-major axis, and only protects the longest period frequencies.
Thus the mean of the semi-major axis is preserved, while the high-frequency evolution is restrained.
The proper semi-major axis is then found by taking the mean of the smoothed semi-major axis time array, similar to the proper $e$ and $I$, but without searching for signals in the frequency space corresponding to the perturbing planets. 

We note that the choice of providing the osculating elements in either the heliocentric or barycentric reference frame {(or even in the Jacobi coordinates of the simulation)} has effectively no impact on the computed proper $e,I$ values of the small body.
{For the eccentricity, the signal added to the osculating eccentricity of the small body and the planets with respect to the orbital center appears almost entirely as a short period term in the eccentricity, which is captured inherently by SBDynT's short-period filter discussed earlier in this Section.
The inclination is entirely detached from the choice of the orbital center, and is only dependent on the choice of the reference plane, whether that be the invariable plane or the ecliptic plane.
In addition, because it is of such short period, the impact of the selected central coordinate frame on the reported planetary proper precession frequencies is below the numerical uncertainty of the frequency determination itself.}

However, as the proper $a$ is computed as simply the mean of the osculating $a$, there is a subtle shift in the reported proper $a$ depending on the selected reference frame. 
SBDynT by default integrates the orbit and computes the proper elements in the barycentric reference frame; users should be aware then that the proper $a$ reported by SBDynT may be slightly offset from the proper $a$ reported by other catalogs which report their proper elements in the heliocentric frame.

We note that certain SSSB's may exhibit orbits that are dominated by the forced terms of the planets, making traditional proper orbital element calculation challenging. 
This is often the case for resonant orbits---both mean-motion and secular---which can strongly influence the long-term evolution of small body orbits.
A discussion of these effects and of how SBDynT reports the ``mean'' elements for resonant objects will be discusses further in Section~\ref{sec:mean_elements}.

\subsection{Proper Element Uncertainties}\label{sec:proper_uncertainties}

\cite{Knezevic:2000} mention that one of the most significant advantages of computing synthetic over analytical proper elements is the ability to compute the uncertainty of the computed proper element.
This is done by looking at the variability of the filtered elements between different time segments of the simulation. 
SBDynT follows the method of computing proper element uncertainties outlined in \cite{Knezevic:2000} (also adopted by \citealt{Nesvorny:2024}): we divide the simulation into five overlapping time windows, each covering 33\% of the simulation length.
SBDynT then recomputes the proper elements for the small body for each of these windows by running the filtering algorithm, re-identifying the proper $g$ and $s$ frequencies, and computing the mean of the filtered $e$ and $I$ values in the same method as described in the previous section.
The uncertainty for the proper element is then taken to be the standard deviation of the reported proper elements from the five windows. 
One can imagine that stable objects will have proper elements that remain very constant over the course of the simulation, producing very small uncertainties, whereas SSSBs with chaotic or strongly evolving orbits will produce larger uncertainties.
In this way, these uncertainties can be seen and used as indicators of the stability of the SSSB's orbit over secular timescales.

We note that these uncertainties do not incorporate observational uncertainties, nor do they include systematic uncertainties due to various reasonable choices by different authors in how proper elements are calculated.
Users that desire to account for the observational uncertainties propagated into a small body's orbital evolution can compute multiple sets of proper elements for the same object by simulating several clones of the object.
While the uncertainty described above gives key insights into the orbital stability of the best-fit orbit, computing proper elements for several clones would similarly measure the variability or chaotic nature of the object's orbit.
We will address effects on the uncertainties due to choices during the filtering method by the authors later on in our discussion of the results in Section~\ref{sec:results}.

\subsection{Mean Elements}
\label{sec:mean_elements}

\begin{figure}
    \centering
    \includegraphics[width=\linewidth]{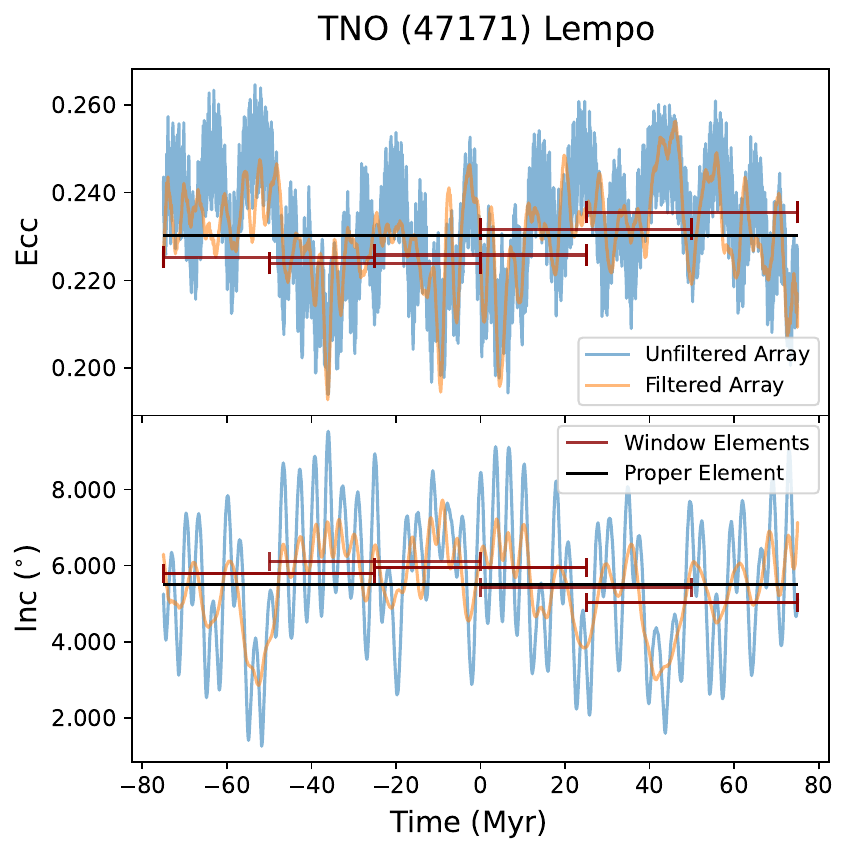}
    \caption{TNO (47171) Lempo's eccentricity and inclination evolution over time; the chaotic evolution is the result of Lempo's occupation within 3:2 mean motion resonance with Neptune.
    Separating the proper elements from the forced elements is therefore noticeably difficult, with perturbations from Neptune dominating the orbit.  
    The final reported proper element resulting from our filtering routine is nearly equal to the simple mean of the unfiltered time array. 
    The individual computed proper element for each of the five time windows is shown, highlighting how chaos affects the uncertainty measurement.}
    \label{fig:mm_chaos}
\end{figure}

\begin{figure}
    \centering
    \includegraphics[width=\linewidth]{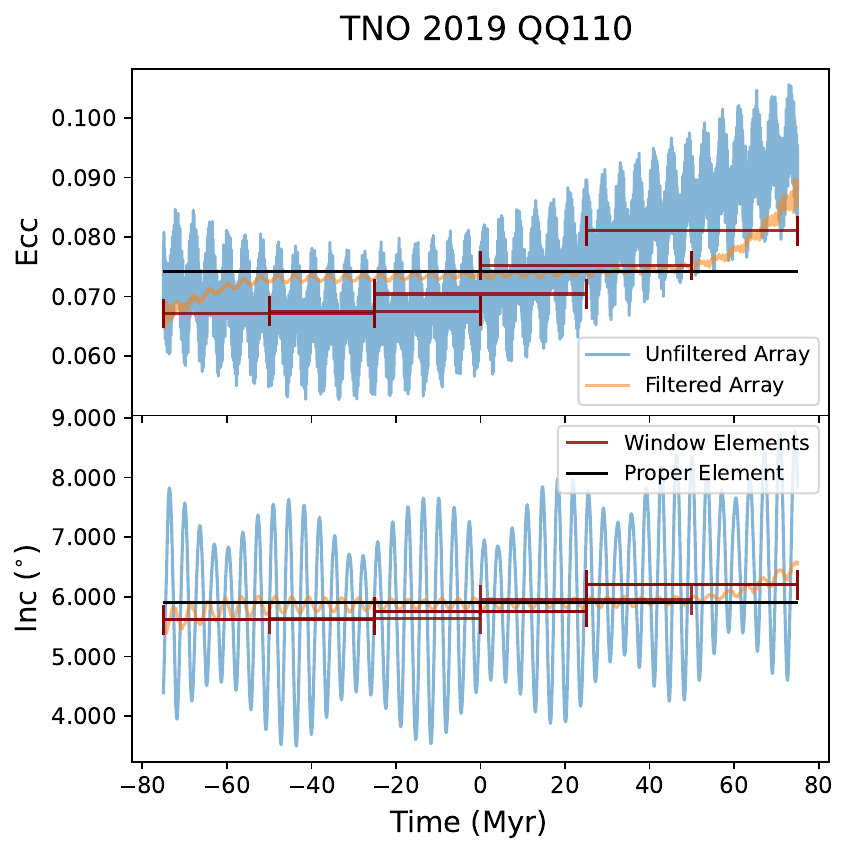}
    \caption{TNO 2019 QQ110's eccentricity and inclination evolution over a 150 Myr integration. 
    The $g+s-g_8-s_8$ resonance produces an $\approx 300$ Myr long-period term in the evolution of this object. While technically periodic, the growth in eccentricity and inclination in this resonance is so slow, it only appears periodic on timescales comparable to the age of the Solar System. Objects even closer to the center of a resonance would have longer period terms, producing apparent constant adiabatic growth in eccentricity and inclination over the age of the Solar System. Even with an additional filter applied to the inclination time-array, this long-term variation cannot be fully filtered out since it is not resolved by the full integration, artificially increasing the computed proper element uncertainties. Increasing the integration length can better handle this long-period term.}
    \label{fig:sec_res}
\end{figure}

As discussed above, there are a number of SSSB's with evolution strongly influenced or dominated by secular and/or mean motion resonant effects. 
Computing the proper or free elements for such objects can be difficult for two reasons:
\begin{enumerate}
    \item Chaotic Motion: Mean motion resonant objects can experience non-periodic (chaotic) evolution in  eccentricity and inclination over a wide range of timescales, including both shorter and longer than the secular timescales described previously (Section~\ref{sec:properelementtheory}). 
Such variability can make it very difficult or impossible to define proper orbital elements for these objects for timescales of interest for dynamical studies. 
\item Long-Period Terms: In some regions of orbital parameters, higher order secular resonances can occur. 
    Examples of these include asteroids for which $g+s\approx g_6+s_6$, or $g+s\approx j(g_6+s_6)$ resonances, where $j$ is an integer \citep[e.g.][]{Milani:1992a,Knezevic:2000,Carruba:2014,Carruba:2018} and TNOs for which $g+s \approx g_8+s_8$ resonance \citep{Knezevic:2003}. Proximity to a higher-order secular resonance can produce long period coupled variations in eccentricity and/or inclination on timescales much longer than the secular timescales described in Section~\ref{sec:properelementtheory}. 
    These coupled long period variations are challenging to filter out, especially when the periods are comparable to the length of the integrations.
\end{enumerate}
Figures~\ref{fig:mm_chaos} and~\ref{fig:sec_res} show examples of a chaotic mean-motion resonant object and a higher order secular resonant object, respectively.

In the case of the mean motion resonant object (Figure~\ref{fig:mm_chaos}), we can see a high frequency component in the eccentricity and an overall chaotic variation. 
The high frequency component is due to the 3:2 mean motion resonance librations, which occur intermittently and with variable amplitude. 
In this case, the proper element computation fails to effectively filter out any significant amount of signal from the eccentricity and inclination evolution over time.
As such, the proper elements computed in each of the windows (indicated in the figure) vary significantly, producing larger uncertainties in the returned proper elements. 

We note that accommodating for mean-motion resonances in computing proper elements has been discussed in the literature, and specialized semi-analytical methods can accurately produce the proper elements for small bodies near resonance \citep{Gronchi:2001}.

Similarly, special methods exist for calculating proper elements for planet-crossing populations \citep{Fenucci:2022}.
{We emphasize that in its current release, the default SBDynT settings are only recommended for computing proper elements for main-belt asteroids and for TNOs with $q>34$ AU, which avoid planet-crossing orbits.
To help users identify potential planet-crossing objects and scattered objects, SBDynT includes two flags in the outputs. 
The first flag, which identifies potential planet-crossing orbits, takes the form of a simple boolean which is labeled as ``True'' if a simulated small body experiences at some point in the integration an orbit with $q<1.7, Q>4.0$ AU for an asteroid, or $q<34$ AU for a TNO.}
{The second flag, to help identify potential scattering events for an object, is computed using the array of $|\Delta a/a|$ values over the small body's orbital simulation.
This flag is reported as ``True'' for asteroids with a $|\Delta a/a| > 2.5\%$, and for TNOs with $|\Delta a/a| > 0.5\%$, indicating that at some point throughout the output orbit, there was some significant energy exchange between the small body at the perturbing planet, greater than any potential variance due to the solar system barycentric motion, or adiabatic energy exchanges which are common, and don't necessarily produce new orbits for the small body.
While not a perfect indicator of scattering effects, this indicator generally performs well at identifying objects which exchange large amounts of energy with perturbing planets.}

{We note that both of these flags are computed from the saved outputs of the simulation, and are not computed at a timestep level during the simulation itself.
Considering the high output cadence of the default simulations, however, these flags perform adequately at identifying potential scattering and planet-crossing orbital evolution for the small bodies. }

{The theory and computation of proper elements is focused on long-term non-resonant secular perturbations and quickly becomes meaningless for objects having close encounters with planets or even just on planet-crossing orbits. 
There are still cases when computing the proper elements using SBDynT for weakly-resonant objects or planet-crossing objects can still be effective, such as in the case of weakly resonant objects, or stable von Zeipel-Lidov-Kozai (vZLK) librators like Pluto, which have their secular motion largely preserved over secular timescales \cite{Ito:2025}.
These secularly-stable objects can generally be identified by measuring the reported numerical uncertainties which result from the proper element computation.
This can only occur when the small body's proper motion is preserved, rather than being overwritten by the mean-motion forced terms of the perturbing planet.
Cases like these are generally the exception to the rule, making the computation of proper elements for mean-motion resonant orbits a challenge.}

In its current stage of development, SBDynT does not produce specialized proper orbital elements for mean motion resonant or planet crossing small bodies, though implementation for these cases is possible, especially considering the open-source nature of our codebase. 

In contrast, the evolution of the secularly resonant TNO (Figure~\ref{fig:sec_res}) is more regular, and the filter adequately accounts for the linear first order $g_8$ and $s_8$ frequencies present in its evolution, as well as a number of higher-order secular frequencies. 
However, because the filtering is performed separately on the complex eccentricity and inclination vectors, the non-linear coupling between the secular frequencies that generates residual long-period modulations in $e$ and $I$ cannot be removed by the filter applied to the $\myvec{e},\myvec{I}$ vectors.
We emphasize the fact that, in the case of 2019 QQ110, the secular resonant terms are of such long period that the default 150 million year integration would only capture approximately half of the full secular period. 
If the length of the integration is extended over a longer timescale that is sufficient to capture the a full secular cycle, the uncertainty measurements can improve from the proper element computation, as is shown in Appendix-\ref{sec:add_filter}.

The long period effects of secular resonance have been discussed in the literature, with methods being proposed to accommodate for these effects when defining the proper elements. 
The two primary methods include (1) reporting the mean of the filtered elements along with the amplitude of variation corresponding to the critical argument \citep{Milani:1992a}, or (2) using semi-analytical methods similar to those proposed for computing proper elements in mean motion resonance \citep{Morbidelli:1993}.

Due to these two variations of orbital evolution, rather than reporting the final output from the filtering algorithm as a proper element, instead SBDynT reports the result as a ``mean'' element, indicating that the provided result is some average of the residual variation that is still present in the filtered element.
To identify ``mean'' elements from proper elements, we search for remaining variability in the filtered eccentricity and inclination arrays.
This is done by taking the FFT of the filtered eccentricity and inclination arrays, rather than the complex vectors.
We remind the reader that the filtered inclination array is defined with respect to the locally forced plane, and so frequency terms within the filtered inclination array should avoid spurious inclusion of motion induced by the reference plane.  

If the object has a non-resonant or non-chaotic orbit, SBDynT's filtering should effectively remove any variability from the eccentricity and inclination over time, leaving only variability within the filtered vector which is related to the $\varpi$ and $\Omega$ evolution, which precess at constant rates after filtering. 
As such, the distribution of power in the Fourier transform of the filtered eccentricity and inclination time arrays should be concentrated entirely within the 0-bin, which represents the mean of the signal.
The fraction of power in the filtered $e,I$ arrays contained outside of the 0-bin can thus be used to identify the level of remaining variability of the eccentricity or inclination, indicating whether the element can be considered mean or proper.

Through empirical testing with comparison to the AstDys 2024 TNO catalog, which only reports proper elements, we define a mean element for cases where the total power contained outside of the 0-bin for the FFT of the filtered $e,I$ arrays is greater than $2.5\times10^{-5}$, where the power represents the square of the associated amplitude of each signal.
In other words, mean elements are defined for objects where the leftover maximum possible amplitude of variation in the filtered $e,sin(I)$ time arrays exceeds a value of $0.005$; as all possible values must line within $0<e,sin(I)<1$, variation as large as $0.005$ indicates some significant variability is present.

Cases may exist where the mean element may is a value very close to 0; in these cases, while the leftover variation may be small relative to the full range of possible values, it may be large relative to the proper element itself.
In these cases, SBDynT also reports a mean element for objects where the the leftover amplitude of variation in the filtered $e,I$ time arrays exceeds 20\% of the mean value of the array.

Using these criteria, SBDynT can define mean elements for all objects with significant leftover variability either relative to the total possible range of eccentricities and inclinations, or relative to the mean element itself, which are reported in a dictionary of various flags and indicators related to possible stability and resonance terms associated with the small body orbits.

We emphasize that the process of filtering for mean elements is identical to that of the proper elements; objects which are defined as having mean elements simply satisfy our criteria of having filtered orbital elements with significant remnant variability.
As the $e,I$ elements are defined separately, cases may exist where only one element is reported as a mean element, while the other is reported as a proper element.
This is often the case for objects which may be stable or non-resonant in $e$, while unstable or resonant in $I$, or vice versa.

We mention that long-term variability which is the result of non-linear secular motion can be considered to be a component of the proper motion of the small body orbital evolution.
To account for this, SBDynT reports the amplitude of the remaining variability in the filtered eccentricity and inclination arrays, as well as the highest-amplitude frequency contained within these now filtered time arrays as additional values contained within a dictionary of flags and indicators relevant to the small body's proper motion. 
For secularly resonant or near-resonant objects, this reported amplitude and frequency are thus components of the actual proper motion of the object.

As such, mean elements may still provide helpful insights into the secular dynamics of many small bodies.

These mean elements are still a useful metric in understanding the secular dynamics of these bodies; for example, such averaging causes objects that are captured in the same resonance to appear tightly clustered in orbital semi-major axis, causing resonances to stand out in plots of the proper elements.
Indeed, objects which are given ``mean'' elements rather than proper elements are almost guaranteed to either have some resonant interaction in their orbits, or display chaotic evolution.
As such, mean elements, combined with an understanding of the orbital dynamics within a region of phase space, can provide context for whether objects lie in or near resonances of any kind.

SBDynT's code for computing proper and mean elements is written by default to perform the process of orbital solution gathering, secular frequency identification, and proper element filtering directly from an SBDynT simulation.
However, more advanced users with their own orbital solutions computed by other methods will still have the option in SBDynT to produce proper elements by inputting the orbital element time arrays directly into the filtering function; examples of this will be included in the Proper Element Tutorial Notebooks found in the SBDynT GitHub repository.

\section{Stability Indicators}\label{sec:chaos}

Chaos refers to a dynamical system's sensitive dependence on initial conditions, where small variations can lead to significantly different long-term behaviors. 
Chaos can be technically defined as an exponential divergence of trajectories due to even infinitesimal changes to the initial conditions of an orbit, with an associated exponential growth timescale. 
%\sout{{While all real systems are somewhat chaotic, chaotic evolution on timescales much longer than the age of the solar system often imply stability in a practical sense.}}
%\sout{{This can often be the case for systems which experience dissipative forces, or balancing torques, which can result from asymmetric interactions between the planets and the small body.}}
%\sout{{We therefore typically contrast chaotic dynamical motion with ``regular'' motion in the solar system, which typically consists of periodic motion which is predictable to some meaningfully-long timescale.}}
In the context of orbital mechanics, ``{stability} indicators'', {(more often referred to in the literature as ``chaos indicators'')}, are quantitative metrics that reveal whether an orbit exhibits chaotic {(exponential divergence)} or regular {(non-exponential divergence)} characteristics. 
These indicators can provide helpful insights into the underlying mechanisms of dynamical evolution.
Including {stability} indicators in the analysis of orbital dynamics serves a critical scientific purpose. 
Objects with chaotic orbits often lose coherence or predictability in their orbital trajectories, making them unreliable in tracing the origins or predicting the future evolution of a system. 
For example, if a system exhibits chaotic divergence in orbital trajectories within just 1 million years, it cannot be used to draw conclusions about solar system formation.  
Furthermore, using multiple {stability} indicators is essential for tailoring analyses to specific scientific objectives. 
Each indicator is sensitive to distinct aspects of chaotic behavior, providing a complementary toolkit that can be customized to interpret dynamical processes across a wide array of analyses.

To assess the chaotic behavior of SSSBs in our simulations, we employ five distinct {stability} indicators in SBDynT. 
These indicators provide insights into the dynamical stability of orbits over different timescales. 
Below, we describe each indicator and the methodology used in our simulations.

\begin{enumerate}[label=\Roman*.]

\item \textbf{Root Mean Square of the Residuals of Orbital Elements}

The first {stability} indicator we use is the root mean square (RMS) of the residuals of the orbital elements of multiple clones sampling an initial orbit and its uncertainty range, providing a measure of how the orbital evolution diverges over time. 
SBDynT initializes a simulation with 5 clones of an observed object with orbital elements drawn from the covariance matrix associated with that object's best-fit orbit (see Section~\ref{sec:sim_ic}); each clone thus has slightly different initial conditions.
    The integration is propagated {for the default length for an asteroid or TNO}, and the RMS value is computed from the residual of each clone's evolution from the best-fit orbit's evolution for the key orbital parameters, semi-major axis, $e$, and $I$. 
Because this indicator relies on the uncertainty in the orbit fit, this indicator effectively measures the unpredictability/chaos over the currently known orbit fit range; orbital evolution could be subject to significant change if the orbit is initially poorly constrained, so caution should be used when interpreting this indicator for newly discovered objects.
We flag these indicators if the RMS exceeds 1\% of the initial semi-major axis, {eccentricity, or sine of the inclination}, indicating a high degree of divergence among the clones, {suggestive of chaotic dynamics}.

\item \textbf{Distance Metric ($d_{met}$) from Proper Element Uncertainties}

This {stability} indicator is produced from the uncertainties of the proper element computation. 
    Large uncertainties in the proper element filtering can result from chaotic evolution; this is often short-term or secular periodic evolution that is so long it is under-sampled by the length of the simulation (see Section~\ref{sec:mean_elements}). 
    Following the example of \cite{Nesvorny:2024}, we can measure the magnitude of the proper element uncertainties using the distance metric between two particles introduced by \cite{Zappala:1990}:
    \begin{equation}
        d_{met} = n_i\, a_i\, \sqrt{\frac{5}{4}\left(\frac{\delta_a}{a_i}\right)^2+2(\delta_e)^2+2(\delta_{inc})^2} 
        \label{dist_met}
\end{equation}
    where $\delta$ indicates the difference in proper elements between the two particles, and $n_i$ and $a_i$ are the mean values of their proper mean motion and proper semi-major axes.
    The value of the metric itself is the relative velocities (reported in m s$^{-1}$) between the two objects in proper orbital phase space. 
    Rather than comparing two separate small bodies, the distance metric can also be modified to indicate the stability of the proper element calculation for an single object by taking each $\delta$ term to be the proper element uncertainty, and $a_i$ and $n_i$ to be the proper semi-major axis and mean motion. 
    \cite{Nesvorny:2024} uses this distance metric to measure the numerical stability of their proper element calculation and to compare their results to the \cite{Novakovic:2022} proper element catalog. 
    We do a similar comparison between catalogs and include a discussion later on in Section~\ref{num_precision}.

    When searching for collisional families in the asteroid belt, a relative velocity cutoff is implemented by the clustering algorithm.
    This cutoff effectively defines a boundary to the collisional family in $a,e,I$ parameter space.
    For an asteroid's collisional family membership to be considered statistically significant, the asteroid must have a computed $d_{met}$ from the proper element uncertainties which lies below the relative velocity cutoff for the family.    
    This cutoff differs depending on the dynamical circumstances of the individual collisional family, but most studies place this cutoff at a value of $10 < d_{met} < 100$ m/s \citep{Milani:2014,Nesvorny:2015,Nesvorny:2024}.

    As for TNOs, an in-depth study on the potential detectability of TNO collisional families by \cite{Marcus:2011} found that, due to the typical dispersion velocities produced by collisions in the classical belt, the typical relative velocity cutoff for a TNO collisional family is expected to lie above $d_{met} > 100$ m/s.
    They conclude that discovering TNO families by clustering of the relative distance metric velocities is unlikely to provide notable results until the TNO catalog is increased significantly. 
    
    However, in the interest of chaos and stability, we find that the $d_{met}$ uncertainties for TNOs in the current catalog are found to generally consist of a similar $d_{met}$ distribution as the asteroid belt, with approximately half of the currently known TNOs having a $d_{met}<10$ m/s (See 
    Figures \ref{fig:hcm_cdf} and \ref{fig:hcm_cdf_tnos}).
    
    SBDynT thus follows this implementation in the resulting {stability} indicator flags produced by the $d_{met}$ for both asteroids and TNOs.
    Using this criteria, SBDynT labels objects with a $d_{met} < 10$ m/s as stable, $10 < d_{met} < 100$ m/s as metastable, and $d_{met} > 100$ m/s as unstable.  

    We note that an object's $d_{met}$ indicator can be reportedly high for both objects in stable secular resonances, which display very predictable but slow orbital evolution, as well as for more chaotic objects.
    These could include mean motion resonant objects, some of which may remain bounded and stable, but evolve unpredictably. 
The user can separate stable secular resonant objects from chaotic mean motion resonance objects by considering the reported $g$ and $s$ frequencies from the proper element computation; for example, TNOs with $g-g_8\approx0$, $s-s_8\approx0$ or $g+s-g_8-s_8\approx0$ would be in or near these secular resonances. 
    
    A visual look at the reported proper semi-major axis of an object can also serve as a clear indication of whether an object exists in or near a mean-motion resonance; this is because the proper semi-major axis of a mean motion resonant object is typically very well aligned with the semi-major axis associated with the mean motion resonance.

\item \textbf{Information Entropy for Orbital Constants (Shannon Entropy)}

    Information entropy, or Shannon entropy \citep{Shannon1949}, provides a statistical measure of the randomness or disorder within a selected dataset $\chi$, and is defined by:
    \begin{equation}\label{entropyeq}
        H = -\sum_{x\in\raisebox{0.8pt}{$\chi$}}^\chi p(x) \, log_{10}\bigl(p(x)\bigr)
    \end{equation}
    where $p$ represents the probability density of a selected member $x$ within the full dataset.
    By extension, then, the information entropy can be defined as the sum of the product of probability densities and their $log_{10}$ values over the entire parameter space being considered.
    Information entropy can be used to measure disorder for any kind of dynamical system; for example \cite{Nunez:1996} measured chaos as the information entropy produced by considering the generalized coordinates of the Hamiltonian equations of motion for a small body's orbit.
    Similarly, we can measure the chaotic evolution of a small body's orbit by computing the information entropy from the specific angular momentum, $h_s= \sqrt{a(1-e^2)}$;
    $h_s$ is a constant over time in linear secular evolution, so deviations in $h_s$ indicate non-linear effects in an object's orbital evolution, which is often associated with chaotic motion.

    SBDynT starts by producing a histogram of the $h_s$ vectors over the length of an entire {\sc rebound} simulation of the best-fit orbit for the small body (10 Myr for an asteroid and a 150 Myr for a TNO). 
    The number of bins used to sample $h_s$ over the full time array is simply $\sfrac{1}{10}$ of the length of the array.
    If the $h_s$ array were perfectly evenly distributed, this would result in exactly 10 time samples per bin in the histogram. SBDynT  produces a probability density for each bin by normalizing the sample over the $n_{bins}$. 
    The information entropy value can then be calculated from these percentages using Equation~\ref{entropyeq} and compared to the the maximum possible value for the given simulation length,  $log_{10}(n_{bins})$; 
the final reported entropy value is 
$J = H/log_{10}(n_{bins})$, sometimes referred to as the relative entropy \cite{Shannon:1949}.

Minor changes in $h_s$ over the simulation are expected due to normal dynamical and numerical limitations, even for regular motion;
    however, such changes are expected to be randomly distributed across the very small range, which should lead to an even distribution.
    Large values of this indicator reflect relatively constant $h_s$, indicating stability; smaller values indicate a skewed sample, indicating instability.

    \item \textbf{Autocorrelation Function Index (ACFI)}

    The Autocorrelation Function Index (ACFI) was introduced as a potential {stability} indicator by \citep{Carruba:2021}. 
    ACFI is calculated by taking the Pearson correlation coefficient (or Pearson index) for the semi-major axis of a small body against itself at a number of time shifts. 
    The Pearson index ranges between 1 and -1, where 1 indicates perfectly correlated values, and -1 indicates perfectly anti-correlated values. 
In SBDynT, we use the ACFI as a long-term {stability} indicator considered over the entire integration span. 
To calculate it, we compute the Pearson coefficient for the first 10\% of the semi-major axis array to sliding 10\% segments within the range of 70\%-90\% of the array (incrementing segments by single array elements).
    \cite{Carruba:2021} found that the specific choice of the selected ranges does not significantly affect the computed ACFI for chaotic orbits with bounded motion;
    we chose our ranges to capture chaotic evolution that tends to present itself later in the simulations such as scattering events or long-term diffusion. 

    We take the ACFI to be the mean of the absolute value of the full set of the computed Pearson indices. 
    Objects that are well correlated, or have low chaos, will have ACFI values closer to 1, while more chaotic objects will have smaller values. 
    In SBDynT, the ACFI flag will be triggered for objects with an ACFI of 0.75 or less, indicating that the orbit demonstrates non-periodic evolution for more than half of the selected timescale.

    \item \textbf{Eccentricity and Inclination Vector Power Distribution }

    Chaos is sometimes described as unpredictable evolution, so chaotic motion is inherently non-periodic.
    Therefore, any method which measures the level of periodicity in motion over time can also be a measure of the chaos.
    The Fourier transform is an effective means of measuring the amplitude distribution of the frequencies present in a signal.
    While the Fourier transform of a perfectly periodic signal would contain all of its signal in a single frequency, non-periodic motion would show the mixing of many signals at different frequencies.
    We can thus use the power spectra of the eccentricity and inclination vectors (already calculated to determine proper elements) to measure the level of non-periodicity, or chaos, contained in the signal.
We define this {stability} indicator as the fraction of the total power of the unfiltered eccentricity and inclination vectors that is contained in the small body's proper $g$ or $s$ frequencies; we take this to be the sum of the four bins nearest to $g,s$ (to account for spectral leakage) divided by the sum of the entire power spectrum.

    For stable objects with regular motion, the total power should be overwhelmingly concentrated in the proper $g$ and $s$ frequencies; in contrast, objects which are unstable or are near resonances, will have less of their total power contained within their proper frequencies.
    
    We note that similarities appear at first glance between this indicator and the power distribution values discussed in Section~\ref{sec:mean_elements} used to define mean elements.
    However, this {stability} indicator differs from the mean element power criteria in two significant ways.
    First, this power distribution indicator measures the relative fraction of power contained in the primary or dominant $g,s$ frequencies, rather than in the 0-bin.
    Second, this indicator is computed on the unfiltered $\myvec{e},\myvec{I}$, rather than the filtered $e,I$ time arrays. 
This indicator can thus be considered a measure of how coherent the orbit’s full secular motion is, including both the linear and non-linear effects from the planets; in contrast, the mean element power criteria captures the amplitude of variability that exists when linear effects have been filtered out.
    
    Non-linear secular motion involves coherent evolution between $e,\varpi$ and $I,\Omega$, meaning the power distribution for secularly resonant objects will be primarily contained in the dominant $g$ and $s$ frequencies of the $\myvec{e},\myvec{I}$ vectors; so while both this indicator and the mean element power comparison measure chaotic motion, this indicator will less frequently identify objects in or near non-linear secular resonances as chaotic.
    As such, this indicator is preferred in defining chaotic orbits which display instability or are unpredictable.

\end{enumerate}

We note that because SBDynT produces integrations with the {\sc rebound} software, SBDynT integrations produced with the WHFast or IAS15 integrators have built-in support for the popular Lyapunov exponent and MEGNO {stability} indicators, which the user may implement if they wish in their integrations \citep{Rein:2016}. 
However, we note that those indicators are most useful for relatively short integrations for a number of reasons including machine precision limitations with repeated normalization over thousands of timesteps that can cause both of these indicators to exponentially diverge.
{Considering that the Lyapunov timescale of the solar system planets themselves is ${\cal O}(10^7)$~years,  \citep{Hayes:2007, Batygin:2008, Brown:2020}, the default SBDynT integrations, which have timescales of 10+ Myr, are not well-suited for the native Lyapunov exponent and MEGNO indicators contained in {\sc rebound}.} 
{Users interested  in short-term chaos and instability among small bodies may find better success and efficiency implementing those native chaos indicators within {\sc rebound}, whereas users interested in stability on secular timescales may find better success using SBDynT's stability indicators.}

{We note, however, that the longest-term secular evolution of small body orbits, (which induce chaotic effects in the orbital evolution, like scattering from enhanced eccentricity), can occur on these very long timescales.
For context, note that Neptune's apsidal and nodal precession occurs on the order of $\approx 2$ Myr, and the combined $g_8+s_8$ precession rate of Neptune circulates at a rate of $\approx70$ Myr.
TNOs which are strongly dynamically impacted by Neptune's evolution can be subject to significant forcing on $\approx70$ Myr timescales, exceeding the length of the typical Lyapunov timescale which can be computed for a system.}

While the indicators discussed here can be helpful for identifying chaos in a system, it's important to note that these only act as diagnostics for understanding the evolution of the system; they do not necessarily reveal what that chaos may look like in that particular case. 
In small-body dynamics, the definition of chaos can change depending on the needs of the researcher; one study may be interested in searching only for objects subject to ejection or collisional events, while another study may be interested in any objects in orbits which are bound to migrate to new regions of dynamical space. 
This relates directly to SSSBs which experience ``bounded'' chaos, where they only explore a small part of parameter space, but do so unpredictably \citep[e.g.,][]{mandd1999}. 
Bounded orbits of this type often occur in mean motion resonances, with the objects bound to some range of semi-major axis, eccentricity, and inclination, with forced terms causing the orbit to vary unpredictably within those ranges.
An example of this was observed in \cite{Milani:1992b}, who found that the asteroid Helga was stably contained in the 7:12 mean motion resonance with Jupiter, yet had a very short Lyapunov timescale.
This object was termed to experience "stable chaos" within the bound orbit, and many more such cases exist among the resonant small bodies throughout the Solar System. 
On the other hand, unbounded chaotic orbits involve some type of migration through phase space, often through scattering events or certain resonant encounters with the large planets. 
Such orbits may remain fixed in one orbital element, like semi-major axis or inclination, but will migrate to new orbital phase space among the other orbital elements.
Unbounded orbits can even lead to ejection or collision within the time-frame of the simulated trajectory. 
As such, providing different {stability} indicators can help to differentiate and even identify the different kinds of chaotic evolution present in small-body orbits.

We will demonstrate examples of the variety of {stability} indicators later in Section~\ref{sec:chaos_results}.

\section{Results}\label{sec:results}

Here we demonstrate the capabilities of SBDynT by comparing its calculated proper elements to the published catalogs of \citet{Nesvorny:2024} for asteroids and \citet{Knezevic:2000} for TNOs; we also include the Asteroid Families Portal catalog \citep{Novakovic:2022} in the asteroid comparison to benchmark the consistency of the previously published catalogs and their methods of proper element computation.
We also present a new catalog of proper elements for TNOs, alongside an additional new catalog of {stability} indicators for these objects. 
Additional work will be published in an upcoming paper describing some of the results of the new proper element catalog and {stability} indicators for the TNO region (Spencer et al., in prep).

\begin{figure*}[!htb]
    \centering
    \vspace*{-0.75cm}\hspace*{-1.2cm}\includegraphics[width=1.18\textwidth, keepaspectratio,clip]{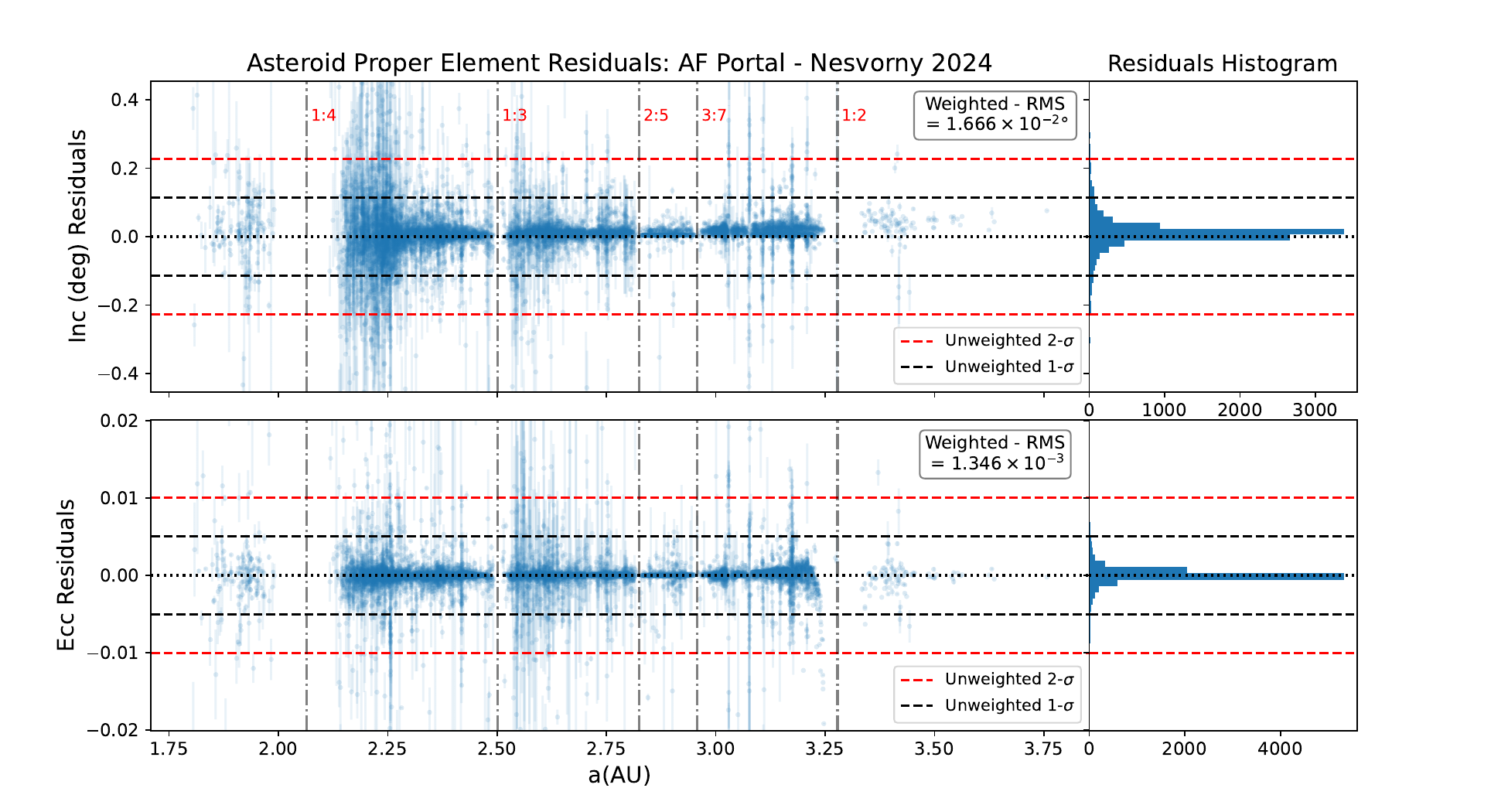}

    \vspace*{-0.85cm}\hspace*{-1.2cm}\includegraphics[width=1.18\textwidth, keepaspectratio,clip]{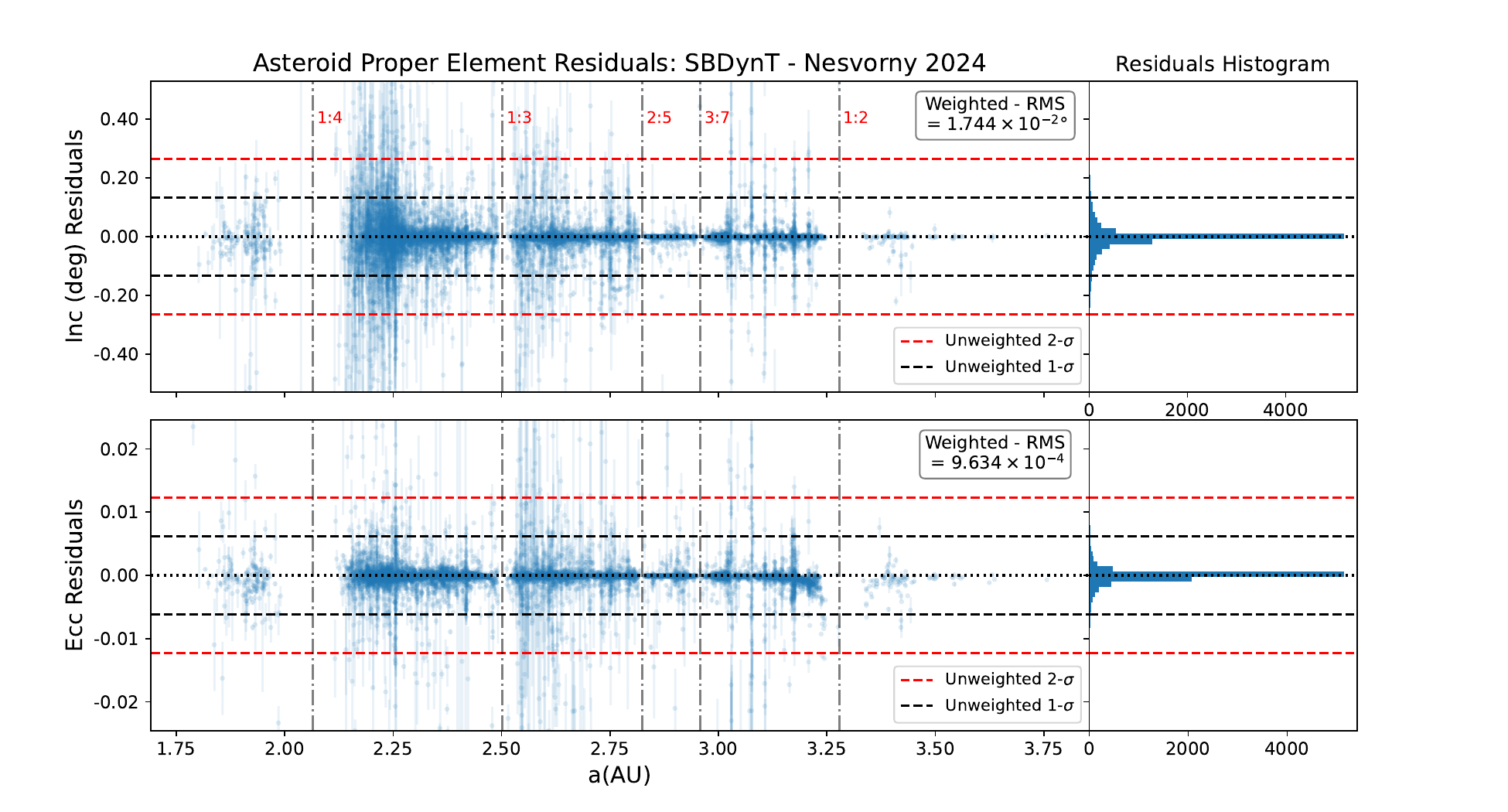}
    \caption{Top: Residuals comparing the Asteroid Families Portal Proper Elements against the \cite{Nesvorny:2024} Catalog of the first 10,000 numbered asteroids.
    Bottom: Residuals comparing SBDynT Proper Elements against the \cite{Nesvorny:2024} Catalog of the first 10,000 numbered asteroids. 
    The error-bars represent the uncertainties reported by \cite{Nesvorny:2024} in both figures. 
    The unweighted $\sigma$ is simply the standard deviation in the proper element residuals, and the error-Weighted RMS is described in Equation \ref{wRMS}.
    }
    \label{fig:CatalogResiduals}
\end{figure*}

\subsection{Asteroid Proper Elements}
\label{sec:nes_comp}

Because it is the most recent asteroid catalog with the most up-to-date orbits, we will make our first asteroid comparisons to the \cite{Nesvorny:2024} catalog. 
However, it is prudent to first briefly discuss the differences in methods between SBDynT and \cite{Nesvorny:2024} in calculating proper elements:
\begin{itemize}
    \item Integrator: \cite{Nesvorny:2024} uses the SWIFT integrator \citep{swift}, while SBDynT uses {\sc rebound}.
    \item Frequency Identification: \cite{Nesvorny:2024} uses the Frequency Modified Fourier Transform method \citep{Sidlichovsky:1996} to identify peak frequencies of the planets for filtering, while SBDynT makes use of the \texttt{numpy} FFT package and various masking and smoothing techniques to do so.
    \item Planets Included: \cite{Nesvorny:2024} includes all 8 planets in their integration. 
    While SBDynT is capable of including all 8 planets, in this comparison, we fold Mercury into the Sun, and integrate with the remaining 7 planets. \item Filtered Frequencies: \cite{Nesvorny:2024} filters just the planetary frequencies $g_j$ and $s_j$, while SBDynT also accounts for a number of non-linear secular frequencies, which can be reviewed in Appendix \ref{freq_list_app}.
    \item Filtering Method: \cite{Nesvorny:2024} filters out the chosen frequencies by subtracting the sinusoid using the amplitude and frequency returned by FMFT. 
    SBDynT smooths out all regions in Fourier space not associated with the small body's secular frequency instead. 
    \item Short-Period Terms: \cite{Nesvorny:2024} does not filter any short-period terms from their integration, while SBDynT does (see Section~\ref{sec:calc_proper}). 
\item Proper element window: \cite{Nesvorny:2024} calculates the proper elements from the mean of a 5 Myr window of the corresponding filtered osculating element. 
    SBDynT uses a 9.5 Myr window instead (discarding the first and last 0.25 Myr to remove edge effects). 
    \item Data Cadence: \cite{Nesvorny:2024} uses an integration stepsize of 1.1 days and  outputs data every 600 years, while SBDynT uses a stepsize of 5.6 days and outputs data every 1000 years. 
\end{itemize}

While most of these differences are very minor, others could result in noticeable differences in the proper elements in certain regions of orbital space, such as regions where non-linear secular frequencies are particularly strong or regions where short-period terms become much more substantial.
The residuals between the compared proper elements are shown in Figure~\ref{fig:CatalogResiduals}, with corresponding error-weighted RMS included for each orbital element being:
\begin{equation}
    \label{wRMS}
\sqrt{\frac{\sum(N^{24}_i-S_i)\times\sigma_{N^{24}_i}}{\sum\frac{1}{\sigma_{N^{24}_i}}}}
\end{equation}
where $N^{24}_i$ indicates the proper element of object $i$ in the \cite{Nesvorny:2024} catalog, $S_i$ indicates the comparison catalog proper element for the same object, and $\sigma_{N^{24}_i}$ indicates   \cite{Nesvorny:2024}'s proper element uncertainty for that object; the sums are performed over the overlapping sets of objects. 
The error-weighted RMS value is a better indication of the statistical variance between the catalogs than the regular RMS, as highly chaotic objects with large proper element uncertainties can unfairly increase the reported RMS.
To benchmark the consistency between SBDynT and the published catalogs, we also compare the proper elements of the sample of the first 10,000 numbered asteroids which are common to both \cite{Nesvorny:2024} and the Asteroid Families Portal, which consists of 9544 asteroids in total.

The error-weighted RMS values between the \cite{Nesvorny:2024} catalog of proper elements and the Asteroid Families Portal catalog are comparable to the same values contrasting SBDynT to the \cite{Nesvorny:2024} catalog. 
We note that the largest contributors to the weighted RMS in both proper $e$ and $I$ in the catalog comparisons is driven by a small sample of outliers with reported uncertainties in the range $10^{-4}$--$10^{-6}$ with residuals between catalogs on the order of $10^{-1}$. 
The inflated weight from the very low uncertainties amplifies these residuals in the weighted-RMS calculation, as is further discussion in Appendix-\ref{add_figs_app}.

We note a few features in Figure~\ref{fig:CatalogResiduals}, which shows how the residuals between the computed proper elements of the catalogs differ as a function of semi-major axis.
The most noticeable spread in the residuals between SBDynT and \cite{Nesvorny:2024} as well as between the Asteroids Family Portal and \cite{Nesvorny:2024} occurs between the 1:3 mean motion resonance with Jupiter, and the Saturnian $s_6$ resonance (around 2.2 AU), which is known to be primarily responsible for shaping the dynamical architecture of the inner main belt.
This region also corresponds to much larger uncertainties in the proper elements, so there is still good overall agreement in the proper elements in this region.

We also point out a specific feature visible in the proper $e$ residuals in both catalog comparisons that occurs near the 1:2 mean motion resonance with Jupiter. 
In an investigation of these objects, we find that proximity to that resonance generally produces strong short-period frequencies in their $e$ evolution. 
For some of these near-resonant objects, the short-period terms in Fourier space can be responsible for as much as 1-2\% of the total power of the full osculating $e$.
An example is shown in Appendix \ref{sec:short_period}, which is representative of the objects in this observed region.
Filtering out the short-period terms introduces a systematic decrease in the proper eccentricity of about 1\%, 
explaining the small deviation from the \cite{Nesvorny:2024} values.
Small differences among the catalogs such as these should serve as a reminder that while the numerical variations within a specific proper element computation may be small, the choice of frequencies to filter out can still produce systematic differences in the proper elements that exceed the numerical uncertainties of the method itself.

\subsubsection{Numerical Precision Comparison}
\label{num_precision}

Having established the accuracy of SBDynT’s proper elements,  we next consider the numerical precision of our method by comparing the distribution of the computed distance metric {stability} indicators (Equation \ref{dist_met}) for each catalog, as done by \cite{Nesvorny:2024}.
Mentioned previously in Section~\ref{sec:chaos}, asteroids with a $d_{met} > 100$ m/s are typically not considered for family finding, and so any method of computing proper elements should theoretically achieve uncertainties that typically lie below this cutoff to be useful for family finding.

\begin{figure}
    \centering
   \hspace*{-0.85cm}\includegraphics[width=1.2\linewidth]{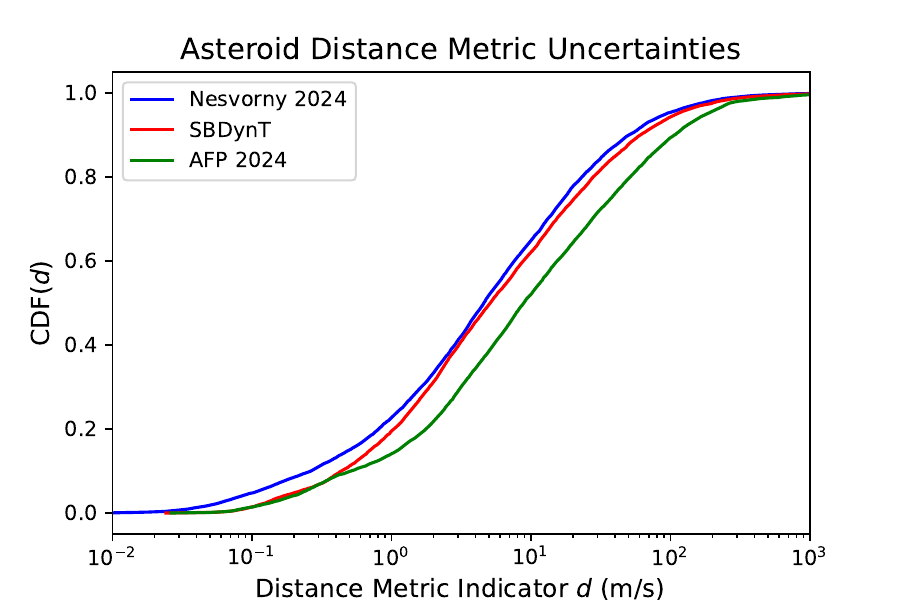}
    \caption{Cumulative distribution functions of the distance metric used in family finding computed using the uncertainties reported by the asteroid catalogs. Proper elements for the purpose of family finding are considered trustworthy if the uncertainties correspond to relative velocities below 10 m/s. SBDynT's uncertainties have a nearly identical distribution as the \cite{Nesvorny:2024} catalog, with nearly 60\% of the objects we consider falling below this threshold, and outperform the Asteroid Families Portal, though the difference is considered minor. Compare to Figure 2 in \cite{Nesvorny:2024}. See also Appendix \ref{add_figs_app} for a related figure.}
    \label{fig:hcm_cdf}
\end{figure}

A comparison of the uncertainties measured using this distance metric for each catalog sample is shown as cumulative density functions in Figure \ref{fig:hcm_cdf}. 
The uncertainties reported by SBDynT are generally smaller than the uncertainties reported by the Asteroid Families Portal; the latter is considered sufficient for family finding, so SBDynT will be as well. 
In comparing both accuracy and precision, we find that there is closer agreement between SBDynT and the recently published \cite{Nesvorny:2024} catalog than between the \cite{Nesvorny:2024} and \cite{Novakovic:2022} catalogs; discrepancies between SBDynT and established catalogs are comparable to the differences between the catalogs themselves. 
This indicates that SBDynT's proper element results are comparable to the methods employed by other proper element catalogs. In addition, differences between the catalogs are almost entirely focused in chaotic regions near mean motion resonances or are systematic, meaning the primary uses of proper elements such as family finding or family age estimates are still valid within each catalog.

We also note that while our sampling frequency resolution was primarily coarser than both of the compared catalogs, the closely equivalent results indicate that our method compensates for some of the drawbacks of that, effectively producing good proper elements for a smaller computational cost in both memory and CPU-time.
Further discussion and diagnostic plots comparing the three proper element catalogs and a general discussion of the accuracy of proper element computation can be found in Appendix \ref{add_figs_app}.

\subsection{TNO Proper Element Comparison}\label{ss:tno_pe}

The only published full catalog of TNO proper elements is the AstDys catalog \footnote{\url{https://newton.spacedys.com/astdys2/index.php?pc=5}}, which was calculated using the OrbFit software \citep{Knezevic:2000}. 
The most recent update to the TNO catalog in AstDys was December 2024, with a selection of 2126 TNOs. 
These TNOs do not represent the entire observed set of small bodies orbiting beyond Neptune; they were selected to fit the criteria for proper element computation.
\cite{Knezevic:2003} specifically mentions that they do not compute proper elements for objects with $e>0.3$, as these objects are typically very chaotic.
However, there are still a number of objects in the current AstDys catalog which satisfy having $e<0.3$ which are not included.
An additional criteria is referenced with respect to asteroids in \cite{Knezevic:2019}, in which they mention removing asteroids from their analysis for which $LCE > LCE_{rc}$ and for which $\sigma_a < 0.0003$ AU. 
It could be that a similar criteria were applied in the case of the TNO catalog, though with a higher $\sigma_a$ criteria, as TNO's are expected to have a more diffusive semi-major axis than the asteroids.

\begin{figure*}[!htb]
    \centering
    \vspace*{-0.75cm}\hspace*{-0.6cm}\includegraphics[width=1.1\textwidth, keepaspectratio]{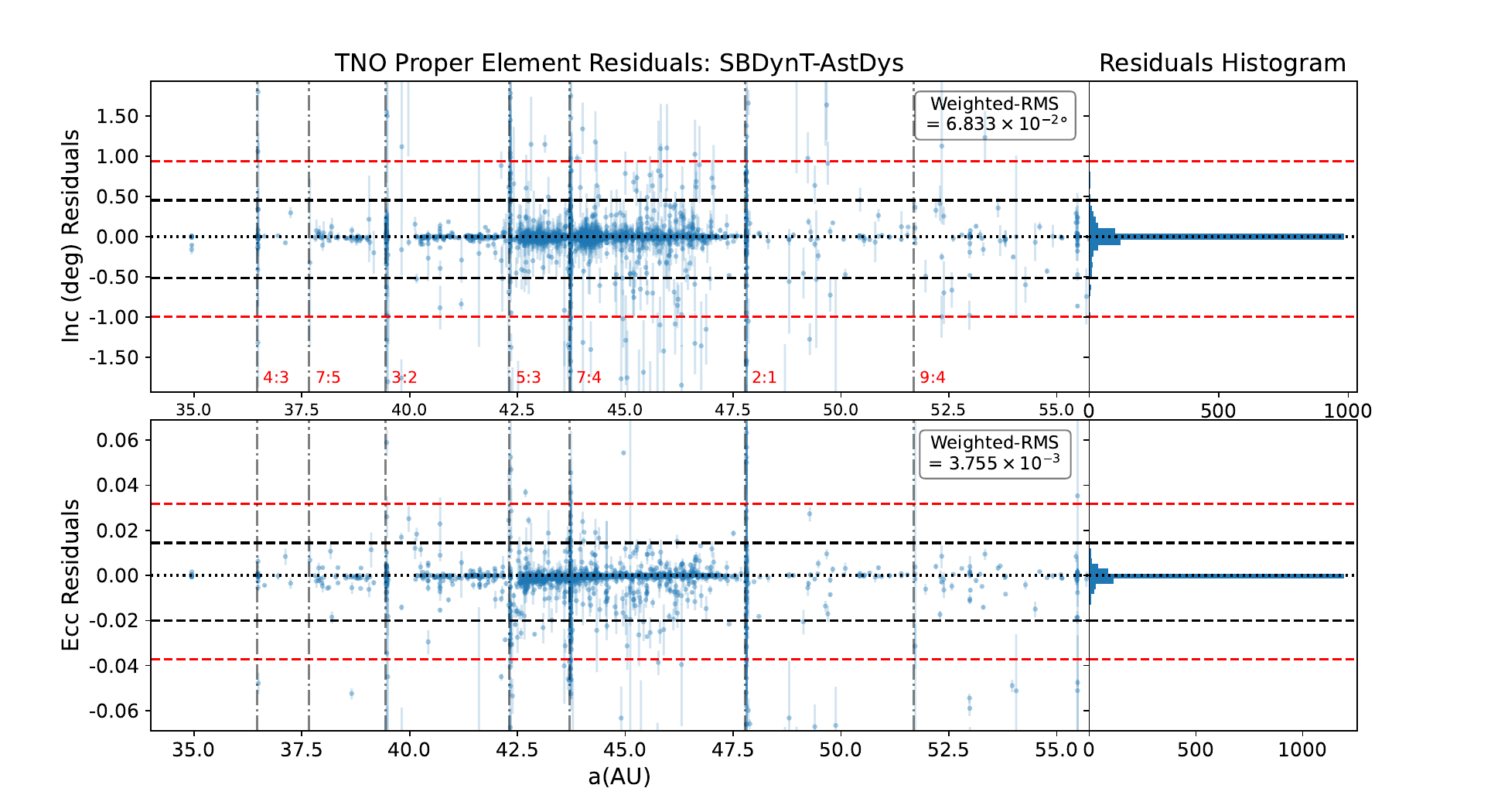}
    \caption{Residuals comparing SBDynT Proper Elements against the AstDys Catalog of TNOs, with uncertainties included.  This plot restricted to $a<56$ AU to display the most relevant details in the TNO belt, even though the AstDys catalog contains objects with $a_{max}\approx65$ AU. Similar to Figure~\ref{fig:CatalogResiduals}, but with different ranges, the largest deviation between the two catalogs occurs at mean motion resonances with Neptune, which is to be expected. }
    \label{fig:TNOResiduals}
\end{figure*}

Figure~\ref{fig:TNOResiduals} shows the residuals between SBDynT's proper elements and those reported by AstDys, including the weighted RMS (as described in Section~\ref{sec:nes_comp}). 
Similar to the asteroid comparison, the largest residuals between SBDynT and the AstDys catalog occur at mean motion resonances (in this case with Neptune rather than Jupiter), which is to be expected. 
However, there are also some significant residuals in the Cold Classical belt region (low-$e$, low-$I$ orbits with $a\approx42-46$~au) that are large when compared to the associated uncertainties from the proper element computation. 

We find that the largest residuals contributing to the weighted RMS measurement are all Cold Classical objects that are strongly affected by the secular $g+s-g_8-s_8$ commensurability with Neptune.
Amongst these objects, there are also a number of cases where the reported proper $g$ or $s$ frequencies differ between the catalogs by significantly more than the frequency resolution of the FFT methods. 
As discussed in Section \ref{sec:mean_elements}, objects that exist in or near non-linear secular resonances can librate over very long timescales.
In the case of these Cold Classical TNOs, they all experience periodic evolution on the timescale of $g_8+s_8\approx66$ Myr. 
Because of this, defining the filtered elements as ``proper'' is difficult, and these objects
fall into the ``mean'' element prescription by SBDynT based on the magnitude of the uncertainties in their filtered elements. 
In other words, the bulk of the large proper element residuals between SBDynT and the AstDys TNOs are due to secular resonant objects identified by AstDys as having very precise uncertainties, despite having large-amplitude long-period terms in the orbital evolution.
Examples of such objects can be found in Appendix \ref{sec:longperiod_tnos}, where we further discuss these effects.

We note that while the comparison is not shown here, we did make a comparison between SBDynT's computed proper inclinations and the \cite{Huang:2022} catalog of proper inclinations.
We find that there is similar agreement between SBDynT and the \cite{Huang:2022} free inclinations as to the catalog comparisons we make above. 
Overall, we find that the filtering done by SBDynT performs adequately in case of the TNOs.
We will include a full exploration of TNO proper elements in a follow-up paper (Spencer et al. in prep).

\subsubsection{TNO Catalog Uncertainties}

\begin{figure}
    \centering
    \includegraphics[width=1\linewidth]{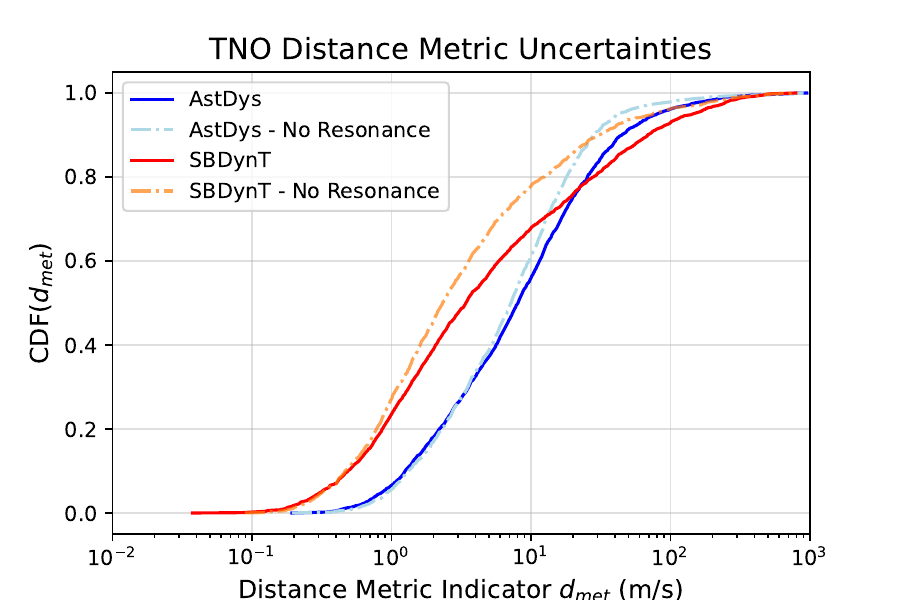}
    \caption{Cumulative distribution functions of the distance metric computed from the proper element uncertainties reported by the AstDys and SBDynT TNO catalogs. 
    Both distributions only represent the TNOs reported by the AstDys catalog as of December 2024, which consists of 2124 TNOs. An additional line shows the distribution of proper elements for all 5060 known TNOs.
    Note that the SBDynT CDF becomes more shallow around 10 m/s, which is primarily due to higher uncertainties reported by SBDynT for mean-motion resonant objects within the sample.}
    \label{fig:hcm_cdf_tnos}
\end{figure}

We present the cumulative distribution functions of the $d_{met}$ for the TNOs as retrieved from the proper element uncertainties for both the AstDys catalog and SBDynT in Figure~\ref{fig:hcm_cdf_tnos}.

As mentioned in Section~\ref{sec:chaos}, we find that more than half of all known TNOs have a $d_{met}<10$ m/s, indicating that generally, TNO orbits are well-constrained.
We note that below a $d_{met}<20$ m/s, SBDynT generally outperforms the AstDys catalog for the produced TNO proper element uncertainties. 
However, for $d_{met}>20$ m/s, the AstDys catalog begins to outperform SBDynT, though very slightly.

We find that this is largely due to mean-motion resonant objects within our sample which experience more chaotic evolution, especially near the end of the integration, and objects with long-period secular terms which are not fully captured by the integration.
Indeed, if we remove objects in the $3:2$, $5:3$, $7:4$, and $2:1$ mean-motion resonances, and objects which are identified as being within the $g+s-g_8-s_8$ resonance from both catalogs, we find that the improvement is more exaggerated for SBDynT than for the AstDys catalog.
In addition, SBDynT's integration for 150 Myr rather than 100 Myr has the added impact of allowing for more potential migration for the resonant TNO's; this causes some objects which were initially stable within a resonance to become ejected near the end of the simulation, producing a larger uncertainty measurement.
Therefore, several of the objects in our longer integrations may not have fulfilled the criteria for stability originally used for selection when producing the present-day AstDys catalog. 

While this does highlight the differences between SBDynT's method versus the OrbFit method of proper element computation, we emphasize that this effect occurs for objects which already have more poorly defined proper elements, and the overall uncertainties reported by each method remain close to each other, while also remaining accurate.
Thus, SBDynT proper elements achieve mostly similar numerical accuracy as the OrbFit code, and could improve if the integration length is handled more precisely in specific cases.

\subsection{Chaos Indicator Results}
\label{sec:chaos_results}

As discussed in Section~\ref{sec:chaos}, the {stability} indicators provided by SBDynT are diverse in terms of their sensitivity and relation to different types of dynamical evolution.
To illustrate this, we compute stability maps for the proper power distribution, ACFI, entropy, and Distance Metric indicators for the Classical belt region of TNO space. 
This is done by first producing a fine grid of 102,951 individual test particles with ranges and spacings in semi-major axis, eccentricity, and inclination as given in Table~\ref{tab:grid_params}. All of these particles were inserted into simulations at the default SBDynT epoch of JD $= 2459580.5$ with $\omega = \Omega = M = 0^{\circ}$.

\begin{table}[!ht]
    \centering
    \hspace*{-1.5cm}\begin{tabular}{|c|c|c|}
    \hline
    &  Range & $\Delta$x\\
    \hline
    Semi-major axis & 40 - 48 AU & 0.1 AU\\
    Eccentricity & 0 - 0.3 & 0.01\\
    Inclination & $0^{\circ}$ - $40^{\circ}$ & $1^{\circ}$\\
    \hline
    \end{tabular}
    \caption{Barycentric osculating ecliptic parameter space sampled by the grid of particles to compute stability maps of the Classical Belt TNOs.}
    \label{tab:grid_params}
\end{table}

The best-case sampled grid would additionally include a range of sampled $\omega$, $\Omega$, and $M$ points in our distribution of particles. 
However, in the interest of memory and efficiency, we limit ourselves to a 3-dimensional sample rather than a full 6-dimensional sample of orbital parameter space. 
We note that sampling our grid in osculating ecliptic elements inherently limits the sample of low proper inclination orbits because the the locally forced plane of the Classical belt is offset from the ecliptic plane by $\approx1.7^{\circ}$; in other words, this grid inherently undersamples objects with proper inclinations near $0^{\circ}$.
However, our grid is sufficient to demonstrate the broad nature of the chaos in these regions; a more detailed analysis which will account for the offset of the invariable plane will be approached in a future paper. 

We integrate the grid of particles for the default 150 million years with SBDynT and save outputs every 50,000 years; this is more coarse than the default setting, but is done for disk space and memory purposes.
While setting a lower time resolution does affect how well we sample shorter-term secular dynamics for the particles, the {stability} indicators we will showcase with this grid are more sensitive to secular effects, and should still respond well to coarser time-arrays.
The final result of the integrations produced 42 Gb simulated data in our REBOUND binary files and required approximately 5000 CPU-hours to integrate, parallelized across 81 individual AMD EPYC 7763 2.45 GHz processors.

The associated {stability} indicators for each particle were then computed much more quickly.
Computing the distance metric for every particle took the longest time, as this requires first computing the proper elements for every particle with the same hardware; it took $\approx 250$ CPU-hours, averaging 9 CPU-seconds per particle. 
The rest of the {stability} indicators can be computed quickly enough that they were produced locally from a personal laptop on a single Intel Core Ultra 9 185H 2.30 GHz processor.
The fastest indicator to compute was the entropy indicator, which took less than a minute to compute for the entire grid.
Computing the ACFI indicator took 12 minutes, and the power concentration indicator took 135 minutes.

\begin{figure*}
    \centering
    \hspace*{-1cm}
    \includegraphics[width=1.1\linewidth]{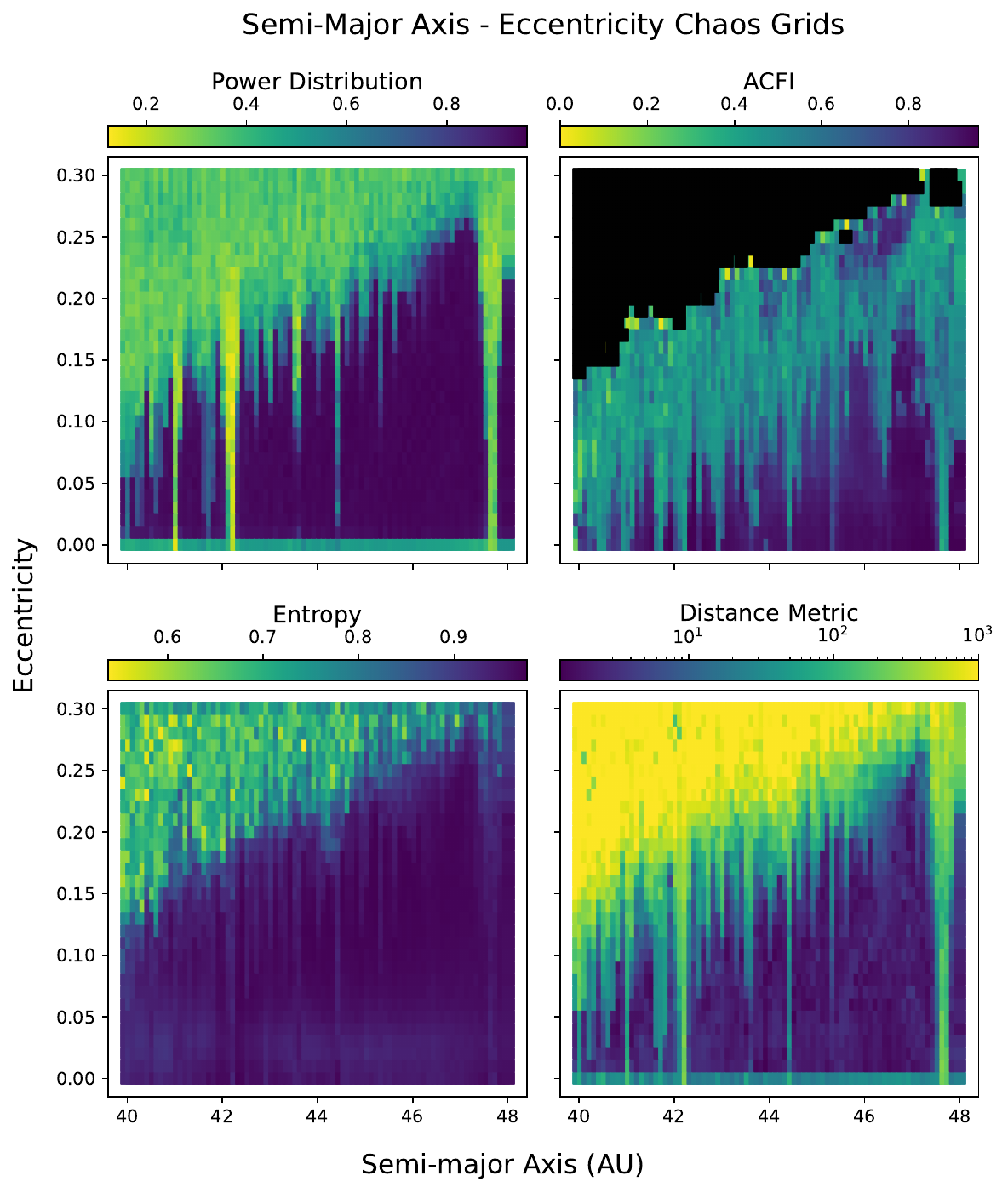}
    \caption{{Stability} indicators computed across a grid of particles as a function of semi-major axis and eccentricity. The color of each point gives the median value of the indicator across all inclinations for the particles in each $a$-$e$ bin. Colorbars are scaled so that yellow/brighter colors indicating choas/instability with bluer/darker colors indicating regularity/stability. Scattered objects have unusual ACFI values and are colored black.}
    \label{fig:sma_ecc}
\end{figure*}

\begin{figure*}
    \centering
    \hspace*{-1cm}
    \includegraphics[width=1.1\linewidth]{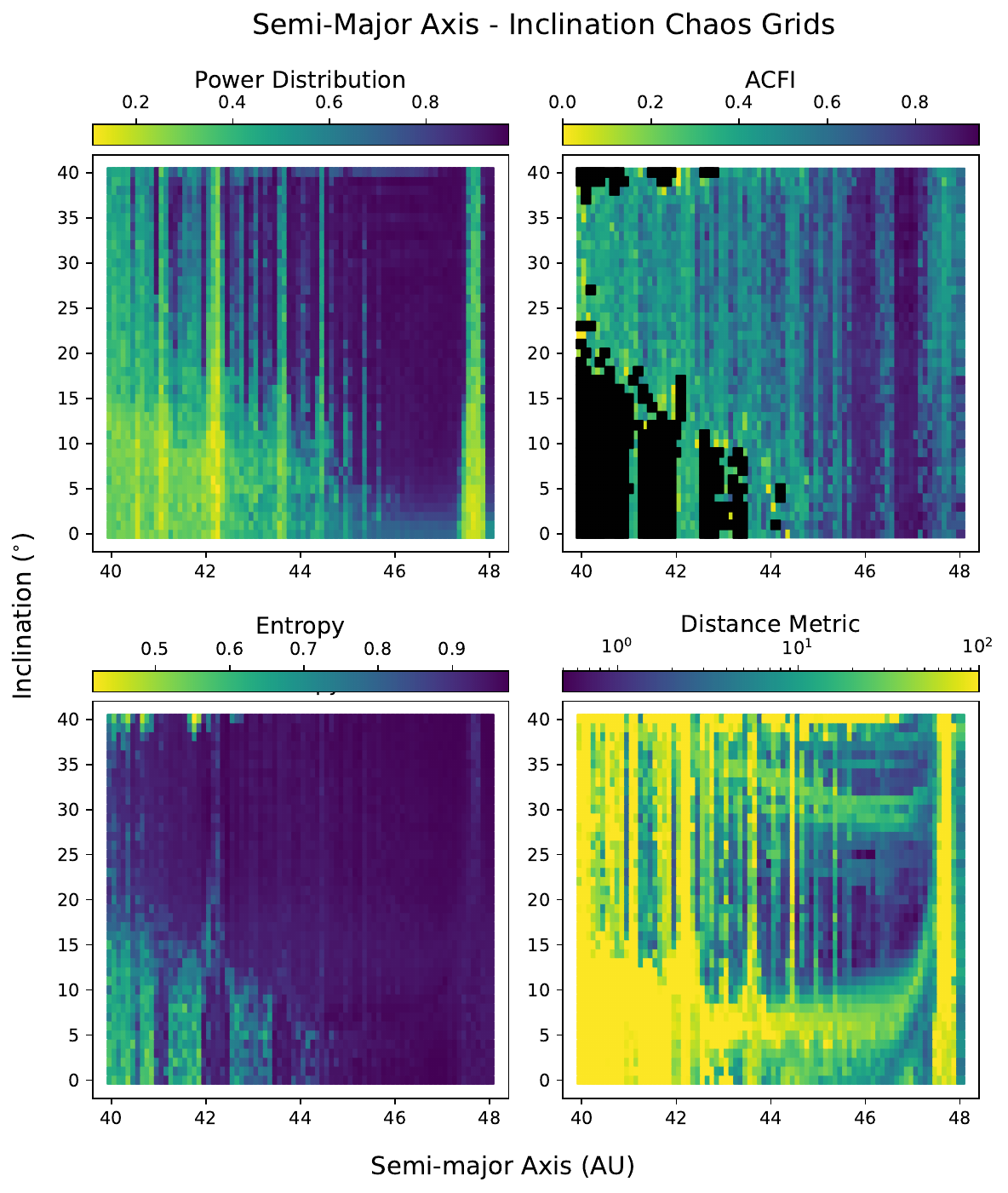}
    \caption{{Stability} indicators computed across a grid of particles as a function of semi-major axis and inclination. The color of each point gives the median value of the indicator across all eccentricities for the particles in each $a$-$I$ bin. Colorbars are scaled slightly differently that in Figure \ref{fig:sma_ecc}, but with the same sense (brighter is more chaotic).}
    \label{fig:sma_inc}
\end{figure*}

Because this grid of particles is distributed throughout 3 dimensions, we display the {stability} indicators in 2-dimensional grids in Figures~\ref{fig:sma_ecc} and \ref{fig:sma_inc}.
The color of each point in the grid represents the median of the {stability} indicator along the orbital element from Table~\ref{tab:grid_params} not represented on the x- or y-axes. 
These chaos maps thus only indicate a relative measure of the instability of the Classical belt TNOs in each of the two dimensions represented in the Figures; we will further explore the full 3-dimensional impact space of this region in a future paper \citep{Spencer:inprep}.

The first 3 indicators each have a range of possible values from 0--1, where a value of 1 indicates stability and 0 indicates instability. 
In contrast, the value of the distance metric {stability} indicator is not bounded and large values indicate instability.
For visual consistency between colors indicative of unstable vs. stable regions, we have inverted the color-map for the distance metric in Figures~\ref{fig:sma_ecc} and \ref{fig:sma_inc}. We also note that we do not restrict the color-map range for the {stability} indicators to match identically between the figures or panels because these chaos maps are meant to display the relative differences in the stability of the Classical belt.

There are some obvious visual features in Figures~\ref{fig:sma_ecc} and \ref{fig:sma_inc} that highlight the differences between the {stability} indicators. 
For example, each indicator suggests higher chaos for orbits with perihelion values $q\leq 38$ AU in Figure~\ref{fig:sma_ecc}. 
Easily noticeable is the black region for ACFI for orbits with $q\lesssim 34$, where black indicates a median value of $ACFI\lesssim0$.
The reason for this is related to the definition of ACFI, which effectively describes how correlated the early orbital evolution of a particle is to later orbital evolution nearer to the end of the simulation. 
Not surprisingly, we find that objects with initial conditions above this value of $q$ are scattered within the first several outputs of the simulation archive, typically resulting in either an ejection or a collision with the Sun.
If the particle is not removed from the simulation (because the ejection does not trigger the removal protocol within {\sc rebound}) the early simulation and the late simulation orbital elements are effectively identical, causing the reported ACFI indicator to appear artificially low. 
Cases also exist where the object remains bound to the Sun, but is scattered so regularly by Neptune, that the standard deviation for the semi-major axis is very high; this causes the Pearson coefficient to be extremely low in both cases, even though chaotic motion is highly present in the orbit.
To account for scattering orbits, we have SBDynT report an $ACFI=-1$ for objects where at any point in the simulation $a\leq0,\,a>10^6$ AU, and/or $\sigma_a>5$AU. 
Orbits which satisfy this criteria are so present for orbits with $q\leq34$ AU, that the median value becomes negative, which we dictate with a separate color. 
Perhaps even more surprising is a small region of stability between $46 < a < 47.5$ AU and $0.22 < e < 0.28$.
We find that orbits in this region still experience scattering, but are distant enough from Neptune that the evolution in $a$ rarely exceeds $a\geq1$ AU over the full 150 Myr integration.
As such, ACFI reports these objects as more stable.

Figures~\ref{fig:sma_ecc} and \ref{fig:sma_inc} also show obvious correlations between the locations of major mean motion resonances and higher chaos.
However, the indicators differ in terms of the reported strength of chaos in the resonances and the width of the resonance. 
For example, the ACFI indicator reports higher chaos for objects within many of the weaker mean motion resonances at lower eccentricities than the other indicators do. 
However, for each indicator, the width of the resonance generally follows the same shape for each indicator, with larger eccentricities and lower inclinations corresponding to a larger width of the resonance.

Another notable feature is a line of higher-chaos objects at $e=0$ for the Power and Distance Metric indicators in Figure~\ref{fig:sma_ecc}.
This is an artifact of collapsing our 3-dimensional grid into a 2-dimensional map and having defined our grid in osculating elements reflecting two primary dynamical effects.
First, because the vertical component of the angular momentum, $L_z = cos(I)*\sqrt{1-e^2}$, is conserved, an object with an $e\approx0$ and a large inclination can experience vZLK oscillations where eccentricity and inclination are exchanged. 
This non-linear secular effect causes the eccentricity to grow and the inclination to shrink, contributing to larger uncertainties being reported in the proper elements for these objects and thus larger distant metrics.
Second, an object with a proper eccentricity very close to 0 will have orbital evolution that is primarily dominated by the forced terms; this is related to the fact that a circular orbit has no apsides, and therefore no true apsidal precession.
Any observed apsidal precession must therefore be forced.
As the power distribution indicator and the distance metric indicator both rely on the strength of the proper motion relative to the forced terms within the power spectra, $e\approx0$ objects thus report higher chaos for these indicators.

Many more features stand out in the chaos maps, including known regions of instability, such as the low-inclination region below 42 AU corresponding to the $\nu_{18}:s-s_8\approx0$ secular resonance.

It is clear from these chaos maps that each {stability} indicator highlights distinct facets of dynamical evolution, from resonant interactions to long-term diffusion and scattering events. 
The variety of these indicators underscores that they are not standalone measures, but rather complementary tools that reveal elements of orbital behavior. 
It falls to the user to interpret these indicators thoughtfully, recognizing the specific insights each offers while being mindful of their limitations.
This can lead to better understanding of the system’s stability, potential chaotic behavior, and evolution over time, allowing for a more comprehensive interpretation of complex dynamical systems.

\section{Computational Efficiency and Performance}\label{sec:comp_efficiency}

SBDynT's development is partially in preparation for the incoming flux of newly discovered objects expected from LSST. LSST is projected to discover $\sim$40,000 TNOs and nearly 5 million new asteroids \citep{lsstsb, Kurlander:2025}, with the majority of discoveries expected to happen within the first two years of operation. 
\citet{eggl2019} reports that the number of main belt asteroid discoveries per night could be up to 100,000, with an average of 2,000 per night. 
Because new observations constantly improve the known orbit of an object, many objects will be dynamically investigated multiple times (e.g., at the times of LSST Data Releases). 
Thus it is useful to provide an estimate here of the computational time required to run SBDynT on the entire LSST catalog.
The time estimates for integration and proper element computation discussed here are computed from integrations performed on cores contained in the marylou12 cluster, which is the largest cluster on the BYU Mary Lou Fulton Supercomputer\footnote{\url{https://rc.byu.edu/documentation/resources}}.
The marylou12 cluster is composed entirely of AMD EPYC 7763 2.45 GHz cores, with 128 cores per node.

As the expected number of asteroids exceeds the expected TNO discoveries by $\approx125\times$, computing proper elements and {stability} indicators for the asteroids will require the most computation time.
As described in Section~\ref{sec:proper_elements}, to produce proper elements for an individual asteroid, the user must run a {\sc rebound} simulation of the asteroid's orbit over 10 Myr, which takes about 20 minutes per asteroid; the calculation of the proper elements from the completed simulation itself takes $\approx3$ seconds per asteroid.
Calculating {stability} indicators is similarly quick. 
For the RMS indicator, producing a 1 Myr integration of 10 clones takes about 2 minutes per asteroid.
The other indicators can be computed from the default SBDynT integration used to compute the proper elements for the small body.
As discussed in Section~\ref{sec:chaos_results}, the time to compute the {stability} indicators is quite fast, with the time required for computation of the proper element far exceeding any of the other indicators.

If LSST discovers 2 million asteroids in the first year, we estimate that it would take on the order of 700,000 CPU hours to dynamically analyze the entire catalog with SBDynT. 
By allocating 10,000 processors to the task, this could be shortened to a 3 day calculation for all SSSBs. 
While not a small amount of CPU time, it is a very feasible task to run such an analysis annually (e.g. at each LSST data release).
We are also exploring methods for propagating uncertainties as orbit fits improve that might allow for more efficient choices in how and how often to re-analyze objects.

We also note that this estimate of the time required to integrate the full asteroid catalog represents the case where every asteroid is investigated using an individual integration containing the asteroid, the seven giant planets, and the Sun, equating to 9 total particles total in the simulation.
This is the default behavior for SBDynT, which we expect will typically be used to integrate singular objects with best-fit orbits and orbital uncertainties defined at a variety of epochs set by external orbit databases and thus not controlled by the user.

However, a large catalog data release like LSST is likely to present the best-fit orbits (and their uncertainties) for discovered asteroids at a common epoch.
This means it would be possible to include multiple asteroids (all as massless particles) within a single integration, which could significantly improve the efficiency of integrating large samples of asteroids.
As the asteroids are always included as test particles, the addition of each asteroid to the simulation scales as $\mathcal{O}(N)$ rather than $\mathcal{O}(N^2)$ and the more computationally expensive propagation of the massive planets can be done just once per large group of asteroids rather than being repeated for each asteroid.
This implies that the improvement to efficiency of including multiple asteroids in a simulation drops off at a number of asteroids approximately equal to the square of the number of active objects in the simulation.
As the asteroid runs include 7 planets and the Sun, it seems reasonable that including up to 50 asteroids in an integration could increase efficiency while balancing the amount of real-time required to run the analysis for all objects.

This exact process was used to improve the efficiency of integrating the grid of particles demonstrated in Section~\ref{sec:chaos_results}.
While a typical TNO integration can take upwards of 15 minutes per particle, merging all of the particles for a single semi-major axis bin into a single {\sc rebound} integration reduced the time to run to $\approx3$ minutes per particle. 
Careful implementation of SBDynT to include many asteroids in a single integration can significantly decrease the required CPU-time to dynamically analyze many objects at once.

We performed a brief test of the improvement to CPU efficiency by including multiple objects in a single integration; this is done by simply integrating a single objects and a varying number of clones.
We find that the improvement to the efficiency drops off after including more than 200 asteroids in a single integration.
At this point, the improvement to total CPU-time is on the order of $\approx95\%$, or in other words, the efficiency is improved by a factor of 20.
This brief estimate would indicate that the total CPU-time of integrating 2 million asteroids with a common best-fit orbital epoch would take $\approx35,000$ CPU-hours, which could be reasonably done in less than a day on a high-performance computer. 

{We reiterate that as SBDynT is built using the REBOUND integrator, many options beyond the default settings exist for the user which may allow the user to improve speed or memory. 
SBDynT's default settings are meant to capture the most accurate orbital simulations, but reasonable modifications, such as choice of integrator or integration timestep, may still produce reasonable results.}

We note that SBDynT is also well-suited for application to simulated datasets, which are often larger than observational datasets but equally critical for exploring theoretical scenarios and testing dynamical hypotheses. 
Simulated data allows for controlled experiments on the evolution of orbital systems, providing insights into processes that may be difficult to observe directly. 
By leveraging simulated data, researchers can investigate the long-term behavior of hypothetical or poorly constrained systems, test the impact of specific perturbations, and explore the dynamics of objects beyond the current observational limits, whether due to size, distance, or reflectivity.

\section{Conclusions}

We describe how SBDynT computes synthetic proper elements for asteroids and TNOs and compare our results with published catalogs. 
We find that the proper elements calculated by SBDynT are comparable to those provided in previous catalogs, confirming the reliability of the code for dynamical studies. 
In addition, the level of precision for the proper element calculation has been confirmed to be sufficient for the process of asteroid family finding and age estimates, which are some of the primary uses of proper orbital elements for solar system small bodies. 
We also show that SBDynT offers sufficient efficiency and computational tractability for analyzing future surveys that will discover millions of new asteroids and tens of thousands of new TNOs.

We also describe SBDynT's calculation of {stability} indicators for chaotic orbital evolution.
We provide a proof-of-concept use of these indicators that highlights how different indicators flag different kinds of chaotic orbital behavior.
The {stability} indicators that are supported by SBDynT provide the dynamicist with a wide range of options for identifying chaos and divergent orbital motion.

Looking ahead, SBDynT is well-positioned as a valuable community tool for dynamical studies of small bodies, offering a framework that can be quickly applied to new discoveries as they are made. 
Ultimately, our goal is to utilize SBDynT to dynamically analyze the LSST catalog of small body discoveries, enabling automated dynamical characterization of newly detected objects with sufficiently long observational arcs. 
As the code continues to evolve, contributions from the broader community will be essential in refining its capabilities and ensuring its long-term impact on planetary dynamics research.

\vspace{12pt}
\noindent{\it Acknowledgements:}
This work was supported by NASA grant 80NSSC23K0886 (PDART). 
DS had additional support from NASA grant 80NSSC23K1374 (FINESST). 
KV acknowledges additional support from NASA grants 80NSSC23K0680 and 80NSSC23K1169.
We thank Josh Bushman for his help producing a graphic design logo for the SBDynT code software, included on the title page of this paper. {We also would like to thank our referee for the excellent comments which helped us to clarify and better explain some nuances of SBDynT's code and numerical accuracy.}

\facilities{ADS, BYU Office of Research Computing, MPC, JPL Horizons}
\software{
{\sc rebound} \citep{rebound},
numpy \citep{numpy}, 
SBDynT \citep{katvolk_2026_19138321},
matplotlib \citep{matplotlib},
scipy \citep{scipy}}
\newpage
\bibliography{AAS73389R1_new.ms}
\bibliographystyle{aasjournalv7}

\newpage
\begin{appendices}
\section{Dominant Frequency Identification}
\label{freq_id_app}

The process of identifying the dominant frequencies in the Fourier transform of the $e$ and $I$ vectors for both the planets and the small body is an important step to properly computing the proper elements of a small body.
As discussed in Section~\ref{sec:calc_proper}, this is more complex than just finding the peak frequency in the Fourier spectrum.
Due to spectral leakage, signals in the Fourier transform of the time array are spread out among several bins.
This can cause dominant short-period frequencies to spread out and appear less prominent than longer-period frequencies in the signal. 
We instead compute the discrete sum of the power for each region of the signal to identify the dominant frequency, while also accounting for the influence of the other planets in the signal.

SBDynT calculates the cumulative sum of the signal's power separately for both the positive and negative region of frequency space in the signal. 
Then the local sum of power for each frequency can be computed by finding the difference in the cumulative sum over a window of frequencies, and saved at every point.
This is equivalent to discretizing the power in each region using a binning method, but is faster because the sums only have to be computed once. 
The peak value in this discretized power spectrum is then tentatively identified as the dominant frequency. 

For the planets, SBDynT performs this process in an order relative to their individual gravitational influence on each other and the rest of the Solar System.
Before beginning the above search for the planetary dominant frequency, we first perform a quick filter of all of the previously identified planet frequencies by simply dividing the region in the power spectrum near that frequency by a value of 500; this value was selected through empirical testing to most accurately account for the influence of the planet's on each other.
For example, if SBDynT begins searching for the $g8$ frequency in Neptune's Fourier spectrum, it first filters out the $g_5$, $g_6$, and $g_7$ frequencies associated with Jupiter, Saturn, and Uranus.

Once each of the $g_i$ and $s_i$ frequencies are identified in this way, the same procedure is then applied to the small body to find its dominant $g$ and $s$ frequencies.
The initial filter of the planetary frequencies is less severe, however, because the dominant frequency for a small-body in or near a secular resonance could be equal to one of the planetary frequencies, indicating presence within the secular resonance. 
We empirically find that a filter dividing the power spectrum near the $g_5$, $g_6$, and $s_6$ frequencies by a value of 50 and all other planets by a value of 20 yields sufficiently accurate initial $g$ and $s$ identifications.

In cases where the small body's peak frequency is in or near a resonance with a planet frequency, this method often skews the determined dominant frequency somewhat in the direction of that nearby planetary resonance. 
To account for this, the final step involves taking a weighted mean of the nearby frequencies using their associated power as the weighting factor.
This final step ensures that the dominant $g$ or $s$ frequency identified by SBDynT is associated with a local peak value in the signal.
There are also some cases where spectral leakage produces a plateau of sorts; in these cases, the peak frequency is located at the center of the flat plateau, rather than at a peak value.
We have confirmed that these corrections do successfully find $g$ and $s$ frequencies for small bodies that agree more closely with the published catalogs.

\begin{figure*}[htb]
    \centering

    \hspace*{-1cm}\includegraphics[width=\textwidth, height=1.25\textheight, keepaspectratio]{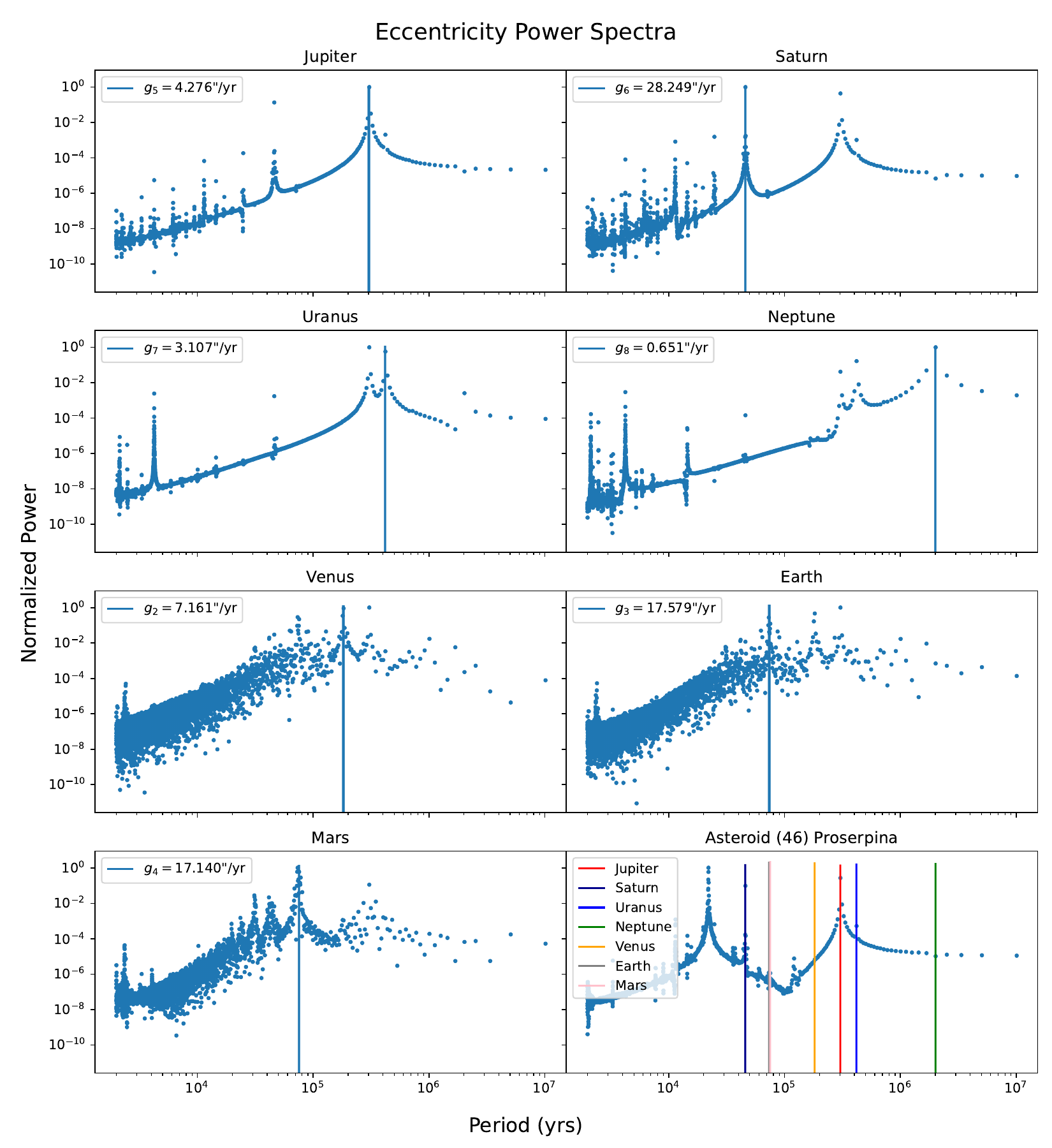}
    \caption{The planetary Fourier transformed power spectrum in $e$, and the identified peak secular frequencies identified by SBDynT from the power spectra. All of the frequencies are also shown for an asteroid signal.}
    \label{fig:planet_ecc}
\end{figure*}

\begin{figure*}[htb]
    \centering
    
    \hspace*{-1cm}\includegraphics[width=\textwidth, height=1.25\textheight, keepaspectratio]{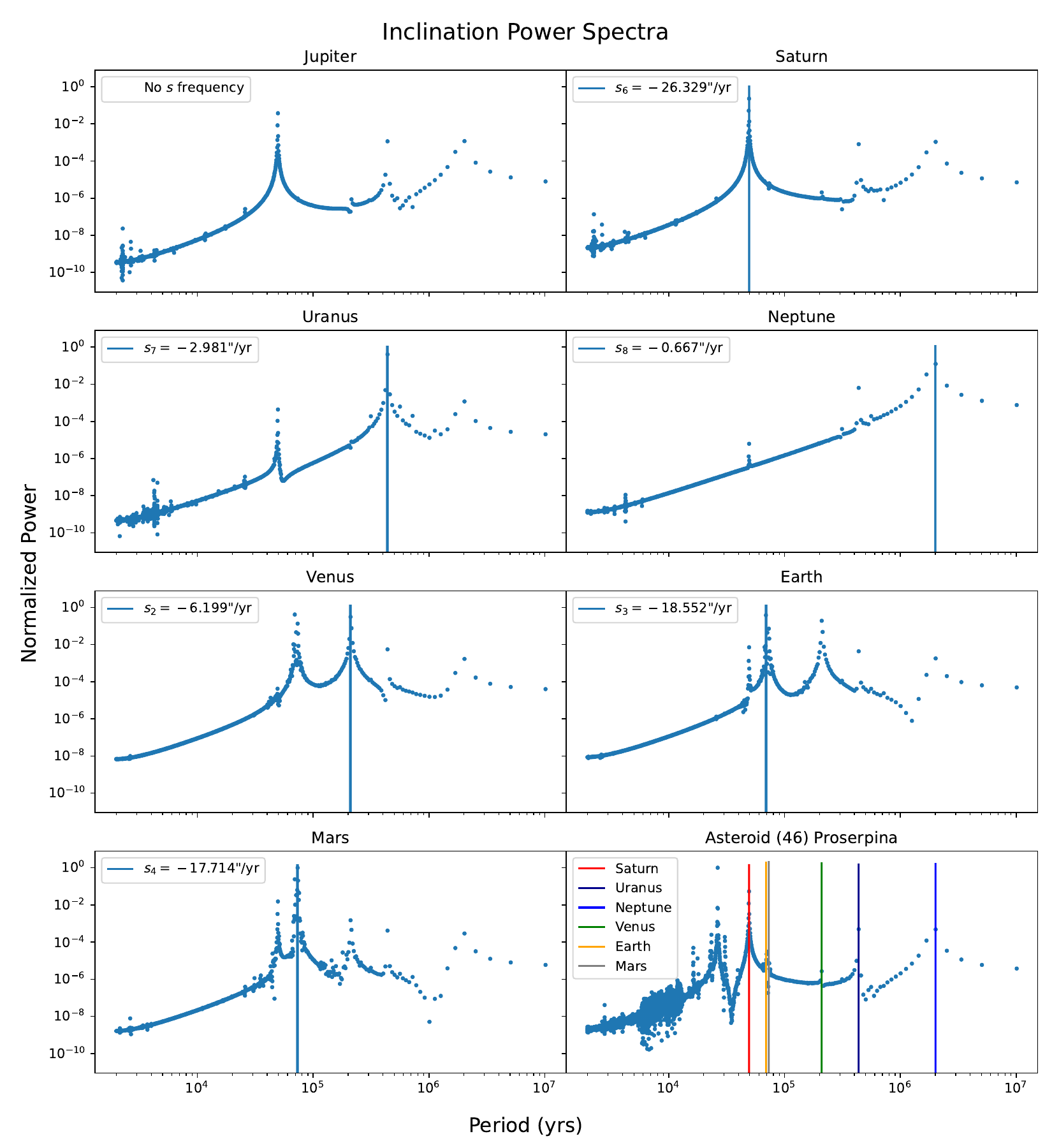}
    \caption{The planetary Fourier transformed power spectrum in $I$, and the identified peak secular frequencies identified by SBDynT from the power spectra. All of the frequencies are also shown for an asteroid signal. The x-axis has been modified to represent the negative frequencies.}
    \label{fig:planet_inc}
\end{figure*}

Examples of the planetary secular frequencies found by SBDynT for the major planets and an example asteroid are shown in Figures~\ref{fig:planet_ecc} and \ref{fig:planet_inc}. 
Table~\ref{tab:sec_freqs} lists the specific values identified in our default integrations compared to prior works.
Note that SBDynT's method of determining peak frequencies identifies the correct dominant signal even when bodies are strongly affected by perturbing planets.
Uranus' $e$ signal is a good example; the peak frequency in the Fourier transform initially corresponds to Jupiter's $g_5$; first determining and then filtering out Jupiter and Saturn's signals allows the correct $g_7$ to instead be identified.

\clearpage

\newpage

\begin{table}[htb]
    \centering
    \begin{tabular}{|c|c|c|c|c|c|c|}
    \hline
        & $g_i$ & BvW 1950 & SBDynT & $s_i$ & BvW 1950 & SBDynT\\
       \hline 
       Venus & $g_2$ & 7.436 "/yr & 7.146 "/yr& $s_2$ & -6.570 "/yr & -6.214 "/yr\\
       Earth & $g_3$& 17.331 "/yr & 17.637 "/yr& $s_3$ & -18.746 "/yr & -18.735 "/yr\\
       Mars & $g_4$& 18.004 "/yr & 16.999 "/yr& $s_4$ & -17.636 "/yr & -17.752 "/yr\\
       \hline 
       & $g_i$ & MK 1994 & SBDynT & $s_i$ & MK 1994 & SBDynT\\
       \hline
       Jupiter & $g_5$& 4.245 "/yr & 4.277 "/yr& $s_5$ & ... & ...\\
       Saturn & $g_6$& 28.236 "/yr & 28.253 "/yr& $s_6$ & -26.314 "/yr & -26.338 "/yr\\
       Uranus & $g_7$& 3.087 "/yr & 3.110 "/yr & $s_7$ & -2.993 "/yr & -2.981 "/yr\\
       Neptune & $g_8$& 0.672 "/yr & 0.648 "/yr & $s_8$ & -0.692 "/yr & -0.656 "/yr\\
       \hline
    \end{tabular}
    \caption{Secular frequencies of the terrestrial planets determined by analytical theory in \cite[][BvW 1950]{Brouwer:1950}, and the giant planets synthetic analysis by \cite[][MK 1994]{Milani:1994}, as compared to values produced by SBDynT in a 10 Myr asteroid simulation. Most differences are minor, and are a consequence of a difference in frequency resolution. \\We note that the frequencies for Uranus and Neptune are strongly affected by the length of the integration, with more accurate values being achieved for the TNO simulations, where their impact is much more significant. We demonstrate in Figures \ref{fig:planet_ecc} and \ref{fig:planet_inc} that these frequencies remain close fits to the secular frequencies as contained in the asteroid signal for this shorter-length simulation, which allows filtering to still remain accurate. }
    \label{tab:sec_freqs}
\end{table}

\newpage
\section{Linear and Non-linear Secular Frequencies in Filtering}
\label{freq_list_app}

Here we describe the linear and non-linear secular frequencies filtered out by SBDynT for each small body, as well as the secular frequencies which are protected for being inherently part of the proper element signal. 

Linear terms are those which are composed of a single precession frequency for the small body and the matching planetary frequency, (i.e. $g-g_5, s-s_6$).
Non-linear terms make up effectively all other terms which combine multiple precession frequencies and/or multiple planetary frequencies (i.e., $g+g_5-2g_6$, $s-s_6-g_5+g_6$, $g+s-g_8-s_8$).

Non-linear secular frequencies thus typically originate from higher-order terms in the expansion of the equations of motion for an orbiting body, specifically from the disturbing function, though many of the most significant non-linear terms in the Solar System are often lower-order. 
One need only look as far as the first three terms of the direct part of the disturbing function to see where several of the terms listed below come from (the second and third zeroth order terms taken from \cite{mandd1999}):
\begin{equation}
(j\lambda - j\lambda' + \bar\omega' - \bar\omega) 
\;\times\; \left(ee'f_{10} + e^3e'f_{11} + ee'^3f_{12} + ee'(s^2 + s'^2)f_{13}\right)
\label{df_2}
\end{equation}
\begin{equation}
(j\lambda' - j\lambda + \Omega' - \Omega) 
\;\times\; \left(ss'f_{14} + ss'(e^2 + e'^2)f_{15} + ss'(s^2 + s'^2)f_{16}\right)
\label{df_3}
\end{equation}
where 
\begin{align*}
    s &= \sin\left(\frac{I}{2}\right), \quad 
    s' = \sin\left(\frac{I'}{2}\right), \quad 
    D \equiv \frac{d}{d\alpha} , \quad 
    \alpha = \frac{a}{a'} \\[5pt]
    f_{10} &= \frac{1}{4}\left[2 + 6j + 4j^2 - 2\alpha D - \alpha^2 D^2\right]A_{j+1} \\
    f_{11} &= \frac{1}{32}\left[
        -6j - 26j^2 - 36j^3 - 16j^4 + 6j\alpha D + 12j^2 \alpha D
        - 4\alpha^2 D^2 + 7j\alpha^2 D^2
        + 8j^2 \alpha^2 D^2 - 6\alpha^3 D^3 - \alpha^4 D^4
    \right]A_{j+1} \\
    f_{12} &= \frac{1}{32}\left[
        4 + 2j - 22j^2 - 36j^3 - 16j^4 - 4\alpha D + 22j\alpha D
        + 20j^2 \alpha D - 22\alpha^2 D^2
        + 7j\alpha^2 D^2 + 8j^2 \alpha^2 D^2 \right.\\
        & \hspace{12pt} - 10\alpha^3 D^3 - \left.\alpha^4 D^4
    \right]A_{j+1} \\
    f_{13} &= \frac{1}{8}\left[-6j\alpha - 4j^2 \alpha + 4\alpha^2 D + \alpha^3 D^2\right](B_j + B_{j+2}) \\
    f_{14} &= \alpha B_{j+1} \\
    f_{15} &= \frac{1}{4}\left[2\alpha - 4j^2\alpha + 4\alpha^2 D + \alpha^3 D^2 \right]B_{j+1} \\
    f_{16} &= \frac{1}{2}\left[
        -\alpha B_{j+1}
        + 3(-\alpha^2)C_j
        + \frac{3}{2}(-\alpha^2)C_{j+2}
    \right]
\end{align*}
There are several non-linear terms in these zeroth order equations, such as $ee's'^2$, which would give rise to coupled $g_i + 2s_i$ terms in the orbital expansion, which should be filtered out if present in the orbital motion.

Non-linear terms may also exist within just the proper motion of the small body, such as the $2g-2s$ term, which is the 5th direct term within the zeroth-order expansion of the disturbing function from \cite{mandd1999}:
\begin{equation}
\cos(j\lambda' - j\lambda + 2\varpi - 2\Omega) 
\;\times\; e^2s^2f_{18}
\label{df_5}
\end{equation}
where 
\begin{align*}
    f_{18} &= \frac{1}{16}\left[12\alpha-15j\alpha+4j^2\alpha+8\alpha^2D - 4j\alpha^2 D+\alpha^3D^2\right]B_{j-1} \\
\end{align*} 
This term implies that while we often treat the eccentricity and inclination evolution as separable at the lowest-order on secular timescales, strong coupling is still present for large enough coefficients of $f_{18}$, for small bodies with large $e$ or $I$, or for small enough cosine arguments of $j\lambda'-j\lambda+2\varpi-2\Omega$.
The assumption that eccentricity and inclination can remain separable when computing proper elements is only appropriate for small bodies which experience a small \ref{df_5} term, and becomes an invalid assumption when considering mean motion resonant objects, vZLK objects (where $2\varpi-2\Omega\equiv\omega\approx0$), or objects which orbit closely to the considered perturber, such as is the case for Centaurs in the outer Solar System.

In their computation of synthetic proper elements, \cite{Knezevic:2000} perform filtering for up to 4th-order secular frequency terms, which are generally weak, but can become significant for objects that have high eccentricities or inclinations, or are inside or very near specific non-linear secular frequencies. 
We follow their approach of considering frequencies up to 4th-order for our filtering. 
Table~\ref{tab:ecc_filt} lists the non-linear secular frequency combinations we filter from the eccentricities of small bodies; we also list example specific values for these frequencies from one of our integrations.
Similarly, Table~\ref{tab:inc_filt} lists the non-linear frequency combinations filtered from the inclinations of small bodies and Table~\ref{tab:ei_filt} lists frequencies filtered from both eccentricity and inclination.
We note that the secular frequencies for planets contained in the simulation are also always targeted for removal, thus accounting for all of the linear secular terms from the planets.

Only the proper frequency secular terms for the small body are protected during filtering.
While many objects may display signal among the higher order $n\times g,s$ frequencies, these are generally forced terms that appear due to commensurability with planetary secular terms, and should be filtered out to properly retrieve the proper motion.

\begin{table}[h]
    \centering
    \begin{tabular}{|c|c|c|c|}
    \hline
        Secular Frequency Terms & Frequency (1/yr) & "/yr & Period (yrs) \\
        \hline
        
$2g_6-g_5$ & $4.030*10^{-5}$& 52.223 & 24816 \\
        $2g_5-g_6$ & $-1.520*10^{-5}$& -19.699 & -65788 \\
        $2g_7-g_6$ & $-1.700*10^{-5}$& -22.032 & -58822 \\
        $2g_6-g_7$ & $4.120*10^{-5}$ & 53.395 & 24271 \\
        $3g_6-2g_5$ & $5.880*10^{-5}$& 76.204 & 17006 \\
$g_6-g_5+g_7$ & $2.089*10^{-5}$ & 27.086 & 47847\\
        $g_6+g_5-g_7$ & $2.270*10^{-5}$ & 29.419 & 44052 \\
        $2g_6-g_5+s_6-s_7$ & $2.230*10^{-5}$ & 28.896 & 44850 \\
        $g_5+s_7-s_6$ & $2.130*10^{-5}$ & 27.610 & 46940 \\

        \hline   
    \end{tabular}
    \caption{Non-linear secular frequencies that are specifically targeted for removal during filtering of the eccentricity vector, as these frequencies take the form of $g-f_i$. Specific frequency values were produced from a 10 Myr simulation of an asteroid with outputs every 500 years; some frequencies may diverge those determined from analytical theory.}
    \label{tab:ecc_filt}
\end{table}

\begin{table}[h]
    \centering
    \begin{tabular}{|c|c|c|c|}
    \hline
    Secular Frequency Terms & Frequency (1/yr) & "/yr & Period (yrs) \\
        \hline
        $s_6+s_7$ & $-2.260*10^{-5}$ & -29.295 & -44,203\\
        $s_6+s_8$ & $-2.082*10^{-5}$ & -26.970 & -48,053\\
        $s_7+s_8$ & $-2.810*10^{-6}$ & -3.637 & -356,370\\
        
        $2s_6-s_7$ & $-3.831*10^{-5}$ & -49.647 & -26,104\\
        $2s_6-s_8$ & $-4.010*10^{-5}$ & -51.972 & -24,936\\
        $2s_7-s_6$ & $1.570*10^{-5}$ & 20.352 & 63,678\\
        $2s_7-s_8$ & $-4.094*10^{-6}$ & -5.306 & -244,263\\
        $2s_8-s_7$ & $1.288*10^{-6}$ & 1.669 & 776,471\\

        $2g_5-s_6$ & $2.690*10^{-5}$ & 34.867 & 37169\\
        $2g_5-s_7$ & $8.900*10^{-5}$ & 11.534 & 112,361\\
        $2g_6-s_6$ & $6.390*10^{-5}$ & 82.819 & 15,648\\
        $2g_6-s_7$ & $4.589*10^{-5}$ & 59.486 & 21,786\\
        
        $g_5 + g_6 - s_6$ & $4.540*10^{-5}$ & 58.843 & 22,024\\
        $g_5 + g_6 - s_7$ & $2.740*10^{-5}$ & 35.510 & 36,496\\
        $g_5 - g_6 + s_6$ & $-3.880*10^{-5}$ & -50.290 & -25,770\\
        
\hline
    \end{tabular}
    \caption{Non-linear secular frequencies that are specifically targeted for removal during filtering of the inclination vector, as these frequencies take the form of $s-f_i$. Specific frequency values were produced from a 10 Myr simulation of an asteroid with outputs every 500 years; some frequencies may diverge from analytical theory.  }
    \label{tab:inc_filt}
\end{table}

\begin{table}[h]
    \centering
    \begin{tabular}{|c|c|c|c|}
    \hline
        Secular Combination & Frequency (1/yr) & "/yr & Period (yrs) \\
        \hline

        $g_6+s_6$ & $1.496*10^{-6}$ & 1.939 & 668,399\\
        $2g_6+s_6$ & $2.330*10^{-5}$ & 30.191 & 42,925\\
        $3g_6+s_6$ & $4.510*10^{-5}$ & 58.444 & 22,174\\
        
        $\pm g_7+s_7$* & $\pm7.175*10^{-8}$ & $\pm0.093$ & $\pm13,936,917$\\
        $\pm g_8+s_8$* & $\pm1.333*10^{-8}$ & $\pm0.017$ & $\pm74,993,665$\\
        
        $g_5+s_6$ & $-1.700*10^{-5}$ & -22.037 & 58,809 \\
        $g_5+s_7$ & $9.999*10^{-7}$ & 1.296 & 1,000,082 \\
        $g_6+s_7$ & $1.948*10^{-5}$ & 25.272 & 51,282 \\
        $g_7+s_8$ &  $1.894*10^{-6}$ & 2.454 & 528,052\\
        
        $g_6-s_6$ & $4.276\times10^{-5}$ & 54.567 & 23,750 \\
        $g_7-s_7$ & $4.700*10^{-6}$ & 6.091 & 212,774 \\
        $g_8-s_8$ & $1.053\times10^{-6}$ & 1.365 & 949,333 \\

        \hline   
    \end{tabular}
    \caption{Non-linear secular frequencies which are specifically targeted for removal during filtering of both the eccentricity and inclination vectors, as these frequencies take the form of $g+s \pm f_i$, or $g-s \pm f_i$. \\*These are instead computed from $g_i$ and $s_i$ frequencies as reported by a 150 Myr TNO integration, as the higher resolution is necessary to accurately represent these very long terms. These secular frequencies are generally only relevant for outer Solar System small bodies, so the precision of these frequencies does not significantly affect the computation of asteroid proper elements.}
    \label{tab:ei_filt}
\end{table}

\begin{table}[h]
    \centering
    \begin{tabular}{|c|c|c|c|}
    \hline
    Proper Secular Frequency & Frequency (1/yr) & "/yr & Period (yrs) \\
    \hline
$g$ & $4.542\times10^{-5}$ & 58.866 & 22,016\\
$s$  &  $-3.779\times10^{-5}$ & -48.981 & -26,459\\
\hline
       
    \end{tabular}
    \caption{Example Proper Secular Frequencies for asteroid (46) Proserpina. This frequencies are respectively protected during filtering of the eccentricity and inclination vectors.}
    \label{tab:ecc_prot}
\end{table}

\clearpage
\section{Additional Diagnostic Proper Element Catalog Comparisons}
\label{add_figs_app}

Here we further discuss the differences between the computed proper elements among the three catalogs considered in this paper. 
We also provide additional figures demonstrating SBDynT's precision. 

\subsection{Precise Uncertainty Computation}
\begin{figure}[!htb]
   \centering
    \hspace{-.8cm}\includegraphics[width=0.8\textwidth]{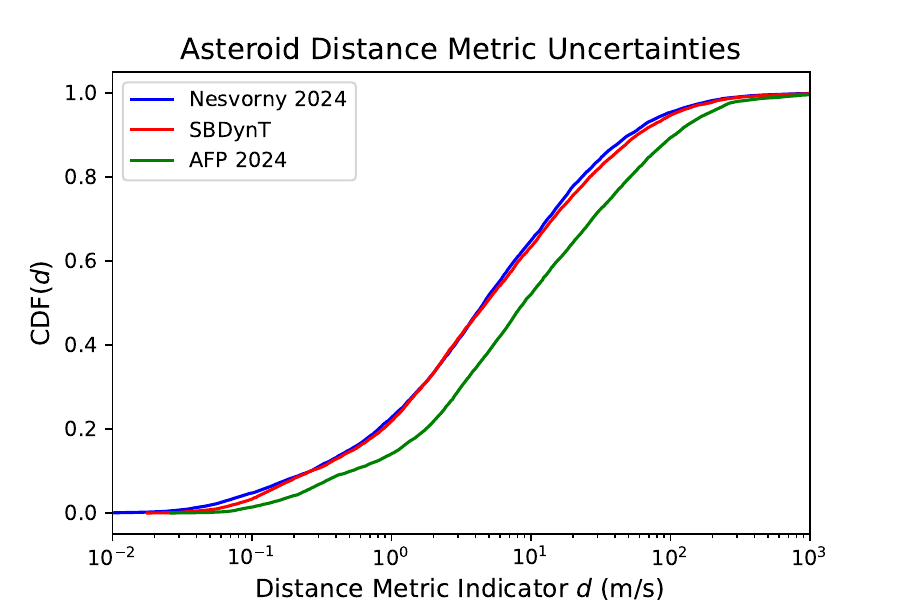}
    \caption{Cumulative distribution of the distance metric computed by calculating the RMS between the five windows in SBDynT's integrations, excluding the proper element of the full length integration. 
    Due to differences in frequency resolution between the entire integration length and the length of the windows, an additional amount of power is filtered out of the $e$ vector, leading to offsets ranging from $10^{-6}-10^{-7}$ in  $e$. 
    Therefore, the lower end of the CDF shrinks somewhat when including the full-length integration in the root-mean square calculation.
    Computing the uncertainties in this way produces a nearly identical distance metric CDF to the \cite{Nesvorny:2024} catalog.}
    \label{fig:hcm_std} 
\end{figure}

In Section~\ref{sec:proper_uncertainties}, we showed the cumulative density function comparing the distance metric computed using the uncertainties from each catalog (Figure \ref{fig:hcm_cdf}).
SBDynT was shown to mostly outperform the Asteroid Families Portal in published uncertainties.
However, they agree for distance metrics below 1 m/s, which require very small uncertainties to achieve (typically on the order of at $10^{-5}$ for $a$, $e$, and $\sin(I)$).
The CDF presented there was computed using the RMS of all five (overlapping) time windows compared to the proper element computed by averaging the filtered values over the full integration. 
In Figure~\ref{fig:hcm_std} we show the same distribution but consider the RMS of the five windows compared the mean of the filtered elements in each window (the filtering is done independently in each window, so the average of the windows is not the same as the average across the entire time-span).
The uncertainties in this case are highly improved for the most stable (smallest distance metric) objects, causing SBDynT to report uncertainties similar to the \cite{Nesvorny:2024} catalog instead. 

A quick investigation of the orbits with the smallest distance metrics indicates that this shift is a feature of our method of computing proper elements and filtering. 
When performing a Discrete Fourier Transform on a time array, the sampling frequency of the Fourier transform is equal to $1 / (dt\times N)$, where $dt$ is timestep, and $N$ represents the length of the array. 
In the case of a full SBDynT asteroid simulation, this corresponds to a sampling frequency resolution of $1/(1000\times 10001) \approx 10^{-7}$.
Each window, however, has a sampling frequency resolution twice that value, because the time array is only half as long; the resulting power spectrum has to be distributed twice as strongly in each bin of the Fourier transform.
The filtering method accounts for this change, but we found that the smoothing process, which redistributes and removes some power from the vector, filters out approximately 1 extra bins-worth of power, perhaps due to the fact that the window is 5001 bins shorter, not 5000.
We have not seen a significant improvement by increasing the length of the windows by one, so we conclude that there is a minor fundamental difference in how the Discrete Fourier transform handles power for like-arrays.

In the end, this causes the windows to consistently have a mean value that is less than the proper element for the full integration window; the difference is typically on the order of $10^{-6}$. 
Because of this, there is no visible effect on the uncertainties reported for most objects.
For objects with highly stable orbits, however, this causes the reported proper element uncertainties we compute to rarely fall below values of $10^{-6}$; there is effectively somewhat of an uncertainty ``floor'' in SBDynT. 

As can be seen in Figure \ref{fig:hcm_cdf}, the reported uncertainties even at this floor match the distribution of uncertainties reported by the Asteroid Families Portal, indicating that the uncertainties are still sufficiently small.
In addition, the number of objects with uncertainties sufficient for family finding remains unchanged, because the uncertainty floor still produces relative velocities for the uncertainties near values of 0.1-1 m/s using the distance metric presented by \cite{Zappala:1990}.
For this reason, we see no reason to further improve the uncertainties. 
However, the stability of our filtering method remains effective, and if only the comparison between like-sized windows is considered, can produce uncertainties that are comparable to the \cite{Nesvorny:2024} catalog.

\subsection{Additional Filtering for the Eccentricity and Inclination Time Arrays}
\label{sec:add_filter}

Mentioned in Section-\ref{sec:calc_proper}, many non-linear frequencies present within the eccentricity and inclination evolution can escape filtering of the $\myvec{e}$ and $\myvec{I}$ vectors.
This is primarily due to phase modulation of those frequencies which occurs during the calculation of the vectors, themselves, as the non-linear frequencies are linear combinations of the proper frequency and the planetary secular frequencies which are present within the small body orbit.

\begin{figure}[!h]
    \centering
    \includegraphics[width=0.6\linewidth]{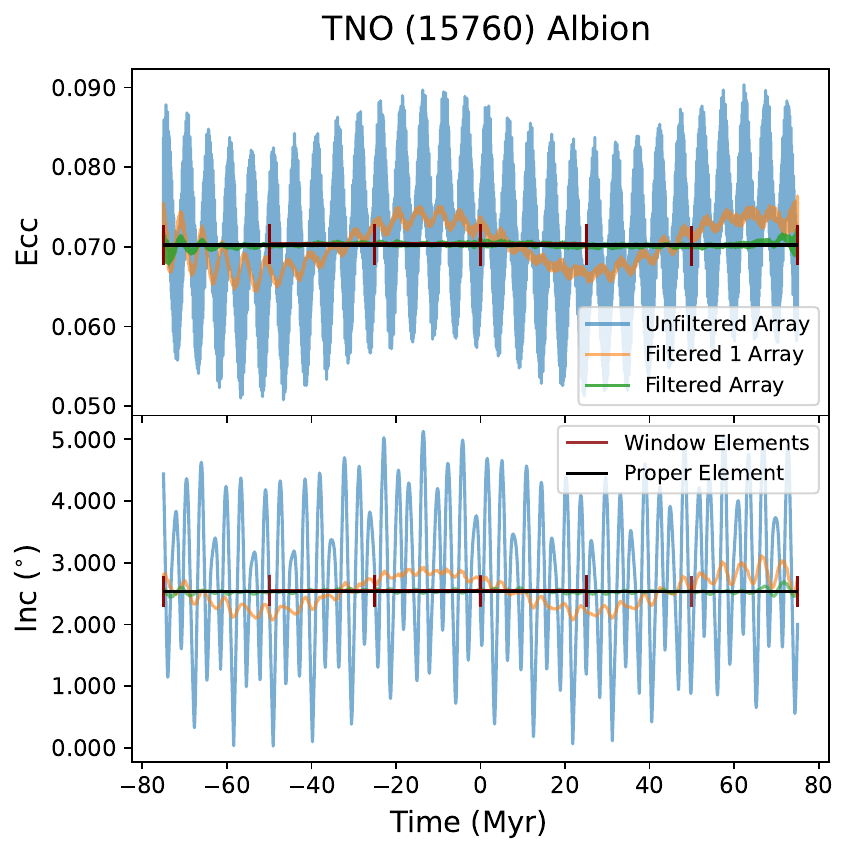}
    \caption{Power spectra of the eccentricity and inclination for TNO (15760) Albion for the unfiltered elements (black), the elements after a single filter on the $e, I$ vectors (blue), and the elements after the additional filter applied directly to the $e,sin(I)$ filtered time arrays (orange). The dominant signals present in the time evolution are primarily contained in non-linear frequencies, as shown by the vertical lines. The initial filter is effective at removing 1st order linear frequencies, but fails to remove higher order signals, especially non-linear signals. An additional filter effectively captures the remaining non-linear terms. Note that signal corresponding most closely with the proper $g,s$ frequencies remains, though the remnant variation is limited.}
    \label{fig:15760_eIarr}
\end{figure}

\begin{figure}[!h]
    \centering
    \includegraphics[width=0.8\linewidth]{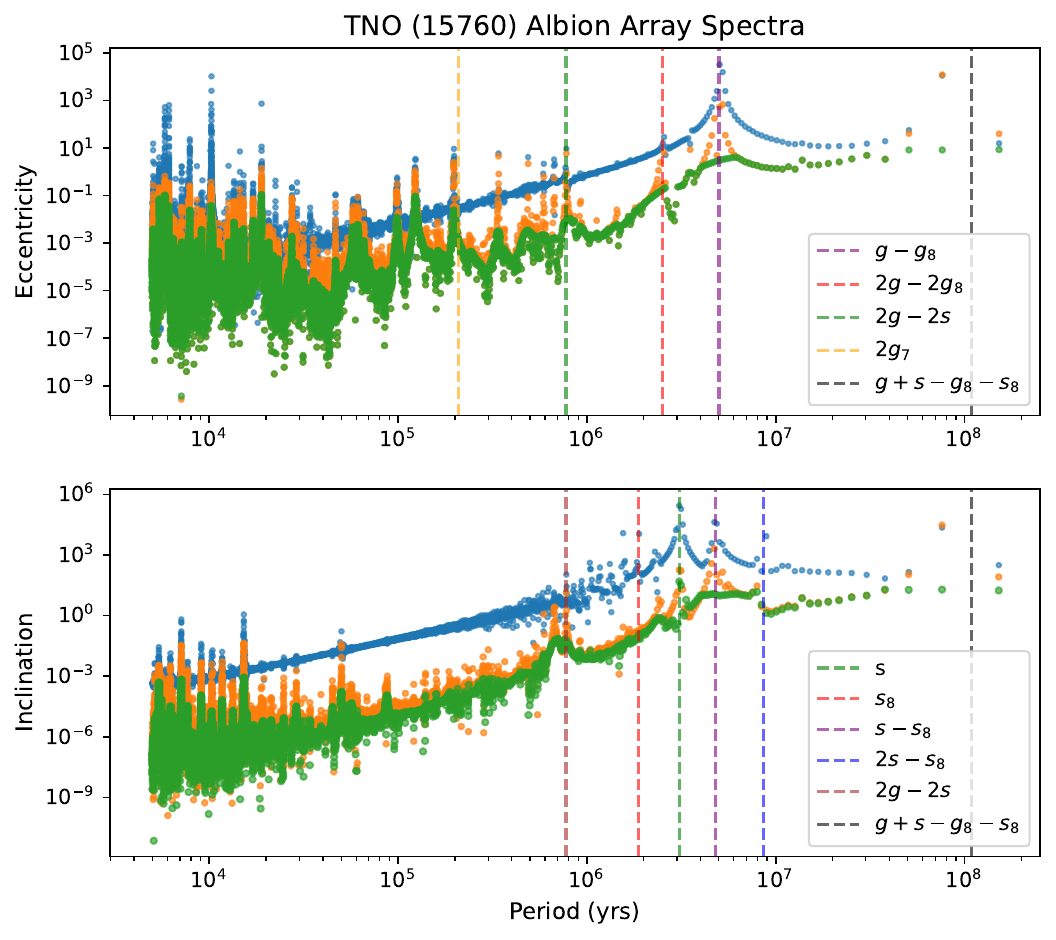}
    \caption{Power spectra of the eccentricity and inclination for TNO (15760) Albion for the unfiltered elements (black), the elements after a single filter on the $e, I$ vectors (blue), and the elements after the additional filter applied directly to the $e,sin(I)$ filtered time arrays (orange). The dominant signals present in the time evolution are primarily contained in non-linear frequencies, as shown by the vertical lines. The initial filter is effective at removing 1st order linear frequencies, but fails to remove higher order signals, especially non-linear signals. An additional filter effectively captures the remaining non-linear terms. Note that signal corresponding most closely with the proper $g,s$ frequencies remains, though the remnant variation is limited.}
    \label{fig:15760_filt_arrs}
\end{figure} 

The is illustrated in the following Figures, which use TNO's (15760) Albion and 2019 QQ110.
We emphasize that while the most dominant non-linear frequency present in the eccentricity and inclination evolution for both objects is near the $g+s-g_8-s_8$, the Arnold web of secular resonances among the TNOs is dense, and so there are many resonances contributing to the overall evolution for these objects. 
For example, we see that before any filtering occurs, Albion contains strong signals in eccentricity at $g-g_8$, $2g-2s$, and $g+s-g_8-s_8$, and in inclination at $s$, $s_8$, $2s-s_8$, $2g-2s$, and $g+s-g_8-s_8$.
There are also a number of integer multiples of these resonant frequencies, which spread from out the non-linear frequency in a sinc function as expected, such as the $2g-2g_8$ frequency in eccentricity.  

In the case of 2019 QQ110, there are many of the same signals present in the eccentricity, with the addition of $g-2g_7$.
In addition, the limited length and resolution of the integration causes the reported $g+s-g_8-s_8$ frequency to not correspond directly with the longest-period term; instead the closest real term is reported as a integer multiple of $g+s-g_8-s_8$.

In both cases, the non-linear resonance causes the relevant long-period terms to be modulated in the full vector form into the proper $g,s$ and planetary $g_8,s_8$ frequencies themselves.
As the proper $g,s$ frequencies are protected during the initial filter on the $\myvec{e}, \myvec{I}$ vectors, the signals corresponding to these long-period terms are protected.

Mentioned previously, SBDynT applies an additional filter to the filtered $e,I$ arrays themselves which still contain these non-linear signals. 
While this has no effect on the final reported proper elements themselves, this does have an impact on the reported uncertainties corresponding to the proper elements for these near-secular resonant objects.

\subsection{Short-Period Filtering Residuals}
\label{sec:short_period}

\begin{figure*}
    \centering
    \includegraphics[width=0.9\linewidth]{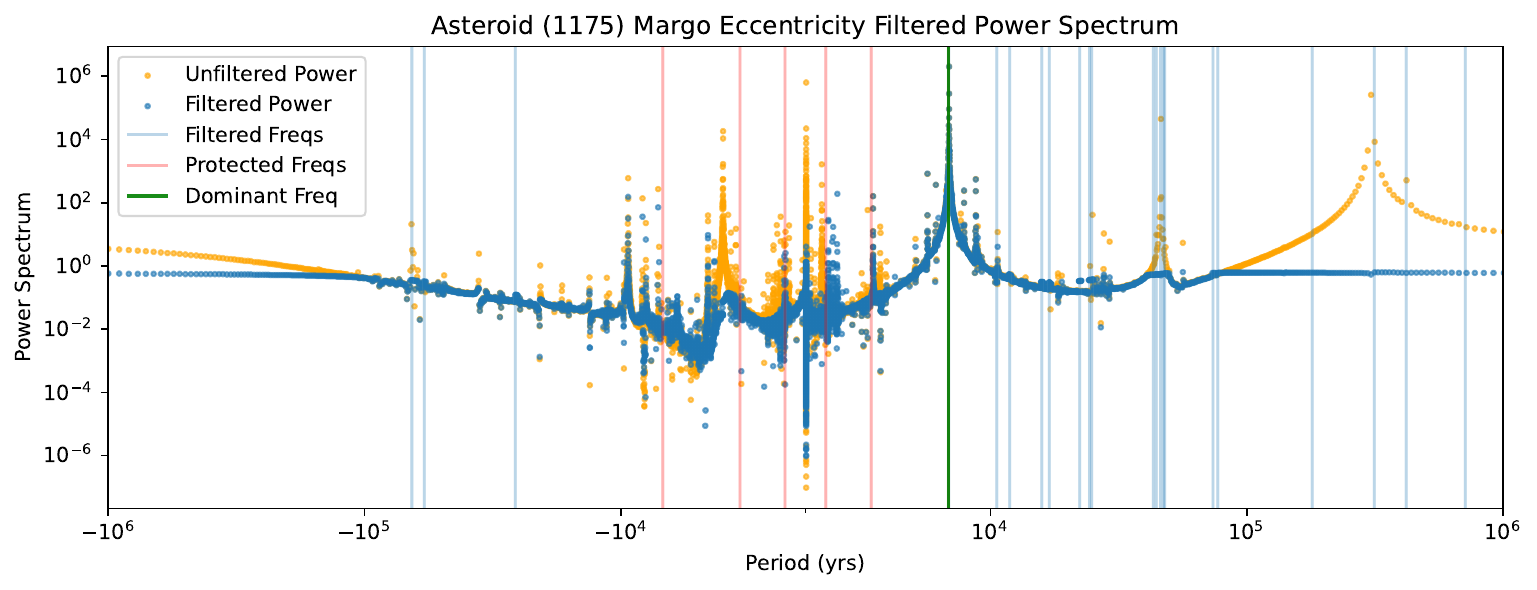}
    \caption{Unfiltered and filtered eccentricity power spectrum for Asteroid (1175) Margo. With an orbit near the 1:2 mean motion resonance with Jupiter, short period terms from Jupiter that lie below the Nyquist frequency are significant contributors to the total power. In this case, a short-period term is produced near -1800 years; this term is the second largest contributor to the power of the eccentricity vector. The x-axis is limited to $\pm 1$ Myr to better display the short period frequencies. Note a number of unfiltered long-period frequencies near -10,000 years, which may also correspond to non-linear forced terms with the planets.}
    \label{shortperiodfreq}
\end{figure*}

As discussed in Section~\ref{sec:calc_proper}, filtering out the short-period frequencies for a small body typically has negligible effects on the final reported proper elements.
However, there are cases near mean-motion resonances where short-period terms are amplified both by the small body's proximity to the resonance and times when they are amplified by the proximity of the short-period terms to the Nyquist frequency of SBDynT's simulation output sampling.
We found specific cases, such as low-eccentricity outer belt asteroids near the 1:2 mean-motion resonance with Jupiter, where short-period terms contribute as much as 1-2\% of the total power of the entire signal and our filtering reduces the proper eccentricity by $\sim1$\% compared to catalogs that don't filter short-period terms.
Figure \ref{shortperiodfreq} shows an example for Asteroid (1175) Margo, which has an orbit with a semi-major axis of 3.211 AU, very close to the 1:2 resonance. 

\begin{figure}
    \centering
    \includegraphics[width=0.6\linewidth]{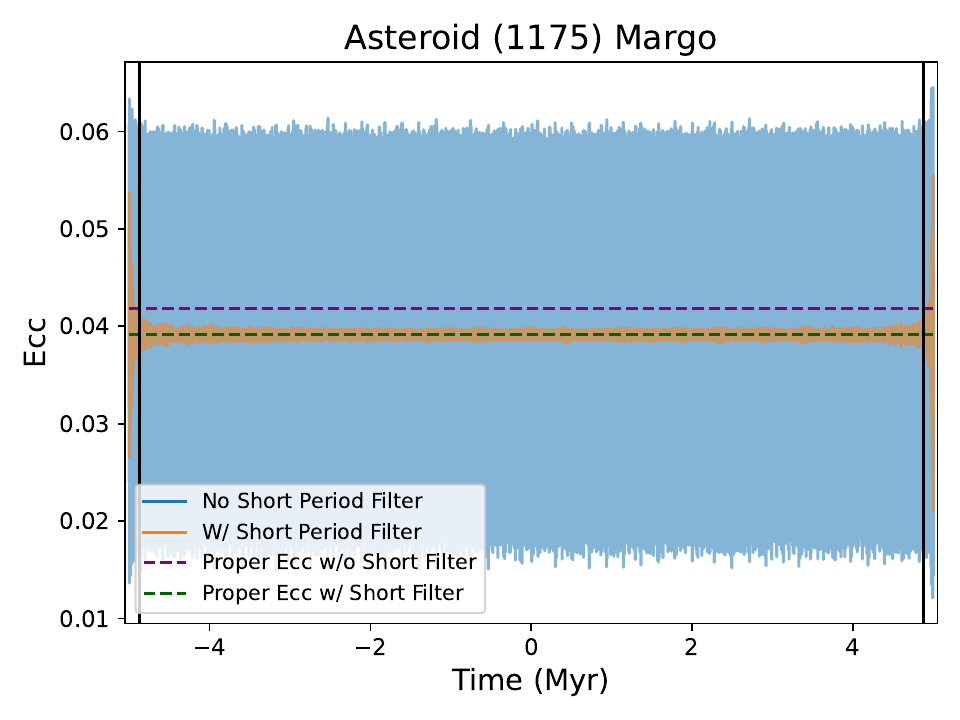}
    \caption{Two different filtered eccentricity time arrays for a forward integration of Asteroid (1175) Margo. Large amplitude terms that might be missed, especially short-period terms related to interactions with Jupiter, can produce small differences in the reported proper eccentricity. Here, filtering out a high-amplitude short-period term lowers the reported proper eccentricity by a value of $2.74\times10^{-3}$. With a reported uncertainty of $2.56\times 10^{-6}$ by \cite{Nesvorny:2024}, not using the short-period filter would result in a large standardized residual of 966 in the reported proper eccentricity between SBDynT and the \cite{Nesvorny:2024} catalog. }
    \label{shortperiodfiltered}
\end{figure}

The proximity to the 1:2 mean motion resonance with Jupiter causes a large short-period term to appear in the power spectra for Asteroid (1175) Margo; in fact, this term is the second largest contributor to the total signal after the dominant, proper term, surpassing even the power added by Jupiter's secular frequency.
This term is therefore the largest forced term in the orbital motion of the asteroid, and needs to be filtered to define the proper eccentricity of the asteroid. 
Figure~\ref{shortperiodfiltered} demonstrates that proper eccentricities calculated with and without filtering out this short-period term will have significant offsets.

\subsection{TNO Residuals for Mean Element Objects}
\label{sec:longperiod_tnos}

As discussed in Section~\ref{ss:tno_pe}, the largest source of residuals between AstDys and SBDynT for the TNOs occur primarily among the Cold Classical objects, which are strongly affected by the $g+s-g_8-s_8$ resonance.
Here we include two examples of such objects, displaying the residuals in the reported eccentricity and inclination due to long-period variation, and discuss how this resonant interaction can produce small offsets in the comptued proper elements.

\begin{figure}
\centering
\begin{minipage}{.48\textwidth}
  \centering
  \includegraphics[width=1\linewidth]{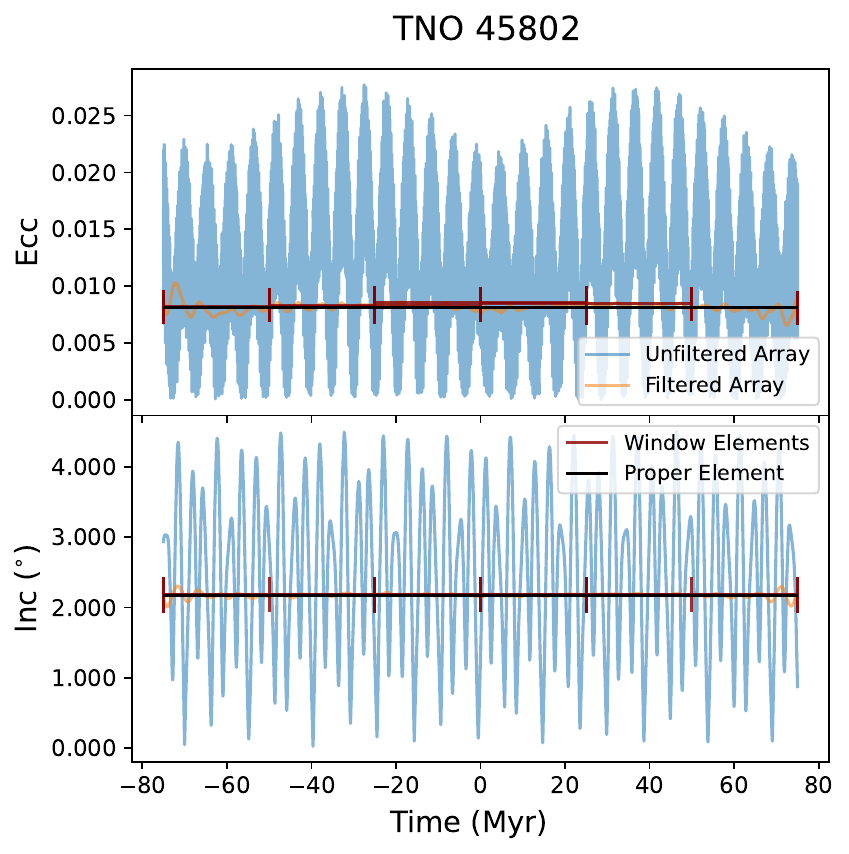}
  \captionof{figure}{Orbital evolution in $e$ and $I$ for TNO (45802) 2000 PV29. Filtering of Neptune's $g_8$ secular frequency and the transformation to the locally forced plane separates the proper $e$ from the mean osculating $e$ significantly.}
  \label{fig:evo_g}
\end{minipage}\hfill
\begin{minipage}{.48\textwidth}
  \centering
  \includegraphics[width=1\linewidth]{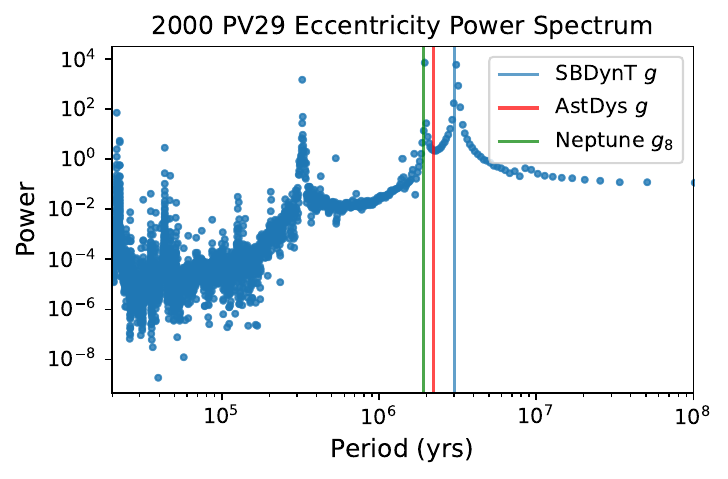}
  \captionof{figure}{The identified g frequencies for TNO (45802) 2000 PV29, with reference to the unfiltered eccentricity power spectrum. The proper $g$ frequency reported by AstDys is much closer to Neptune's $g_8$ frequency than the $g$ reported by SBDynT. This could be the result of differences in the integration sampling frequency, the integration length, or the method of dominant frequency identification between OrbFit and SBDynT.}
  \label{fig:diff_g}
\end{minipage}
\end{figure}

The first case is shown in Figures~\ref{fig:evo_g} for TNO (45802) 2000 PV29; the eccentricity residual between AstDys and SBDynT  is large, with $\Delta e_{prop} \approx 0.0025$.
We find that the reported proper eccentricity from AstDys is actually close to the mean of the unfiltered eccentricity over time, indicating that perhaps the AstDys value avoids filtering out Neptune, or possibly computes the transformation to the locally forced plane in a different way.

We also find difference between the AstDys reported proper $g$ frequency and ours, shown in Figure \ref{fig:diff_g}; AstDys reports a proper g frequency that is nearer to $g_8$ than to the proper $g$ term reported by SBDynT. 
SBDynT avoids filtering out the planetary frequencies when they are too close to the small body's identified proper frequency.
It could be that a similar check occurs in the OrbFit code and was triggered because of its identified $g$ for 2000 PV29, causing it to avoid filtering out Neptune and leading to the large residual between SBDynT and AstDys.
The method for identifying the dominant frequency can lead to different proper elements if a different $g$ or $s$ frequency is identified for a small body.

\begin{figure}
    \centering
    \vspace*{-0.1cm}\hspace*{-0.85cm}\includegraphics[width=0.6\linewidth]{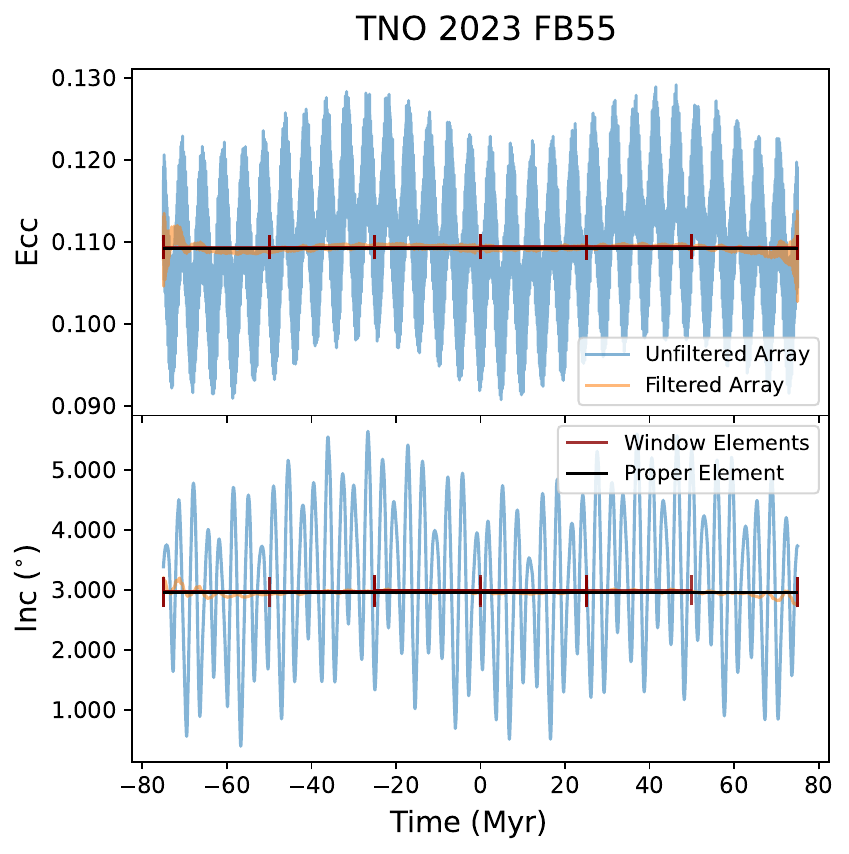}
    \vspace*{-0.25cm}
    \caption{Orbital evolution in $e$ and $I$ for 2023 FB55. Long-period variation occurs in the both the proper inclination and eccentricity due to interactions with the $g+s-g_8-s_8$ non-linear frequency.} 
    \label{fig:2023FB55_evolve}
\end{figure}

\begin{figure}
    \centering
    \includegraphics[width=1\linewidth]{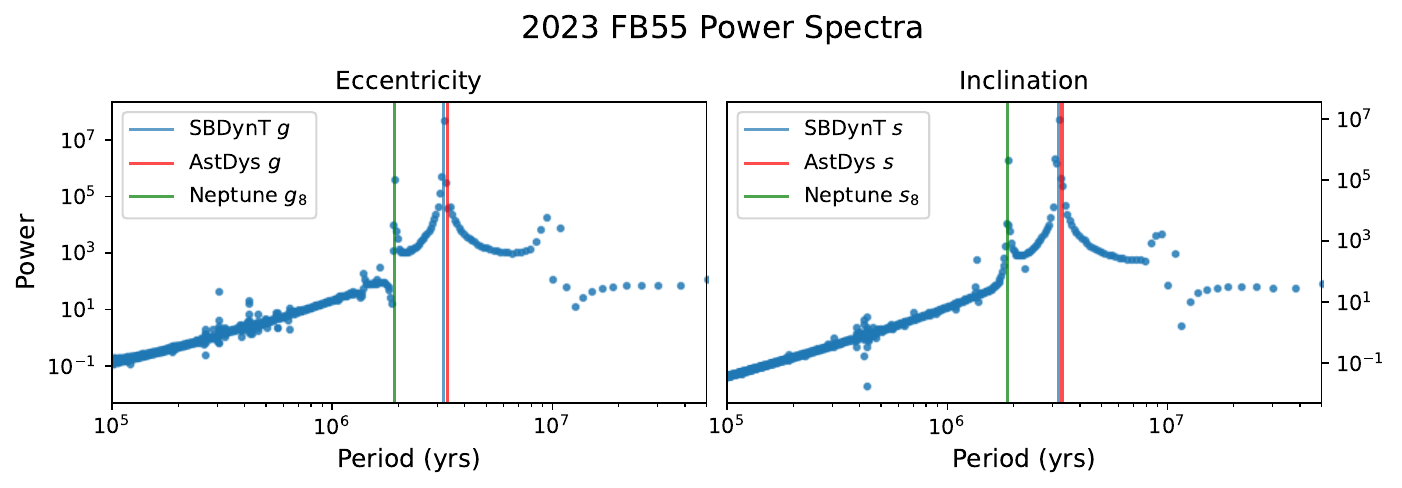}
    \caption{Eccentricity and Inclination power spectra for TNO 2023 FB55. While very close, the $g$ and $s$ proper frequencies reported by SBDynT and AstDys are very slightly offset from each other. SBDynT reports $g= -s\approx0.406$ "/yr, while AstDys reports $(g\approx0.389\, , \, s\approx-0.391)$ "/yr. This approximates to a 1 and 2 bin difference at the resolution of the default SBDynT run.}
    \label{fig:2023fb55_spectra}
\end{figure}

A separate example with long-period terms in inclination can be seen in Figure \ref{fig:2023FB55_evolve}, which experiences long-period variation in both inclination and eccentricity.
Similar to 2000 PV29, there are minor differences between the reported proper $g,s$ frequencies from SBDynT and AstDys, though the differences are much smaller. 
However, we mention that SBDynT reports $g=-s$, while AstDys has a slight separation between the magnitude of the $g$ and $s$ frequencies, likely due to higher spectral resolution. 
However, it is also interesting that the reported AstDys frequencies are both slightly lower frequency values than SBDynT, and that the values are lower by more than a full frequency bin in both cases. 
The uncertainties reported by SBDynT are similar to those reported by AstDys indicating that the resonant interactions induced by the long-period terms are mostly handled to the same degree.
However, minor differences to the reported proper elements which exceed the computed uncertainties themselves indicate some differences remain in our computation of the proper elements.

The last, and perhaps the most significant, factor we find is that the evolution in $a$ produces $44.35 < a_{helio} < 45.05$ AU, with a proper $a_{prop} = 44.68$ AU.
However, AstDys reports an $a_{prop} = 44.21$ AU, which lies outside the range of sampled semi-major axes within the simulation.
We can confirm that sampling the barycentric semi-major axis only decreases the range of sampled $a$, so this could be be the cause of the disagreement.
The most likely explanation for this offset in the reported semi-major axes is that the best-fit orbit used to initialize our simulation was more recently queried from JPL Horizons than the orbit used to compute the synthetic proper elements for AstDys catalog.

This emphasizes how the current method of computing numerical uncertainties for synthetic proper elements is strongly impacted by the certainty of orbital determination. 
While a solar system small body may have an orbit in a very dynamically stable region of space, the numerical certainty of the best-fit orbit proper elements may outperform the orbital certainty.
While asteroid orbits can often be more accurately constrained, this is especially true for TNOs, with orbital periods on the scale of hundreds of years, which require multi-year or even multi-decade long observational arcs to tightly constrain orbital uncertainties.

In fact, querying the covariance matrix for the 2124 objects within the AstDys TNO catalog, we find that only 211 objects have reported orbital uncertainties for $a$, $e$, and $I$ that are all lower than the reported proper element uncertainties by AstDys.
This trend remains true for the full TNO catalog of 5060 objects, with still only 222 objects having reported orbital uncertainties all lower than the reported proper elements uncertainties by SBDynT.

In addition, we find that the reason why the proper element uncertainties exceed the orbital uncertainties in each case is because these objects all have chaotic orbits; as such, the proper element uncertainties are inflated by the scattering orbital evolution of these objects.

This highlights the importance of recomputing proper elements for objects as orbits are more precisely measured, as well as jointly considering the orbital covariance along with the proper element uncertainty when studying an object's dynamical history. The mapping from orbital to proper elements may be smooth enough in some regions of phase space to allow for updates in proper elements without new calculations.

In any case, we report TNO residuals that, on average, filter out the influence of the giant planets more severely than AstDys, leading to proper elements that have been more strongly filtered and thereby lowered. 
This effect is noticeably and primarily systematic, meaning it is consistent across an entire population of dynamically-like objects.
For the purpose of dynamical clustering, then, we consider such a disagreement to be a small issue, as the systematic offset should apply to all objects in a family group, which would cause them to still be clustered together.
We do mention it, however, as an effect to consider in future studies.

\subsection{Standardized Residuals of the Catalogs}
\label{sec:stand_resids}

\begin{figure*}
    \centering
    \hspace*{-1cm}\includegraphics[page=1,width=1.1\textwidth]{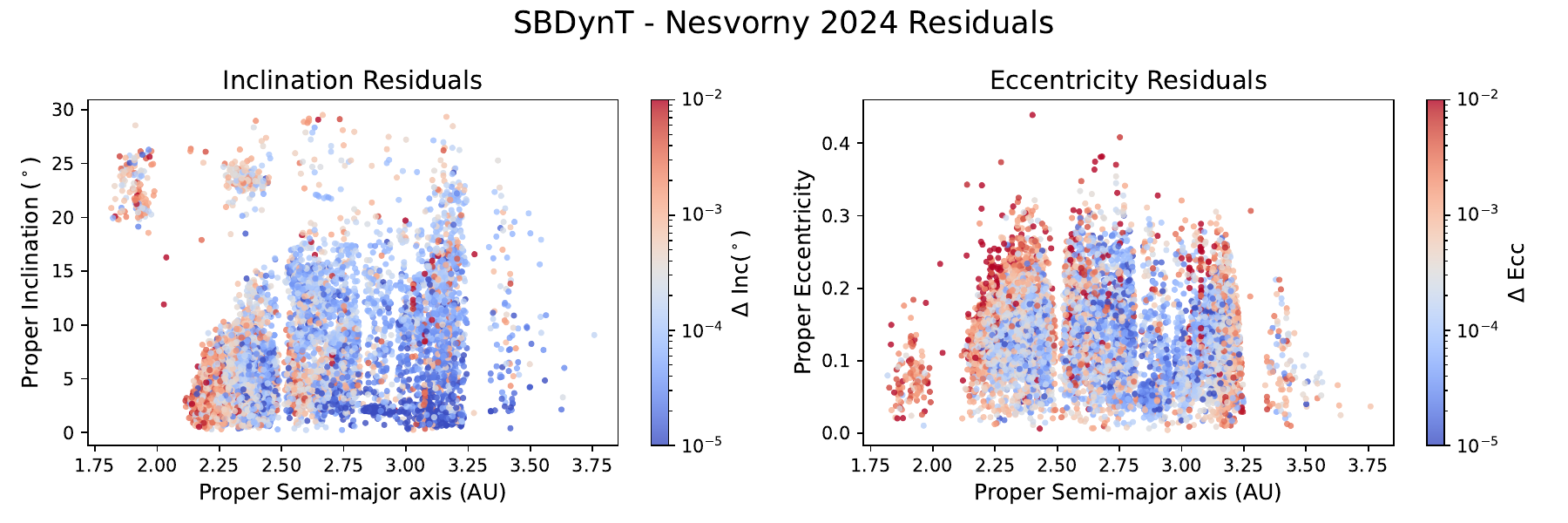}
    \par\vspace{0.75cm}
    \hspace*{-1cm}\includegraphics[page=2,width=1.1\textwidth]{multipage_figures_resid.pdf}
    \par\vspace{0.75cm}
    \hspace*{-1cm}\includegraphics[page=3,width=1.1\textwidth]{multipage_figures_resid.pdf}
    \caption{Residuals comparing the differences in proper $e$ and $I$ between the 3 catalogs.}
    \label{fig:just_resid}
\end{figure*}

\begin{figure*}
    \centering
    \hspace*{-1cm}\includegraphics[page=1,width=1.1\textwidth]{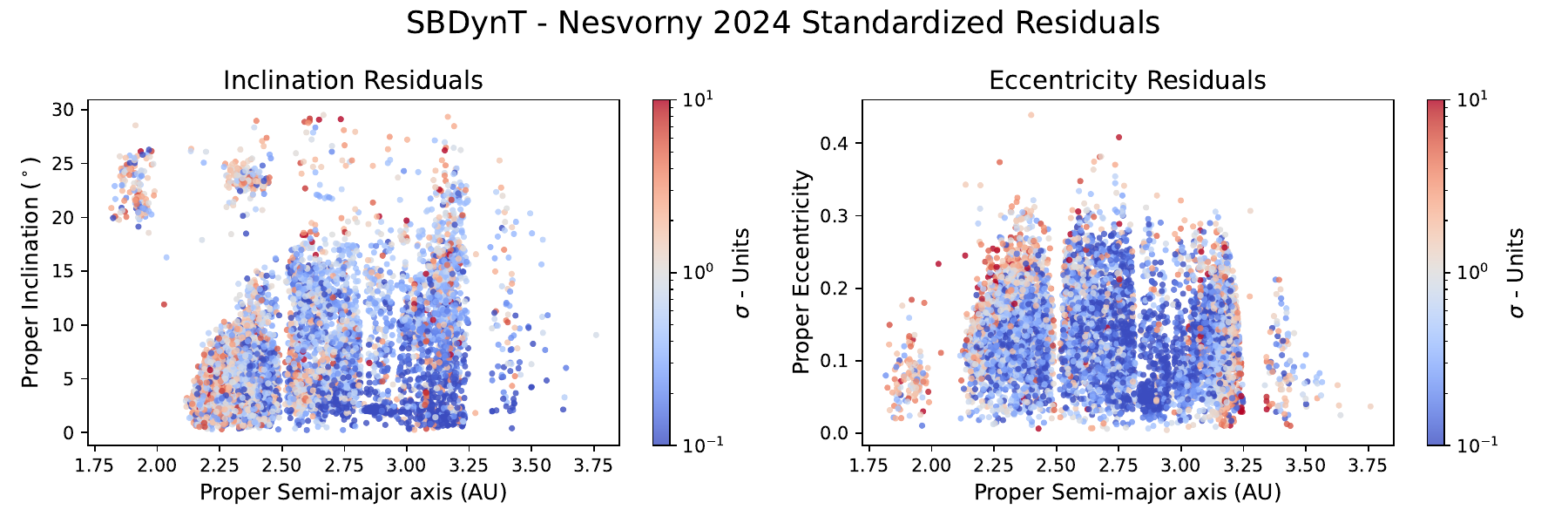}
    \par\vspace{0.75cm}
    \hspace*{-1cm}\includegraphics[page=2,width=1.1\textwidth]{multipage_figures.pdf}
    \par\vspace{0.75cm}
    \hspace*{-1cm}\includegraphics[page=3,width=1.1\textwidth]{multipage_figures.pdf}
    \caption{Standardized Residuals comparing the differences in proper $e$ and $I$ between the 3 catalogs. Colors are limited to show between 0.1 - 10 standard deviations. Several interesting comparisons can be made between each of these figures, as discussed in this appendix. }
    \label{fig:stan_resid}
\end{figure*}

To fully explore the differences between \cite{Nesvorny:2024}, the Asteroids Family Portal \citep{Novakovic:2022}, and SBDynT, we examine the catalog residuals across different regions of the main asteroid belt.
Figures \ref{fig:just_resid} and \ref{fig:stan_resid} show the proper orbital elements of the overlapping set of asteroids color coded by the raw residuals between the catalogs (Figure \ref{fig:just_resid}) and the standardized residuals between the catalogs (the residuals in units of the standard deviation for each individual comparison; Figure~\ref{fig:stan_resid}).

We draw attention to several interesting regions of systematic residuals between the catalogs. 
First, we point out the larger residuals between the catalogs seen in both the $e$ and $I$ residuals in Figure \ref{fig:just_resid} at low inclinations, high eccentricities, from $2.10<a<2.25$ AU. 
This corresponds to the flaring region seen in Figure \ref{fig:CatalogResiduals} where many  proper elements have large uncertainties due to instability in the orbits of asteroids in these regions.
This same region is actually present in Figure 3 of \citep{Nesvorny:2024}, which displays the regions of high uncertainty for the \citep{Nesvorny:2024} catalog, and represents the impact]of the $2g-2g_6+s-s_6$ non-linear secular resonance, which contributes to the residuals at this location.

We note that looking at the standardized residuals for the same region in Figure \ref{fig:stan_resid} comparing SBDynT and the Asteroid Families Portal that the standardized residuals for this region are actually quite well constrained, indicating that this resonance was handled similarly between OrbFit and SBDynT.

We also note in Figure \ref{fig:stan_resid} that there is a reasonably close agreement between the \cite{Nesvorny:2024} catalog and SBDynT results among the general asteroid belt population.
Generally, the SBDynT and \cite{Nesvorny:2024} results seem to agree quite well, barring regions strongly affected by non-linear dynamics, such as the $g+g_5-2g_6$ resonance, which is somewhat visible in the SBDynT - Nesvorny standardized residuals; however, large systematic offsets between the catalogs of the typical asteroids are generally not seen.

The same cannot be said in comparing both catalogs to the Asteroid Families Portal elements, however. 
This is particularly clear in Figure \ref{fig:stan_resid}, comparing both the $e$ and $I$ standardized residuals in the middle and bottom panels.
In the $I$ standardized residuals, there is a clear delineation from 2.5-3.25 AU where the standardized residuals exceed a standard unit for the entire upper-right region.
We find that SBDynT and \cite{Nesvorny:2024} systematically report higher proper inclinations than OrbFit does for this population of asteroids (see Figure \ref{fig:CatalogResiduals}). 
The complex secular resonance structures in this region could contribute to these differences.
Part of this divergent region seems to fall along the edge of the $2g_6-g5$ secular resonance.
This frequency is accounted for in SBDynT's filtering method and is not known to be accounted for in the \cite{Nesvorny:2024} filtering. 
In addition, the $z_2$, $z_3$, $g_5+s_6$, and $g_6+s_6$ frequencies also all fall into this region, which are again all handled by SBDynT's filtering method. 
It could be that the filtering of these frequencies is more severe in the OrbFit code, or that these frequencies regularly lie close to the dominant frequency, in which case they are left untouched.

A similar, but much less prominent, feature is seen in the $e$ standardized residuals for the Asteroid Family Portal from 3.0-3.25 AU, where the upper right population of asteroids differ in their reported eccentricity; this feature is not present in the residuals comparing the SBDynT and \cite{Nesvorny:2024} catalogs. 
It seems like that these two features could be related, as at high inclination, there begins to be some coupling between the eccentricity and inclination evolution.

Despite this, the weighted RMS offsets between the catalogs remain quite good, though this systematic offset in the computed proper inclination between the Asteroid Families Portal and the other catalogs is surprising, because each study filters according to the same general theory.
In addition, while the overall approaches to frequency filtering by each catalog may be somewhat different, they each have a mathematically similar basis rooted in Fourier theory. 

Lastly, we note the region near the 1:2 mean motion resonance with Jupiter, where higher residuals are visible in each comparison plot. 
In particular, we see that the more-stable low-$e$ orbits in this region actually display the highest standardized residuals, which may appear counter-intuitive.
This effect is discussed in Section~\ref{sec:short_period}, and is related to the filtering of the short-period frequency terms caused by the orbital proximity to Jupiter's 1:2 mean motion resonance.

We mention this as a reminder that near resonances, there could still exist many systematic errors that exceed the numerical uncertainties, which should be taken into account when performing family finding. 

\begin{figure*}
    \centering
    \hspace*{-1cm}\includegraphics[width=1.1\textwidth]{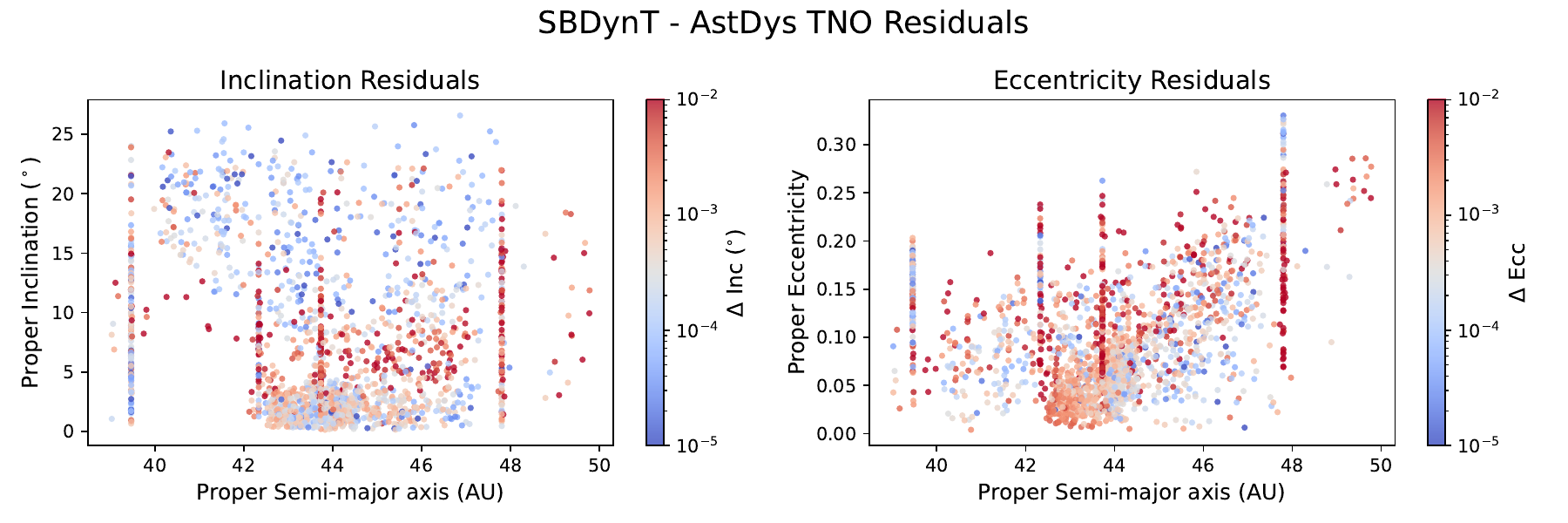}
    \par\vspace{0.75cm}
    \hspace*{-1cm}\includegraphics[width=1.1\textwidth]{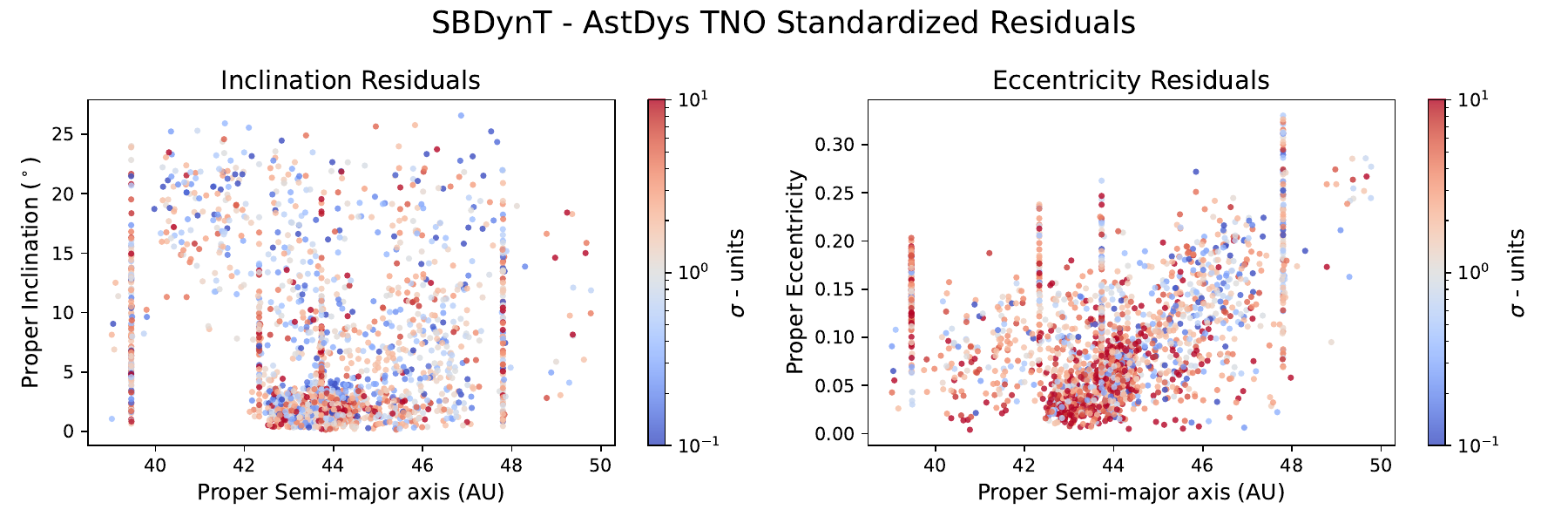}
    \par\vspace{0.75cm}
    \hspace*{-1cm}\includegraphics[width=1.1\textwidth]{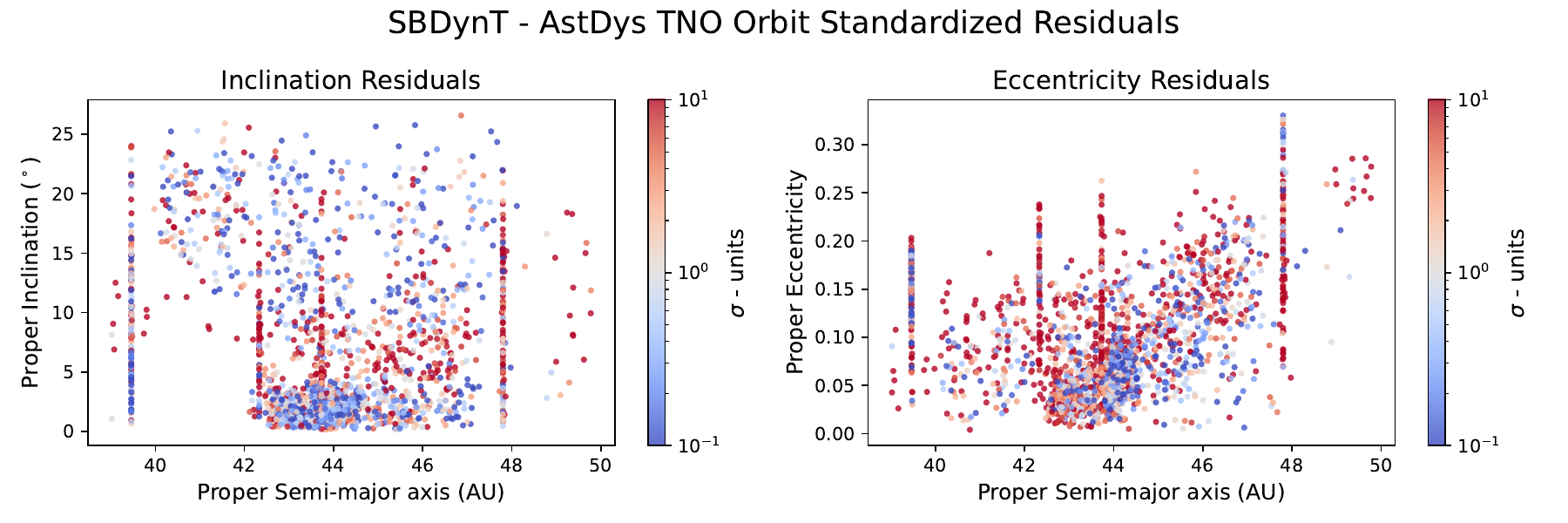}
    \par\vspace{0.75cm}
    
    \caption{Residuals, Standardized Residuals computed from the AstDys uncertainties, and Standardized Residuals computed from the best-fit orbit uncertainties comparing the differences in proper $e$ and $I$ between the AstDys and SBDynT TNO catalogs. Colors are limited to show between 0.1 - 10 standard deviations. Note that the larger orbital uncertainties produce much smaller residuals, indicating that improvements to the best-fit orbits for the TNOs will almost always shift the proper elements significantly to the reported proper element uncertainties.}
    \label{fig:resid_tnos}
\end{figure*}

Similar plots can be made comparing the TNO AstDys catalog to the SBDynT catalog, as shown in Figure-\ref{fig:resid_tnos}, with an additional plot demonstrating the residuals if they are standardized according to the orbital uncertainty for the individual objects, as reported by JPL Horizon \cite{JPLsbdb}, rather than the numerical uncertainty reported by AstDys.

There initially seems to be poorer agreement between the TNO catalogs, especially in the case of the eccentricities.
However, we note that the disagreement among the Cold Classical TNO's in both inclination and eccentricity is generally within the amplitude of long-period variation for secularly resonant objects. 
In addition, the residuals for the Cold Classical TNO's are well within the orbital uncertainties, which are subject to change whenever new observations are made for these objects.
Since we may not have used the same orbital solutions as AstDys,  small changes in initial conditions (relative to the orbital uncertainties) could lead to significant differences (relative to the proper element uncertainties taken from long-term windows in numerical integrations).

Thus we find that the TNO catalogs still agree generally well, but care should be taken to consider both updates to the best-fit orbital certainties and the impact of long-period variation from secular effects, which can cause small changes to the reported proper elements, and which may be large relative to the numerical uncertainties. 

\end{appendices}

\end{document}